\documentclass[journal,twocolumn]{IEEEtran}
\pdfoutput=1
\usepackage{cite}
\usepackage{graphicx}
\graphicspath{{../pdf/}{../jpeg/}}
\usepackage{epstopdf}

\newtheorem{theorem}{\bf Theorem}

\usepackage{amssymb}
\usepackage{dsfont}
\usepackage{multirow}
\usepackage{xcolor}
\usepackage{array}
\usepackage{url}
\usepackage{subfig}
\usepackage{diagbox}
\usepackage{booktabs}
\usepackage{colortbl,booktabs}\definecolor{mygreen}{rgb}{.0,.70,.0}
\usepackage{amsfonts}
\usepackage{verbatim}
\usepackage{setspace}
\usepackage{changepage}
\usepackage{textcomp}
\usepackage{enumerate}
\usepackage{bm}
\usepackage{stfloats}
\usepackage{nomencl}
\usepackage{ifthen}
\renewcommand{\nomgroup}[1]{%
\ifthenelse{\equal{#1}{C}}{\item[\textit{Parameters and Constants}]}{%
\ifthenelse{\equal{#1}{V}}{\item[\textit{Variables}]}{%
\ifthenelse{\equal{#1}{S}}{\item[\textit{Sets and Indices}]}{%
\ifthenelse{\equal{#1}{A}}{\item[\textit{Abbreviations}]}{}}}}
}
\makenomenclature

\makeatletter
\newif\if@restonecol
\makeatother

\let\chapter\section
\usepackage[linesnumbered,ruled,vlined]{algorithm2e}%[ruled,vlined]{
\usepackage{algpseudocode}
\usepackage{amsmath}
\UseRawInputEncoding

\begin{document}
%\title{Bare Demo of IEEEtran.cls\\ for IEEE Journals}
\title{Bidirectional Pricing and Demand Response for Nanogrids with HVAC Systems}
%Energy management and two-way pricing for nanogrids with HVAC units: A Stackelberg game approach
%Optimal energy management and bidirectional pricing for nanogrids with HVAC systems: A Stackelberg game approach
%Price-based demand response for nanogrids with HVAC systems: A Stackelberg game approach
\author{Jiaxin~Cao,
        Bo~Yang,
        %~\IEEEmembership{Senior~Member,~IEEE,}
        Shanying~Zhu,
        %~\IEEEmembership{Member,~IEEE,}
        Kai~Ma,
        %~\IEEEmembership{Member,~IEEE,}
        and~Xinping~Guan
\thanks{This work was supported in part by the National Key Research and
Development Program of China (2016YFB0901900), in part by the NSF of China (61573245). \emph{(Corresponding author: Bo Yang.)}}
\thanks{J. Cao, B. Yang, S. Zhu and X. Guan are with the Department of Automation, Shanghai Jiao Tong University, Shanghai 200240, China, and also with the Key Laboratory of System Control and   Information Processing, Ministry of Education of China, Shanghai 200240, China (e-mail: jiaxincao@sjtu.edu.cn; bo.yang@sjtu.edu.cn; shyzhu@sjtu.edu.cn; xpguan@sjtu.edu.cn).}
\thanks{K. Ma is with the School of Electrical Engineering, Yanshan University, Qinhuangdao 066004, China (e-mail: kma@ysu.edu.cn).}}
%M. Shell was with the Department
%of Electrical and Computer Engineering, Georgia Institute of Technology, Atlanta,

\markboth{LaTex Class Files}%
{Shell \MakeLowercase{\textit{et al.}}: Bare Demo of IEEEtran.cls for IEEE Journals}



\maketitle
\begin{abstract}
Owing to the fluctuant renewable generation and power demand,
the energy surplus or deficit in each nanogrid is embodied differently across time.
To stimulate local renewable energy consumption and minimize the long-term energy cost, some issues still remain to be explored: when and how the energy demand and bidirectional trading prices are scheduled considering personal comfort preferences and environmental factors.
For this purpose, the demand response and two-way pricing problems concurrently for nanogrids and a public monitoring entity (PME) are studied with exploiting the large potential thermal elastic ability of heating, ventilation and air-conditioning (HVAC) units.
Different from nanogrids, in terms of minimizing time-average costs, PME aims to set reasonable prices and optimize profits by trading with nanogrids and the main grid bi-directionally.
In particular, such bilevel energy management problem is formulated as a stochastic form in a long-term horizon.
Since there are uncertain system parameters, time-coupled queue constraints and the interplay of bilevel decision-making, it is challenging to solve the formulated problems.
To this end, we derive a form of relaxation based on Lyapunov optimization technique to make the energy management problem tractable without forecasting the related system parameters.
The transaction between nanogrids and PME is captured by a one-leader and multi-follower Stackelberg game framework.
Then, theoretical analysis of the existence and uniqueness of Stackelberg equilibrium (SE) is developed based on the proposed game property.
Following that, we devise an optimization algorithm to reach the SE with less information exchange. Numerical experiments validate the effectiveness of the proposed approach.
\end{abstract}

\begin{IEEEkeywords}
%%Nanogrid, demand response, HVAC, Stackelberg game, bidirectional pricing
Bidirectional pricing, demand response, HVAC, nanogrid, Stackelberg game
\end{IEEEkeywords}
% \fi
\IEEEpeerreviewmaketitle

%\nomenclature[Cp]{$p_{Di}$}{Active power demand at bus $i$.}
%\nomenclature[Vp]{$p_{Gi}$}{Active power generation at bus $i$.}
\nomenclature[Vp]{$G_{T}^{k}$}{Net energy generation in PME (kWh)}
\nomenclature[Vp]{$E^{k}$}{Energy state of battery unit (kWh)}
\nomenclature[Vp]{$y^{k}$}{Charging or discharging amount of battery unit in PME (kWh)}
\nomenclature[Cp]{$n$}{Total amount of nanogrids}
\nomenclature[Cp]{$C_{b}$}{Battery using cost coefficient (\textcent/(kWh)$^2$)}
\nomenclature[Cp]{$E^{\min}/E^{\max}$}{Minimum/maximum allowable energy state of battery unit (kWh)}
\nomenclature[Cp]{$u^{\rm cmax}/u^{\rm dmax}$}{Maximum charging/discharging rate of battery unit (kWh)}
%\nomenclature[Cp]{$E^{\max}$}{Maximum allowable energy state of battery unit.}
\nomenclature[Cp]{$L_{i}^{\max}$}{Maximum power injection into/exported from nanogrid $i$ (kWh)}
\nomenclature[Cp]{$\varepsilon_{i}$}{HVAC inertial coefficient in nanogrid~$i$}
\nomenclature[Cp]{$\eta_{i}$}{Energy conversion coefficient of HVAC unit in nanogrid $i$ ($^{\circ}$F/kWh)}
\nomenclature[Vp]{$T_{i,out}^{k}$}{Outdoor temperature in nanogrid $i$ ($^{\circ}$F)}
\nomenclature[Vp]{$T_{i}^{k}$}{Indoor temperature in nanogrid $i$ ($^{\circ}$F) }
\nomenclature[Vp]{$e_{i}^{k}$}{Energy consumption of HVAC in nanogrid $i$ (kWh)}
\nomenclature[Cp]{$e_{i}^{\max}$}{Rated power of HVAC unit in nanogrid $i$ (kWh)}
\nomenclature[Cp]{$T_{i}^{\min}/T_{i}^{\max}$}{Lower/upper bounds of comfort temperature level for users in nanogrid $i$ ($^{\circ}$F)}
\nomenclature[Cp]{$T_{i,out}^{\min}/T_{i,out}^{\max}$}{Lower/Upper limits of outdoor temperature of nanogrid $i$ ($^{\circ}$F)}
\nomenclature[Vp]{$RP_{i}^{k}$}{Power generation of small scale uncontrollable DGs in nanogrid~$i$ (kWh)}
\nomenclature[Vp]{$tp_{i}^{k}$}{Power injected into/exported from nanogrid $i$ (kWh)}
\nomenclature[Vp]{$D_{i}^{k}$}{Basic load of nanogrid $i$ (kWh)}
\nomenclature[Vp]{$m_{b}^{k}$}{Buying price of the main grid (\textcent/kWh)}
\nomenclature[Vp]{$m_{s}^{k}$}{Selling price of the main grid (\textcent/kWh)}
\nomenclature[Vp]{$p_{b}^{k}$}{Buying price of the PME (\textcent/kWh)}
\nomenclature[Vp]{$p_{s}^{k}$}{Selling price of the PME (\textcent/kWh)}
\nomenclature[Cp]{$\gamma_{i}$}{Discomfort cost weighting coefficient for users in nanogrid $i$ (\textcent/($^{\circ}$F)$^{2}$)}
\nomenclature[Vp]{$T_{i}^{opt,k}$}{Optimum comfort temperature for users in nanogrid $i$ ($^{\circ}$F)}
%\nomenclature[SO]{$\Omega_{G}$}{Set of generator buses.}
%\nomenclature[SO]{$k$}{Period of the day index in time units [hour].}
\nomenclature[SO]{$k$}{Index of the time slot (hour)}
%\nomenclature[SO]{${\bm{e}}_{i}$}{Strategy space for nanogrid $i$ over entire optimization period}
%\nomenclature[SO]{$\bm{p_{b}}$/$\bm{p_{s}}$/$\bm{y}$}{Strategy space of bidirectional pricing and battery charging over entire optimization period}
\nomenclature[SO]{$\Omega_{ng,i}/\Omega_{\text{PME}}$}{Feasible strategy set for nanogrid~$i$/PME}
\nomenclature[SO]{${\bm{\chi}}^{k}$}{Substitute representation of decision set for PME $\{p_{s}^{k}, p_{b}^{k}, y^{k}\}$}
\nomenclature[Cp]{$V_{i}$}{Weighting parameter for nanogrid $i$ under the Lyapunov optimization framework}
\nomenclature[Cp]{$V_{P}$}{Weighting parameter for PME under the Lyapunov optimization framework}
\nomenclature[Vp]{$H_{i}^{k}$}{State of virtual temperature queue in nanogrid $i$ ($^{\circ}$F)}
\nomenclature[Vp]{$B^{k}$}{State of virtual battery energy queue (kWh)}
\nomenclature[Cp]{$\Gamma_{i}$}{Queue shift parameter related to indoor temperature in nanogrid $i$ ($^{\circ}$F)}
\nomenclature[Cp]{$\theta$}{Queue shift parameter related to battery energy (kWh)}

\nomenclature[AO]{HVAC}{Heating, ventilation and air-conditioning}
\nomenclature[AO]{PME}{Public monitoring entity}
\nomenclature[AO]{SE}{Stackelberg equilibrium}
\nomenclature[AO]{DGs}{Distributed generations}
\nomenclature[AO]{DR}{Demand response}
\nomenclature[AO]{DR}{Demand response}
%\nomenclature[AO]{SG}{Stackelberg game}
\nomenclature[AO]{EMS}{Energy management system}
\nomenclature[AO]{TATD}{Total average temperature deviation}
\printnomenclature[0.71in]
%\printnomenclature
\section{Introduction}
\IEEEPARstart{R}{ecently}, more and more distributed generations (DGs) are integrated into power systems for reducing carbon emissions and long-distance transmission loss~\cite{adefarati2017reliability,jamil2019novel}.
%~\cite{solanki2017sustainable}.
%Microgrid/nanogrid has emerged as an effective energy unit to help transform the power system from a traditional centralized mode to a more distributed mode and improve the reliability and economic effect of the system~\cite{burmester2017review}.
Microgrid/nanogrid has emerged as an effective energy unit with the transformation from a traditional centralized mode into a distributed one making the system more reliable and economic efficient~\cite{burmester2017review}.
%~\cite{6663743,burmester2017review}.
A nanogrid represents a small version of a microgrid, which is a power distribution system for a single house/small building~\cite{sandgani2018energy}. With intelligent communication and power electronics technologies, nanogrid can realize two-way communications and energy flow satisfying users' needs in a more flexible way.
Unfortunately, the intermittent renewable energy and dynamic energy requirements can lead to the mismatch between power supply and demand, which is detrimental to the efficiency of the connected nanogrids \cite{kerdphol2017virtual}.

The existing approaches in maintaining the supply-demand balance are categorized into supply-side management
%(e.g., controlling the output of dispatchable generators \cite{mao2019finite} or dynamic pricing~\cite{karunanithi2017integration}) and demand-side management/demand response (DR)~\cite{wang2017values}.
(e.g.,
scheduling dispatchable generators' output to optimize total generation costs and satisfy users' demand~\cite{mao2019finite} or determining dynamic electricity transaction pricing~\cite{karunanithi2017integration}) and demand-side management/demand response (DR)~\cite{wang2017values}.
%~\cite{haider2016review}.
%The supply-side management usually includes(e.g., controlling the output of dispatchable generators or pricing)
%Having energy management system (EMS) and advanced metering infrastructure installed,
With the emergence of energy management system (EMS) and advanced metering infrastructure,
smart appliances have been developed
%creating more elastic loads
at the consumer side, such as the heating, ventilation and air-conditioning (HVAC) unit~\cite{zheng2014distributed,chen2015modeling,meng2017coordinated}, battery storage system of electric vehicle~\cite{o2018solar}, etc.
%~\cite{mohammadi2018design},
%and~etc.
Their energy consumption can be optimized and adjusted to benefit from dynamic prices set by the external utility.
%and help keep power supply and demand balance, lower carbon emission and reduce the energy generation cost by shifting/shaving the energy consumption from high-peak to low-peak periods \cite{ma2014distributed}.
That is so-called the price-based DR, which has been used in diverse to help maintain the supply-demand balance~\cite{hu2018distributed},
%~\cite{yu2016supply},
lower carbon emissions~\cite{soares2017stochastic} and reduce users' energy bills by shifting/shaving the energy demand from high-peak to off-peak periods \cite{yan2018review}.

In these household appliances, HVAC units account for up to 60\% of total energy consumption, and the elastic nature and the thermal capacity of dwellings signify certain kinds of power storage characteristics of HVAC units. Such features will bring challenges to the implementation of an effective DR.
The reason is that the power demand of HVAC unit is unknown and it introduces the correlation of indoor temperature over time (i.e., the time coupling property).
%The research of this respect is still inadequate.
It has become a meaningful research subject.
Some studies focus on solving such device energy scheduling problems by employing dynamic programming, Monte Carlo~\cite{siano2016assessing} and model predictive control method~\cite{ma2015stochastic}.
For example, \cite{wang2020chance} provides a stochastic model predictive HVAC control scheme cooperating chance constraints to jointly optimize not only the energy use but also thermal comfort with effective utilization of renewables.
These works can minimize the expected energy cost under the assumption that the future parameters can be predicted exactly or the underlying stochastic process is known.
%However it is critical to obtain a good prediction of future state.
However, such makes these works difficult to adapt to the scenarios that exist un-modeled uncertainties or changing probabilities.
%In \cite{ma2014distributed}, to ensure a stable DR, Ma~\textit{et~al}. study the pricing condition and propose an HVAC control algorithm to remove peak load.
Some other works have taken into account the long-term optimal problem for HVAC devices to reduce the variation of energy consumption~\cite{zheng2014distributed}, to minimize the aggregate deviation between zone temperatures and their set points and the total energy cost~\cite{zhang2017decentralized,yu2018distributed} without the system parameter prediction.
%In \cite{zheng2014distributed}, a DR program is developed to reduce the variation of energy consumption by scheduling the on-off status of HVAC unit without the system parameter prediction.
%The authors in \cite{zhang2017decentralized,yu2018distributed} investigate the long-term optimal problem for minimizing the total cost of the energy consumption and the aggregate deviation between zone temperatures and their set points.
It is noted that these related works usually focus on the cost optimization of one side (e.g., the customer side), while any information error of the other side will disturb the predetermined energy strategies and even lead to a new unbalance of power supply and demand.

Alternatively, the existing DR models for both the supply side and demand side are attractive in using market bidding/auction~\cite{li2018optimal},
%~\cite{li2018optimal,nunna2018energy},
game theory~\cite{nekouei2015game,zhang2019credit,motalleb2017non}
%~\cite{nekouei2015game,motalleb2017non,Saad2015Game}
%and other optimization approaches
to investigate the electricity trading behaviors of multi-players.
%behaviors for multi-players both supplier side and demand side.
%In [], an auction model is proposed where the market players are classified into buyers and sellers with an established role.
%For microgrid/nanogrid or similar power systems, they may fail to form a complicated bidding market since the system scale is small.
Recently, Stackelberg game has become a popular approach to handle the sequential decision-making in two-stage problems for independent participants
with different objectives by using the leader-follower structure~\cite{motalleb2019networked}.
Such an approach has been widely used
%for handling the problem of pricing and demand response between an end-user/smart building and an aggregator/utility/system operator.
for modeling the energy trading process
%between an end-user/smart building and an aggregator/utility/system operator
between an end-user and external utility
to solve the problem of pricing and energy management in microgrid or similar systems~\cite{lu2019nash}.
%Previous game theory-based DR studies can be classified into two main streams. The studies in the first stream aim to sequentially optimize the profit of each individual.
For example, Maharjan~\textit{et~al}.~\cite{maharjan2013dependable} have studied the complicated interactions between multiple utility companies and multiple users and aim to maximize the payoffs for both sides in one slot.
Likewise, a real-time price-based energy scheduling problem is formulated as a Stackelberg game model with the objective of balancing supply and demand as well as flattening the aggregated load; the pricing model is given directly with a function of marginal cost~\cite{yu2016supply}.
As an extension, \cite{yu2017incentive} provides a hierarchical structure for a grid operator, multiple service providers, and corresponding customers and proposes a two-loop Stackelberg game to help the operator obtain the required energy from the supply and demand sides with the lowest cost.
These works focus on short-term objectives and may not guarantee the long-term interests of overall systems owing to the uncertainties related to random power generation, demand and etc.
%Besides these applications, the effects of storage unit such as batteries in electric vehicles~\cite{tushar2012economics,zhao2018generalized} or in power generation side~\cite{tushar2015three} are studied by the corresponding game models. And the results indicate that proper charging strategies based on DR programs can increase the flexibility of demand regulation and counter the electrical production fluctuation.
%The studies in the second stream has designed a global objective function with the aim of maximizing the social benefits, which is not suitable for the scenario where the leader and follower belong to different interest groups and pursue their own interests selfishly.
Consequently, several recent works have investigated stochastic dynamic decision processes with game-theoretic framework to tackle these uncertainties in time-coupling problems~\cite{garcia2019modelling}.
%stochastic dual dynamic programming
In~\cite{wu2016stochastic}, the effects of storage units such as batteries on energy management are studied by the corresponding game models.
The electricity cost minimization problem is proposed based on Markov decision process and then solved by the stochastic dynamic programming approach.
But the solution may suffer from the curse of dimensionality when it is implemented in the large-scale user community.
Besides these applications, authors in \cite{shakrina2021stackelberg} have studied a stochastic formulation of game model with a one-leader and N-follower under a real-time pricing demand response scheme where a certain probability function of energy load is adopted.
A scenario-based stochastic energy management with bonus pricing optimization problem has also been proposed in \cite{9078044} to maximize the matching level of users' load and forecasted power generation.
%The results indicate that proper charging strategies based on DR programs can increase the flexibility of demand regulation and counter the electrical production fluctuation.
Differently, authors in~\cite{zhou2017online} have designed a special Stackelberg game model with the receding horizon control strategy to optimize the social benefit and minimize the devices' operation cost concurrently for networked distributed energy resources and customers during each sample time.
Note that the above energy management problems with game model in a long-term optimization period explicitly/implicitly require the statistics information of future parameters or need parameter forecasting and usually ignore a two-way trade pattern.
The energy entities in these works are supposed to play a single kind of predefined role possessing abundant energy or lacking energy all the time.
In fact, the renewable generation is stochastic and the users' demands are dynamic, such that entities may switch back and forth between energy consumers and suppliers across time.
%Note that existing energy management problems with game model optimize in a long-term horizon usually require the statistics information of future parameters and ignore the two-way trade pattern where
%where the energy entities are supposed to play a single kind of predefined role possessing abundant energy or lacking energy all the time.
It is indeed a two-way trade pattern.
However, how to model and solve the corresponding bidirectional pricing problem between players with unfixed roles across time taking account of the residents' different comfort requirements is difficult.
%To the best of our knowledge, there are no existing results reported so far.
The challenges are mainly twofold.
On one hand, the decision-making is coupled among different players across time intervals.
%In addition, the roles of players is unfixed which is affected by market prices, stochastic power generation and load.
Specifically, as mentioned before, the power demand of HVAC units in nanogrid is unknown.
%On the other hand, there are time coupling constraints related with the HVAC consumption and battery  storage, and the future status of system is usually unknown or is difficult to get the accurate value.
On the other hand, there are time coupling constraints and the future status of system is usually unknown or is difficult to get the accurate value.

In this work, to cope with the above issues, we investigate the bilevel energy management problem about two-way real-time pricing and DR in a long period for a public monitoring entity (PME) and nanogrids that can be both a consumer and a supplier during different time slots.
%In this work, to cope with the above issues, we investigate the two-stage problem of bidirectional real-time pricing and energy management in a long period for a public monitoring entity (PME) and nanogrids that can be both a consumer and a supplier during different time slots.
Different from nanogrids, in terms of minimizing the total cost, the PME who has the ability to coordinate the energy demand of nanogrids, aims to set electricity prices and optimize the trading profit.
%Specifically, we propose a bidirectional real-time pricing scheme and energy management strategy in a long period for market participants who can make decisions distributively and independently,
%taking into account the players' economic efficiency and HVAC users' comfort level
%and with no need for any system parameter forecasting and any related statistical knowledge.
The main contributions of this paper are summarized as follows.
\begin{enumerate}[1)]
\item In the setting of a two-way trade pattern, we propose a new three-layer framework where PME can trade energy with nanogrids and the main grid bi-directionally.
    We develop novel individual energy cost and trading profit functions for nanogrids and PME taking into account the bidirectional real-time pricing, random two-way power injection and the thermal discomfort cost of residents in nanogrids.
    %And a one-leader and multi-follower Stackelberg game model is built to capture the interaction between PME and nanogrids that can make decisions independently.
\item With the consideration of uncertainties in system status, the optimization problem is formulated in a long-term horizon where the time-coupling constraints and inter-constraint decision-making\footnote{It indicates the coupling interaction relationship in decision-making between the PME and $n$ nanogrids in the energy management problem, which is specified in \eqref{eq15} and Section~\ref{sec3}.} between nanogrids and PME make the time-average expected model complicated.
    To make such model tractable, we introduce virtual queues and utilize the Lyapunov optimization approach to obtain a relaxed form. Rigorous analysis is provided to show that the solutions to the relaxed one are still feasible to the original one.
    We point out that the proposed approach does not need the knowledge of the prior system statistics.
\item %A one-leader and multi-follower Stackelberg game framework is built to capture the transaction interaction between PME and nanogrids that can make decisions independently.
    The transaction interaction between PME and nanogrids that can make decisions independently is captured by a one-leader and multi-follower Stackelberg game framework.
    The existence and uniqueness of the Stackelberg equilibrium (SE) are proved theoretically. Moreover, we develop an energy management algorithm with only a little of information exchanged between nanogrids and PME, to find the equilibrium iteratively.
\end{enumerate}

The rest of this paper is organized as follows. In Section~\ref{sec2}, we present the system architecture and then formulate the optimization problem.
%The game approach is proposed and further processed in Section~\ref{sec3}, where its performance is also analyzed.
%Solution process for the two-stage energy management problem is developed in Section~\ref{sec3}, where its performance is also analyzed.
Solution process for the bilevel energy management problem is developed in Section~\ref{sec3}, where its performance is also analyzed.
The devised optimization algorithm is shown in Section~\ref{sec4}.
%we process the problem and propose an online algorithm based on game theory.
The simulation results with practical data are provided in Section~\ref{sec5}. Finally, conclusions are given in Section~\ref{sec6}.
%\vspace{-0.35cm}
\section{System Framework and Problem Formulation}\label{sec2}
%\vspace{-0.35cm}
\subsection{System Model}
\begin{figure}[!t]
\centering
\includegraphics[width=3.2in]{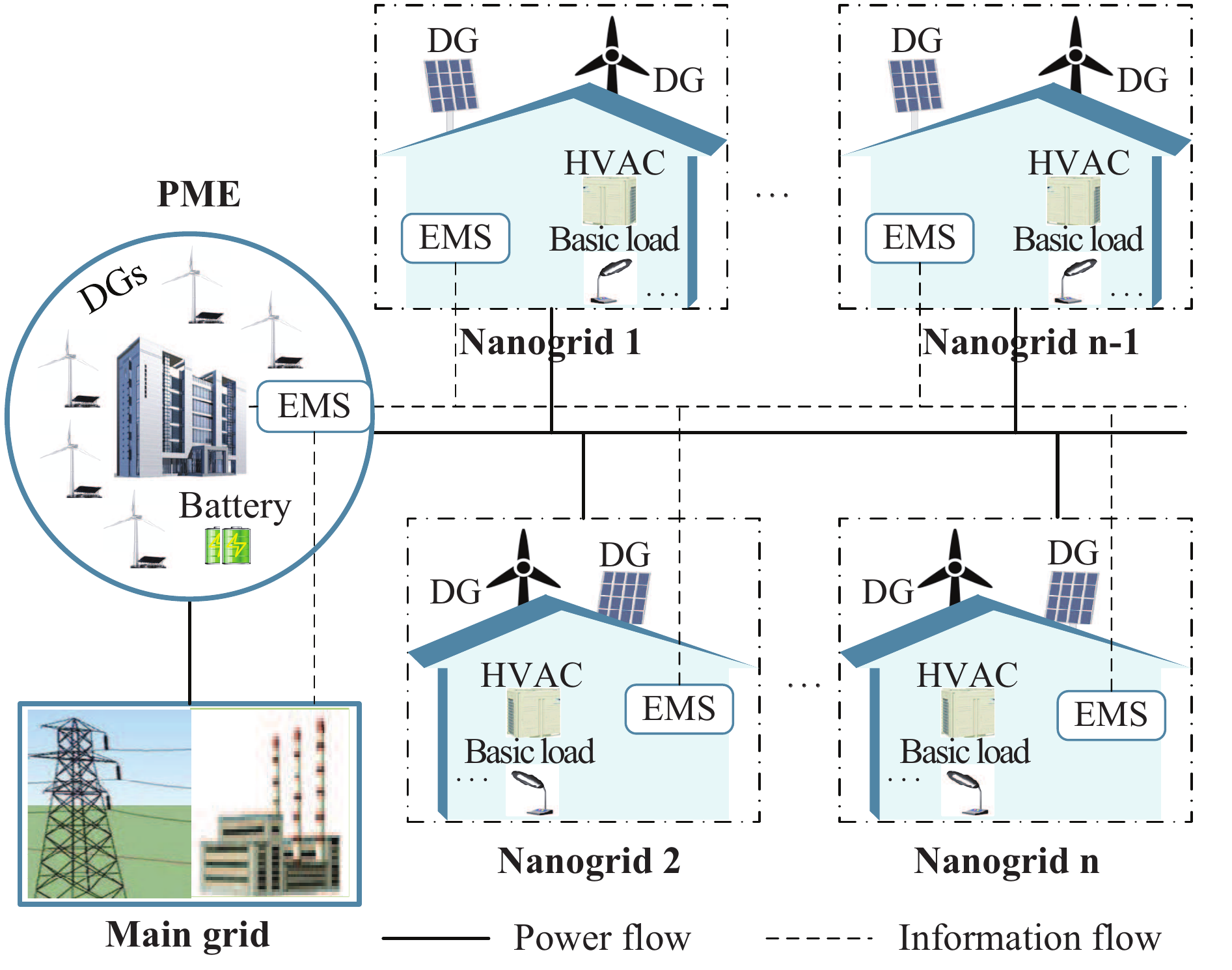}
% where an .eps filename suffix will be assumed under latex,
% and a .pdf suffix will be assumed for pdflatex; or what has been declared
% via \DeclareGraphicsExtensions.
\caption{Schematic of a residential power system.}
\label{fig1}
\vspace{-0.36cm}
\end{figure}

In this paper, we consider a residential power system consisting of nanogrids, PME and main grid shown in Fig.\ref{fig1}. In the context, each nanogrid corresponds to one smart house which is equipped with small-scale uncontrollable DGs (e.g., roof-top photovoltaic systems or small wind turbines), electricity load and house EMS.
Each nanogrid consumer, in this work, is supposed to have two kinds of electricity load. They are the critical basic electricity demand\footnote{In this paper, we focus on HVAC-like thermal elastic demand appliances which need to meet users' satisfaction, and model other appliances simply as a certain inelastic basic load.} which should be maintained under any circumstances and is deemed as a random parameter,
and the flexible electricity demand that could be adjusted for the purpose of demand response.
Specifically, note that the thermostatically controlled devices acknowledged as fast response and universal thermal inertia such as HVAC units occupy a larger fraction of demand response program. This kind of load would have been able to maintain users' comfort level in an acceptable range even with a curtailed consumption.
Under the circumstances, in this work, HVAC units are considered as adjustable loads owing to their higher power consumption and elastic nature.
%an HVAC unit, inelastic basic load\footnote{In this paper, we focus on HVAC-like thermal elastic demand appliances which need to meet users' satisfaction, and model other appliances simply as a certain inelastic basic load.}.
For PME, it has its own generation units, local load and a storage device.
As a regulator, equipped with an EMS, PME can gather and receive data from nanogrids and main grid.
%Besides, PME enables to trade energy with nanogrids to provide supply-demand balance for them and obtain more revenue by making wiser decisions of pricing and storage charging.
Besides, PME is responsible to purchase energy from nanogrids with renewable power surplus and sell energy to nanogrids short of power.
The residual unbalanced energy of PME, if any, can be offset by trading with the main grid in the spot balancing market.

For convenience, we introduce the net generation concept $G_{T}^{k}$ for PME. It is equal to the difference between the power output of generation units and the local load in the PME during slot $k$.\footnote{This paper considers a long-term horizon with a time-slotted model indexed by $k=\{0,1,...\}$. In addition, all power quantities ($G_T^k$, $y^k$, $e_{i}^k$, etc,) are in the unit of energy per slot.}
%As for the storage battery of PME, denote by $E^{k}$ the stored energy state, and we have
As for the storage battery in PME, the stored energy state is denoted by $E^{k}$. Assume that the storage battery unit is ideal with unit efficiency. Then we have the following battery dynamics
%\footnote{Note that, with some slight modifications in the model (i.e., $E^{k+1}=E^{k}+\eta^{c}y^{c,k}-\frac{1}{\eta^{d}}y^{d,k}$, where $\eta^{c}$ and $\eta^{d}$ ($y^{c,k}$ and $y^{d,k}$) are the battery charging and discharging conversion efficiencies (amounts), respectively), the subsequent analysis and proposed method can also cope with the general model, since the conversion efficiency parameters are two constant parameters. In order to simplify the notations and variables, however, we assume that the conversion efficiencies $\eta_{c}=1$ and $\eta_{d}=1$ with $E^{k+1}=E^{k}+y^{k}$.}
\begin{eqnarray}
{{E}^{k+1}}={{E}^{k}}+{{y}^{k}},\
\label{eq9}\\
%\end{equation}
%\begin{equation}
{{E}^{\min }}\le {{E}^{k}}\le {{E}^{\max }},\
\label{eq11}
\end{eqnarray}
where $E^{\max}$ is the maximum battery capacity, $E^{\min}$ the minimum residual capacity to preserve battery life, and
%$\vert y^{k} \vert$
$y^{k}$ is the charged amount (if $y^{k}>0$) or discharged amount (if $y^{k}<0$)
during slot $k$. Considering the finite maximum charge rate ($u^{\rm cmax}$) and discharge rate ($u^{\rm dmax}$), $y^{k}$ should satisfy
\begin{equation}
-{{u}^{\rm dmax }}\le {{y}^{k}}\le {{u}^{\rm cmax }}.\
\label{eq10}
\end{equation}
Besides, in pratice, the using cost of battery should be considered in view of the limited charging/discharging service life. Over the course of charging/discharging, conversion loss and energy leakage may occur which are usually affected by the factors, such as the speed/amount/frequency of charging/discharging. Instead of accurately modeling of these factors, an amortized cost function $f_{b}^{k}=\frac{1}{2}C_{b}(y^{k})^{2}$ is adopted to model the effect of charging/discharging process on battery unit within one slot. In this function, $C_{b}$ is a constant coefficient and we denote $\mathcal{C}^{\max}$/$\mathcal{C}^{\min}$ as the maximum/minimum first derivative of $f_{b}^{k}$ versus $y^{k}$.

During each slot $k$, the basic load $D_{i}^{k}$ of nanogrid~$i$ (e.g., lighting, elevator), is unadjustable and should be first satisfied.
Let $e_{i}^{k}$ be the elastic heat load of HVAC unit in nanogrid $i$.
It is well known that $e_{i}^{k}$ is related with the indoor temperature $T_{i}^{k+1}$ under heating mode of HVAC unit\footnote{The subsequent analysis developed in the paper can be easily adjusted to deal with the cooling mode, where the evolution function is revised by changing the last plus sign in~\eqref{eq1} to a minus sign.}~\cite{thatte2012towards}, satisfying
%~\cite{hao2017transactive,thatte2012towards}, satisfying
\begin{equation}
T_{i}^{k+1}=\varepsilon_{i} T_{i}^{k}+(1-\varepsilon_{i})(T_{i,out}^{k}+\eta_{i}e_{i}^{k}),
\label{eq1}
\end{equation}
with the constraint
\begin{equation}
%T_{i}^{\min }\le T_{i}^{k}\le T_{i}^{\max },\forall k,
T_{i}^{\min }\le T_{i}^{k}\le T_{i}^{\max },
\label{eq6}
\end{equation}
where $T_{i,out}^{k}$ is the outdoor temperature in slot $k$; $\varepsilon_{i}\in(0, 1)$ is the inertial coefficient; $\eta_{i}$ is the energy conversion coefficient related with the heat-conversion efficiency and the thermal conductivity of nanogrid $i$; $T_{i}^{\min }$ and $T_{i}^{\max }$ are the lower and upper bounds of comfort temperature for users in nanogrid $i$, respectively.

In this paper, the HVAC load consumption is assumed to be regulated continuously in a certain range, i.e.,
\begin{equation}
%0\le e_{i}^{k}\le e_{i}^{\max },\forall k,
0\le e_{i}^{k}\le e_{i}^{\max },
\label{eq7}
\end{equation}
where $e_{i}^{\max}$ is the rated power of HVAC unit.
Specially, when HVAC units are directly controlled by the on and off cycles, the power consumption $e_{i}^{k}$ satisfies $e_{i}^{k}\in \{e_{i}^{\max},0\}$.
%This case
This case involving the binary variable can also be tackled by extending the proposed Lyapunov approach in this paper, and see our previous work~\cite{cao2017residential} for details.
%To avoid dealing with such integer nonlinear programming model directly, some techniques (e.g., convex hull\cite{zhou2017online}) can be used to relax the constraint set into a convex one~\eqref{eq7}.
%The proposed Lyapunov approach in this paper can also be extended to include such on and off cycles see~\cite{cao2017residential} for details.

Due to the intermittent and stochastic nature of the renewable energy generation and random power demand,
nanogrids may have surplus energy during off-peak times or, conversely, lack energy during high-demand periods.
Under this circumstance, each nanogrid can be both an energy supplier and consumer across a long-term horizon.
Thus a two-way trade pattern with corresponding bidirectional pricing is needed to keep the balance of power demand and supply.
We denote the power injected into nanogrid $i$ from PME as $tp_{i}^{k}$, which could be positive or negative. The negative value means that there exists power exported from nanogrid $i$ in slot~$k$.
Moreover it satisfies
\begin{eqnarray}
RP_{i}^{k}+tp_{i}^{k}=D_{i}^{k}+e_{i}^{k},\label{eq2}\\
-L_{i}^{\max }\le tp_{i}^{k}\le L_{i}^{\max },\label{eq8}
\end{eqnarray}
%\begin{equation}
%RP_{i}^{k}+tp_{i}^{k}=D_{i}^{k}+e_{i}^{k},
%\label{eq2}
%\end{equation}
%\begin{equation}
%-L_{i}^{\max }\le tp_{i}^{k}\le L_{i}^{\max },
%\label{eq8}
%\end{equation}
where $RP_{i}^{k}$ is the power generation of DGs in nanogrid~$i$ and $L_{i}^{\max }$ is the maximum injection power from PME.
\subsection{Problem Formulation}
%Generally, considering higher selling and lower buying prices of the main grid, nanogrids are inclined to trade with PME by purchasing energy at a lower price or selling their redundant energy at a higher price. First, to enable this process, we assume without loss of generality that
Generally, given higher selling and lower buying prices of the main grid, nanogrids are stimulated to optimize their consumption and trade with the PME by purchasing energy at a lower price or selling their redundant energy at a higher price. In this paper, PME is in charge of providing supply-demand balance for nanogrids with procuring more revenue by making wiser decisions of pricing and storage charging. First, to enable this process, we assume without loss of generality that
\begin{equation}
m_{b}^{k}\le p_{b}^{k}< p_{s}^{k}\le m_{s}^{k},
\label{eq14}
\end{equation}
where $m_{b}^{k}$ ($m_{s}^{k}$) and $p_{b}^{k}$ ($p_{s}^{k}$) are the buying (selling) prices of the main grid and the PME\footnote{The assumption about $p_{s}^{k}\le m_{s}^{k}$ is rational for PME with limited storage capacity. Otherwise, nanogrids are inclined to buy energy from the main grid directly. And then the residual energy of PME has to be bought by the main grid at lower prices. Note that this setting also ensures that the determined selling price is less than the average selling price of PME.} in time slot $k$, respectively.
%where $p_{b}^{k}$ and $p_{s}^{k}$ are the buying and selling price of PME in slot $k$, respectively.

In this context, each nanogrid aims to minimize its average long-term individual cost by scheduling the HVAC energy consumption in each time slot.
Note that considering the maintenance and operation costs of HVAC simultaneously is more realistic in the practical case.
As mentioned in \cite{wang2013modeling}, the maintenance of HVAC is usually done with a regular period or when the equipment is failed.
Indeed, there are some studies that adopt the lifetime maintenance cost which can be allocated to the annual or even daily operation cost. For example, Ref.~\cite{avgelis2009application} has used an amortized annual maintenance cost of HVAC.
It is noted that this amortized maintenance cost is usually related to the year and can be deemed as a constant value within a certain operation horizon (e.g., one day).
In this case, the maintenance cost of HVAC is omitted in this paper.
In addition, our work employs the electricity consumption cost and accompanying virtual thermal discomfort as the operation cost, which is dependent on the bidirectional electricity prices, energy supply and temperature conditions.
A more complicated case can be extended by including the startup and shutdown operation costs with the corresponding on-off control.
The potential solution method can refer to our previous work~\cite{cao2017residential}, the direction of which is not elaborated here.
To sum up, the individual cost of nanogrid~$i$ includes the bidirectional energy trading cost (involving electricity consumption expense) and thermal discomfort cost\footnote{Note that the operation and maintenance cost of renewable generators can also be included in the system. However, due to negligible order of magnitudes~\cite{su2010microgrid}, the cost of this kind can be relatively neglected.
%and complex cost factors~\cite{baldi2019passive} related to binary operation status, historical service condition, temperature condition and so on, a multi-period non-convexity energy management model needs to be coped with that would complicate problem formulation.
%This direction is not elaborated here.
}.
But recall that nanogrids will dynamically switch the role between the energy consumer and supplier and the injection power may be positive or negative in response to the varying prices during different time slots. In this case, the comprehensive cost achieved by nanogrid $i$ under this two-way trade pattern necessitates the following form,
%\begin{equation}
%\begin{aligned}
%UN_{i}^{k}=&{p_{s}^{k}}\!\cdot\!\max (tp_{i}^{k},0)+{p_{b}^{k}}\!\cdot\!\min (tp_{i}^{k},0)\\
%&+\gamma_{i} {{(T_{i}^{k+1}-T_{i}^{opt,k+1})}^{2}},
%\end{aligned}
%\label{eq3}
%\end{equation}
\begin{equation}
%UN_{i}^{k}\!=\!{p_{s}^{k}}\!\cdot\!\max (tp_{i}^{k},0)+{p_{b}^{k}}\!\cdot\!\min (tp_{i}^{k},0)+\gamma_{i} {{(T_{i}^{k+1}\!-\!T_{i}^{opt,k+1})}^{2}},
\resizebox{0.99\hsize}{!}{$UN_{i}^{k}\!=\!{p_{s}^{k}}\cdot\max (tp_{i}^{k},0)\!+\!{p_{b}^{k}}\cdot\min (tp_{i}^{k},0)\!+\!\gamma_{i} {{(T_{i}^{k+1}\!-\!T_{i}^{opt,k+1})}^{2}},$}
\label{eq3}
\end{equation}
where the last term is thermal discomfort cost which is modeled by the the Taguchi Loss Function with a quadratic form~\cite{tran2015coordinated,ma2016residential};
%where the last term is thermal discomfort cost which is modeled by the the Taguchi Loss Function with a quadratic form ~\cite{ma2014distributed,tran2015coordinated};
%$p_{s}^{k}$ and $p_{b}^{k}$ are the selling and buying prices of PME in time slot $k$,
$\gamma_{i}$ is the discomfort weighting coefficient;
%\textcolor{blue}{$T_{i}^{opt,k+1}$ is the optimum comfort temperature  which can be set in advance at slot $k$ by users in nanogrid $i$.}
$T_{i}^{opt,k+1}$ is the optimum comfort temperature for users in nanogrid $i$

Now, as energy management is performed on each slot separately, the overall cost of nanogrid~$i$ can be assessed by minimizing the long-term value of \eqref{eq3}. Nevertheless, real-time energy management has no idea about the future power generation, demand and temperature, which are highly required in minimizing the long-term value of \eqref{eq3}. Consequently, the optimization problem \textbf{P1} of nanogrid $i$ in this paper is formulated as a long-term stochastic optimization problem as follows
%The optimization problem \textbf{P1} of nanogrid $i$ is given as
\begin{eqnarray}
\min_{e_{i}^{k}}\quad &\overline{U{{N}_{i}}}=\underset{T\to \infty}{\mathop{\lim }}
\tfrac{1}{T}\sum\nolimits_{k=0}^{T-1}{\mathbb{E}\left\{ {{UN}_{i}^{k}}\right\}}\label{eq4}\\
{\rm s.t.}\quad&\eqref{eq1}, \eqref{eq6}, \eqref{eq7}, \eqref{eq2}, \eqref{eq8},\  \forall k. \notag
\end{eqnarray}

For PME,
%its objective is to maximize the average long-term profit by determining the bidirectional prices and battery charge ($p_{s}^{k}$, $p_{b}^{k}$, $y^{k}$) (denoted as $\bm{\chi}^{k}$ for brevity).
taking two-way trade pattern and battery using cost into consideration, the obtained net profit during slot $k$ is formulated as follows,
\begin{equation}
\resizebox{0.9\hsize}{!}{$
\begin{aligned}
pro^{k}\!=\!\;&[\sum\nolimits_{i=1}^{n}{p_{s}^{k}}\cdot\max (tp_{i}^{k},0)+\sum\nolimits_{i=1}^{n}{p_{b}^{k}}\cdot\min (tp_{i}^{k},0)]\\&-\frac{1}{2}C_{b}(y^{k})^{2}-
[m_{s}^{k}\!\cdot\!\max (\sum\nolimits_{i=1}^{n}{tp_{i}^{k}\!-\!G_{T}^{k}\!+\!{{y}^{k}}},0)\\&+\!m_{b}^{k}\!\cdot\!\min(\sum\nolimits_{i=1}^{n}{tp_{i}^{k}\!-\!G_{T}^{k}\!+\!{{y}^{k}}},0)],
\end{aligned}
$}
\label{eq12}
\end{equation}
where the first two items represent the revenue procured by the trading with all nanogrids;
the third item is the aforementioned amortized battery using cost;
the last two items denote the cost incurred in offsetting the residual unbalanced energy of PME with the main grid at the prices of $m_{s}^{k}$ and $m_{b}^{k}$ which generally need to be forecast in the optimization problem with an infinite horizon.

Similarly, the objective of PME is to maximize the average long-term profit.
% by determining
The decision variables are
the bidirectional prices and battery charge $\{p_{s}^{k}$, $p_{b}^{k}$, $y^{k}\}$ (for brevity, such decision set is denoted as $\bm{\chi}^{k}$). Then we have the following problem \textbf{P2} of PME,
\begin{eqnarray}
\max_{\bm{\chi}^{k}}\quad &\overline{pro}=\underset{T\to \infty}{\mathop{\lim }}
\tfrac{1}{T}\sum\nolimits_{k=0}^{T-1}{\mathbb{E}\left\{ {{pro}^{k}}\right\}}\label{eq13}\\
{\rm s.t.}\quad&\eqref{eq9}, \eqref{eq11}, \eqref{eq10}, \eqref{eq14},\  \forall k,\notag \\
%&tp_{i}^{k}=\arg \ \min\ \overline{U{{N}_{i}}}.\ \forall \textit{k}\label{eq15}
&e_{i}^{k}(\bm{\chi}^{k})=\mathop{\arg\min}_{e_{i}^{k}} \ \overline{UN_{i}},\ \forall k, \forall i.\label{eq15}
\end{eqnarray}
%Constraint \eqref{eq15} indicates that the exchanged energy is determined by nanogrids and affected by the strategy set of PME.
Constraint \eqref{eq15} indicates the interaction relationship between
the PME and $n$ nanogrids in the decision-making process. To be specific, the energy consumption is determined by each nanogrid and affected by the strategy set of PME.

In this paper, we aim at devising a two-way pricing and DR scheme to optimize the long-term profit of PME and individual cost of each nanogrid with a guarantee of users' comfort level.
Meanwhile, we expect to obtain the optimized result in a distributed way and
without forecasting future time-varying prices, power generation, demand and outdoor temperatures.
%\section{Stackelberg game approach to price-based DR}\label{sec3}
\section{Solution strategy of price-based DR}\label{sec3}
%In this section, we develop a Stackelberg game model to analyze DR process between PME and nanogrids in \textbf{P1} and \textbf{P2}.
%Next to solve the original problem, we introduce virtual queues and obtain a relaxed form with Lyapunov optimization technique. After that, the feasibility of the proposed approach is demonstrated.
In this section, to solve price-based DR problems described in the previous section, we first introduce virtual queues and obtain a relaxed form with Lyapunov optimization technique. Then we develop a Stackelberg game model $\mathcal{G}$ to analyze the interaction procedure between PME and nanogrids. After that, the feasibility of the proposed approach is demonstrated.

%The bidirectional pricing scheme set by PME will induce how nanogrids schedule their power consumption, which will conversely affect the planning of price mechanism through the total profit obtained by PME.
%Motivated by this observation, we use a one-leader and multi-follower Stackelberg game model $S$ to capture the coupling decision-making process between nanogrids and PME.
%\begin{equation}
%\begin{aligned}
%S&=\big\{(\{n\ \text{nanogrids}\}\cup \{\text{PME}\});{{\mathbf{e}}_{\mathbf{1}}}, ... ,{{\mathbf{e}}_{n}};\\
%&\phantom{=\;\;}\overline{{UN}_{1}}, ... ,\overline{{UN}_{n}};{{\mathbf{p}}_{\mathbf{b}}};{{\mathbf{p}}_{\mathbf{s}}};\mathbf{y};\overline{pro}\big\},\
%\end{aligned}
%\label{eq32}
%\end{equation}
%where nanogrids are followers, who decide their energy management actions from the strategy space $\{{{\mathbf{e}}_{\mathbf{1}}},...,{{\mathbf{e}}_{n}}\}$ in response to the prices and the battery charging strategy set by the leader (PME) from the space $\{{{\mathbf{p}}_{\mathbf{b}}};{{\mathbf{p}}_{\mathbf{s}}};\mathbf{y}\}$.
%$\overline{{UN}_{1}},...,\overline{{UN}_{n}}, \overline{pro}$ are the utility functions of followers and leader, respectively.
%
%It is observed that the problem of seeking the best strategies is equivalent to optimizing the utility functions of nanogrids and PME sequentially.

It is observed that, in problems \textbf{P1} and \textbf{P2}, the indoor temperature \eqref{eq1} and battery storage level \eqref{eq9} are both time-coupled which means the antecedent decision-making will influence the decisions in the subsequent time slots. Similar issues are usually resolved by dynamic programming, which are computationally intensive in large-scale implementation. In addition, the future parameters (e.g., electricity prices, random power generation, load and outdoor temperatures) in the long-term optimization problems vary over time with unknown statistics, which is a barrier for accurate energy management and pricing.

\begin{figure}[!t]
\centering
\includegraphics[width=3in]{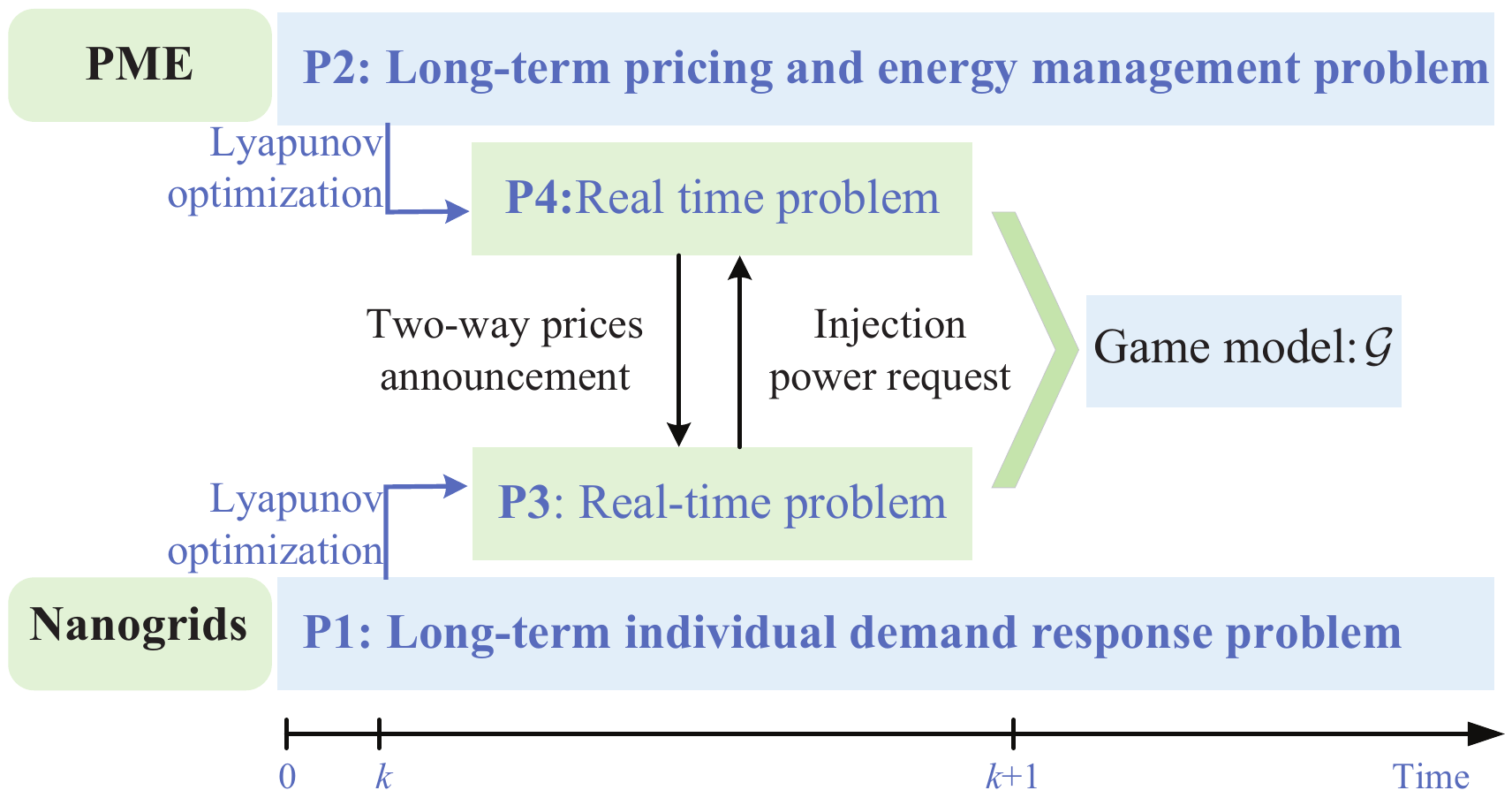}
\caption{Problem formulation flow diagram.}
\label{fig517}
\vspace{-0.36cm}
\end{figure}

In the following, we will develop a method based on Lyapunov optimization technique.
Different from dynamic programming, this method uses an alternative approach based on minimizing the drift of a Lyapunov function. This is done by defining an appropriate set of virtual queues.
Subsequently, the drift-plus-penalty is obtained with the expectation over the system state and the drift bound is minimized greedily~\cite{neely2010stochastic}.
After the conversion, the original time-average problems are finally transformed into some real-time subproblems, which can allow nanogrids and PME to interact dynamically without the knowledge of the stochastic system dynamics and HVAC demand information.
For clarity, the above problem formulation process is summarized as in Fig.\ref{fig517}.
It can be observed that P1 and P2 within a long-term optimization period are finally converted as the real-time online problem based on Lyapunov optimization method. In practice, the time scale in the scheduling is one hour and it helps to meet the reality.
%\subsection{Lyapunov Optimization for Nanogrids}
\subsection{Solution to Nanogrids Optimization problem}
\subsubsection{Virtual temperature queue design}
Instead of solving the time-coupling constraint \eqref{eq1} directly, one way is to study its relaxed form where the average indoor temperature $\overline{T_{i}^{k}}$ is bounded over time, i.e.,
\begin{equation}
\setlength{\abovedisplayskip}{3pt}
\setlength{\belowdisplayskip}{3pt}
T_{i}^{\min}\le \overline{T_{i}^{k}}\le T_{i}^{\max}.
\label{600}
\end{equation}
It is noted that \eqref{600} only ensures the average thermal comfort for nanogrid $i$. However indoor temperatures at some time points might exceed the comfortable range. Thus, the indoor temperature in such worst-case should also be guaranteed.

For this purpose, we introduce a virtual temperature queue $H_{i}^{k}$ with a shift parameter $\Gamma_{i}$ in Lyapunov optimization framework~\cite[Sec. 4.4]{neely2010stochastic}
%citep[chap. 2]{jon90}
%~\cite{neely2010stochastic}
to ensure that \eqref{eq6} is feasible all the time,
\begin{equation}
\setlength{\abovedisplayskip}{3pt}
\setlength{\belowdisplayskip}{3pt}
H_{i}^{k}=T_{i}^{k}+\Gamma_{i},\
\label{eq16}
\end{equation}
%where $\Gamma_{i}$ is a constant within a logical range $[\Gamma_{i}^{\min},\Gamma_{i}^{\max}]$ which will be specified later. Incorporating \eqref{eq16} into \eqref{eq1}, we have the following dynamics
%where $\Gamma_{i}\in[\Gamma_{i}^{\min},\Gamma_{i}^{\max}]$
where $\Gamma_{i}$ is a real constant.
%within a certain range which will be specified later in Theorem 3 and Appendix D.
Actually, the intuition of this design is that the thermal demand requests adding shift parameter $\Gamma_i$ are buffered in virtual queues when the actual backlog is nonempty. In this way, the virtual queue $H_{i}^{k}$ would incur a larger backlog if thermal loads in queues $T_{i}^{k}$ have not been served for a long period of time.
In the later section, Theorem~3 and Appendix D prove that we could regulate the system to enable queues $H_{i}^{k}$ and $T_{i}^{k}$ to have finite bounds when $\Gamma_{i}$ is within a certain range, and then the users' temperature comfort level can be satisfied.
Besides, incorporating \eqref{eq16} into \eqref{eq1}, we have the following dynamics
\begin{equation}
\setlength{\abovedisplayskip}{3pt}
\setlength{\belowdisplayskip}{3pt}
H_{i}^{k+1}={{\varepsilon }_{i}}H_{i}^{k}+(1-{{\varepsilon }_{i}})({{\Gamma }_{i}}+T_{i,out}^{k}+\eta_{i}e_{i}^{k}).\
\label{eq17}
\end{equation}
\subsubsection{Obtaining the drift-plus-penalty}
%In order to minimize the time-averaged energy cost while maintaining the indoor temperature in a stable context,
Firstly, in order to maintain the above temperature queue in a stable context,
we define a Lyapunov function $L(H_{i}^{k})=\tfrac{1}{2}{{(H_{i}^{k})}^{2}}$ for nanogrid $i$. Subsequently, the one-slot conditional Lyapunov drift is given as
\begin{equation}
{\Delta_{i}^{k}}=\mathbb{E}\{L(H_{i}^{k+1})-L(H_{i}^{k})\left| {H_{i}^{k}} \right.\},
\label{eq19}
\end{equation}
where the expectation is with respect to the random power generation, basic load, outdoor temperatures, optimum comfort temperature
and stochastic selection of power consumption strategy.
Then, to stabilize the queue and minimize the nanogrids' time-averaged comprehensive cost simultaneously,
we design a drift-plus-penalty term $\Delta_{v,i}$ by adding a weighted cost function to $\Delta_{i}^{k}$, as following
\begin{equation}
\Delta_{v,i}={\Delta_{i}^{k}}+V_{i}\mathbb{E}\{UN_{i}^{k}\left| {H_{i}^{k}} \right.\},
\label{eq20}
\end{equation}
where the weighting parameter $V_{i}$ is a constant which denotes the trade-off between the temperature queue stability and the decrease in comprehensive energy cost of nanogrid $i$.
When $V_{i}=0$ is chosen, only the Lyapunov drift is minimized which means it does not provide any guarantees on the resulting time average comprehensive energy cost of nanogrid~$i$.
In contrast, with a properly designed $V_{i}$, it can be shown that whenever the HVAC unit consumes energy, the indoor temperature is always in a feasible region (see Theorem~\ref{th1} and Appendix D for details).
\subsubsection{Minimizing the upper bound of drift-plus-penalty}
It can be shown that the objective value of \textbf{P1} is determined by the upper bound of the drift-plus-penalty term $\Delta_{v,i}$ \cite[Sec. 4.5]{neely2010stochastic}.
Squaring both sides of \eqref{eq17} and combining with \eqref{eq19}, we derive that
\begin{equation}
{{\Delta}_{i}^{k}}\le \Omega_{i}^{\max }+\mathbb{E}\{{{\varepsilon }_{i}}(1-{{\varepsilon }_{i}})H_{i}^{k}({{\Gamma }_{i}}+T_{i,out}^{k}+\eta_{i}e_{i}^{k})\left| {H_{i}^{k}} \right.\},
\label{eq601}
\end{equation}
where $\Omega_{i}^{\max}={\tfrac{1}{2}}(1-\varepsilon)^{2}\max\{(\Gamma_{i}+T_{i,out}^{\min})^{2}, (\Gamma_{i}+T_{i,out}^{\max}+\eta_{i}e_{i}^{\max})^{2}\}$,
and $T_{i,out}^{\min}$ and $T_{i,out}^{\max}$ are respectively the lower and upper limits of outdoor temperature.

After plugging \eqref{eq601} into \eqref{eq20}, we obtain that
\begin{equation}
\resizebox{0.88\hsize}{!}{$
\begin{aligned}
{{\Delta }_{v,i}}\le \Omega_{i}^{\max }&\!+\!\mathbb{E}\{{{\varepsilon }_{i}}(1\!-\!{{\varepsilon }_{i}})H_{i}^{k}({{\Gamma }_{i}}\!+\!T_{i,out}^{k}\!+\!\eta_{i}e_{i}^{k})\left| {H_{i}^{k}} \right.\}\\
&\!+\!{{V}_{i}}\mathbb{E}\Big\{p_{s}^{k}\!\cdot\! \max (tp_{i}^{k},0)\!+\!p_{b}^{k}\!\cdot\! \min (tp_{i}^{k},0)\\
&\!+\!{{\gamma }_{i}}{{(T_{i}^{k+1}\!-\!T_{i}^{opt,k+1})}^{2}}\left| {H_{i}^{k}} \right.\Big\},
\end{aligned}
$}
\label{eq21}
\end{equation}

By minimizing the upper bound of $\Delta_{v,i}$ shown in right-hand-side of \eqref{eq21} based on the theoretical framework of `opportunistically minimizing an expectation'~\cite[Sec. 1.8]{neely2010stochastic}, we can obtain the following simplified problem \textbf{P3} after several manipulations (refer to the Appendix~\ref{App0}),
%(refer to the Appendix A),
%we can minimize the total cost of nanogrid~$i$ concurrently with stabilizing the temperature queue.It allows to minimize the following simplified problem (after several manipulations, refer to the Appendix A) via the observed current state of system and virtual queue on each slot $k$.
%As $\Omega_{i}^{\max}$ is a constant, and $T_{i}^{k}$, $T_{i,out}^{k}$, $T_{i}^{opt,k+1}$ have been specified at the beginning of each slot $k$, with incorporating \eqref{eq1} into \eqref{eq21},
%it is equivalent to minimizing the simplified objective function $UN_{i}^{'}$, with $UN_{i}^{'}\!=\!V_{i}\gamma_{i}{{(1\!-\!\varepsilon_{i})}^{2}}{{(\eta_{i}e_{i}^{k})}^{2}}\!+\!\Big\{{{\varepsilon }_{i}}(1\!-\!{{\varepsilon }_{i}})H_{i}^{k}\!+\!2{{V}_{i}}{{\gamma }_{i}}(1\!-\!{{\varepsilon }_{i}})[(1\!-\!{{\varepsilon }_{i}})T_{i,out}^{k}\!+\!{{\varepsilon }_{i}}T_{i}^{k}\!-\!T_{i}^{opt,k+1}]\Big\}\eta_{i}e_{i}^{k}\!+\!V_{i}[p_{s}^{k}\cdot \max (tp_{i}^{k},0)+p_{b}^{k}\cdot \min (tp_{i}^{k},0)]$.
%After several manipulations (refer to the Appendix A), we obtain the following simplified optimization problem \textbf{P3},
\begin{align}
{\min}_{e_{i}^{k}}&\;\,\; UN_{i}^{'}\label{eq23}\\
{\rm s.t.} \;\;\, &\max \{-L_{i}^{\max }-D_{i}^{k}+RP_{i}^{k},0\}\le e_{i}^{k}\notag\\
&\le \min \{L_{i}^{\max }-D_{i}^{k}+RP_{i}^{k}, e_{i}^{\max }\}\label{eq24}
\end{align}
where the objective
$UN_{i}^{'}\!=\!V_{i}\gamma_{i}{{(1\!-\!\varepsilon_{i})}^{2}}{{(\eta_{i}e_{i}^{k})}^{2}}\!+\!\Big\{{{\varepsilon }_{i}}(1\!-\!{{\varepsilon }_{i}})H_{i}^{k}\!+\!2{{V}_{i}}{{\gamma }_{i}}(1\!-\!{{\varepsilon }_{i}})[(1\!-\!{{\varepsilon }_{i}})T_{i,out}^{k}\!+\!{{\varepsilon }_{i}}T_{i}^{k}\!-\!T_{i}^{opt,k+1}]\Big\}\eta_{i}e_{i}^{k}\!+\!V_{i}[{\tfrac{1}{2}}(p_{s}^{k}\!-\!p_{b}^{k})\left| D_{i}^{k}\!-\!RP_{i}^{k}\!+\!e_{i}^{k} \right|\!+\!{\tfrac{1}{2}}(p_{s}^{k}\!+\!p_{b}^{k})(D_{i}^{k}\!-\!RP_{i}^{k}\!+\!e_{i}^{k})]$.
The feasible strategy set of nanogrid $i$ is given as $\Omega_{ng,i}=\{e_{i}^{k}|e_{i}^{k}\in R, \text{subject to}~\eqref{eq24}\}$.

In this way, we can decide the strategy at each slot $k$ purely as a function of the current system state while guaranteeing the time-coupling constraint, which will be shown in Theorem~\ref{th1}.
After obtaining the optimized power consumption $e_{i}^{k,*}$ of
%\eqref{eq23}, \eqref{eq24},
\textbf{P3},
the optimal injection power of nanogrid $i$ is
\begin{equation}
tp_{i}^{k,*}=D_{i}^{k}+e_{i}^{k,*}-RP_{i}^{k}.\
\label{eq333}
\end{equation}
%We postpone the analysis of optimal value $e_{i}^{k,*}$ solving process to Appendix D.

The following theorem has provided insight into the analysis of optimal value $e_{i}^{k,*}$ under different prices.

\begin{theorem}\label{th11}
%\begin{lemma}\label{le1}
The optimal consumption strategy of HVAC in nanogrid $i$ is given by
\begin{equation}
e_{i}^{k,*}=\left\{ \begin{aligned}
  & 0,\,\,\qquad\text{if}\;V_{i}p_{b}^{\min}>-\varepsilon_{i}(1-\varepsilon_{i})H_{i}^{k}\eta_{i}-\alpha_{i}^{k}; \\
 & e_{i}^{\max},\ \, \,\,\text{if}\;V_{i}p_{s}^{\max}<-\varepsilon_{i}(1-\varepsilon_{i})H_{i}^{k}\eta_{i}-\beta_{i}^{k}; \\
 & f({\bm{\chi}}^{k}),\,\text{otherwise};
 \\
\end{aligned} \right.\
\label{eq36}
\end{equation}
where $p_{b}^{\min}$ is the minimum buying price, $p_{s}^{\max}$ is the maximum selling price of PME; and
%$\alpha_{i}^{k}=2V_{i}\gamma_{i}(1-\varepsilon_{i})^{2}\eta_{i}(T_{i,out}^{k}+\tfrac{\varepsilon_{i}T_{i}^{k}-T_{i}^{opt}}{1-\varepsilon_{i}})$, $\beta_{i}^{k}=2V_{i}\gamma_{i}(1-\varepsilon_{i})^{2}\eta_{i}(T_{i,out}^{k}+\tfrac{\varepsilon_{i}T_{i}^{k}-T_{i}^{opt}}{1-\varepsilon_{i}}+\eta_{i}e_{i}^{\max})$.
$\alpha_{i}^{k}\!=\!2V_{i}\gamma_{i}(1\!-\!\varepsilon_{i})^{2}\eta_{i}[T_{i,out}^{k}\!+\!(\varepsilon_{i}T_{i}^{k}-T_{i}^{opt,k+1})/(1-\varepsilon_{i})]$, %$\beta_{i}^{k}=2V_{i}\gamma_{i}(1-\varepsilon_{i})^{2}\eta_{i}[T_{i,out}^{k}+(\varepsilon_{i}T_{i}^{k}-T_{i}^{opt})/(1-\varepsilon_{i})+\eta_{i}e_{i}^{\max}]$.
$\beta_{i}^{k}\!=\!\alpha_{i}^{k}+2V_{i}\gamma_{i}(1-\varepsilon_{i})^{2}(\eta_{i})^{2}e_{i}^{\max}$.
%\end{lemma}
\end{theorem}
%The main idea of the proof of Theorem \ref{th11} is that when the purchasing price offered by the PME exceeds a certain threshold, the nanogrid is willing to consume HVAC power as few as possible to maximize its profit. Inversely, when the selling price is low, the nanogird tends to inject the maximum HVAC power.

%The main idea of the former two results in \eqref{eq36} is
The former two cases with the explicit formulation in \eqref{eq36} are obtained by the method of reduction to absurdity which is given as the first part of Appendix B.
The results mean that when the buying price offered by PME exceeds a certain threshold, the nanogrid is willing to consume HVAC power as few as possible to maximize its profit. Inversely, when the selling price is low, the nanogird tends to inject the maximum HVAC power from PME.
Note that, the implicit function $f({\bm{\chi}}^{k})$ in the third case includes several different kinds of classification which is difficult to obtain a precise calculated formulation directly. In addition, the value that $f({\bm{\chi}}^{k})$ may take is also discussed in the second part of Appendix~\ref{App1}
through the method of portrayal.
%\subsection{Lyapunov Optimization for PME}
\subsection{Solution to PME Optimization Problem}
For PME, it dynamically makes decisions to solve its long-term profit maximization problem (\textbf{P2}).
Note that the battery constraints \eqref{eq9} and \eqref{eq11} bring the time-coupling characters which complicate the optimization problem.
To avoid such coupling, a time-average expected constraint is considered, i.e.,
\begin{equation}
\underset{T\to \infty}{\mathop{\lim }}
\tfrac{1}{T}\sum\nolimits_{k=0}^{T-1}{\mathbb{E}\left\{ {{y}^{k}}\right\}}=0.
\label{eq602}
\end{equation}
We can prove that \eqref{eq9} and \eqref{eq11} signify \eqref{eq602}. Summing both sides of \eqref{eq9} over all time slots and taking expectation yields
\begin{equation}
{\mathbb{E}\left\{ {{E}^{T}}\right\}}-E^{0}=\sum\nolimits_{k=0}^{T-1}{\mathbb{E}\left\{ {{y}^{k}}\right\}}.
\label{eq603}
\end{equation}
Then dividing them by $T$ and taking $T\to \infty$, we have \eqref{eq602} since the initial storage state $E^{0}$ and storage capacity are all finite.
After eliminating the dependency property between storage energy state across time slots owing to the limited battery storage capacity, \textbf{P2} can be resolved by following the Lyapunov optimization framework in a similar way.

%Denote by $B_{k}$ the virtual queue associated with battery energy level to guarantee the feasibility of \eqref{eq9}, and we have
First, we introduce a virtual battery energy queue $B^{k}$ with $B^{k}=E^{k}+\theta$,
%\begin{equation}
%B^{k}=E^{k}+\theta,\
%\label{eq25}
%\end{equation}
where the constant $\theta$ is the shift parameter and will be presented in the later section. Besides, %combining \eqref{eq25} and~\eqref{eq9},
$B^{k}$ is updated as
%we have the dynamic update process of $B_{k}$ as follows,
\begin{equation}
B^{k+1}={{B}^{k}}+{{y}^{k}}.\
\label{eq26}
\end{equation}

The constraint \eqref{eq602} can be transformed into the virtual queue stability constraint as shown in~\cite[Chap. 2]{neely2010stochastic} to guarantee the feasibleness of \eqref{eq11} even in the worst case.

Following that, the one-slot conditional Lyapunov drift is given by
\begin{equation}
{\Delta_{P}^{k}}=\mathbb{E}\{L(B^{k+1})-L(B^{k})\left| {B^{k}} \right.\},\
\label{eq28}
\end{equation}
where $L(B^{k})=\tfrac{1}{2}{{(B^{k})}^{2}}$.
%For PME, it expects to maximize its time-averaged profits and stabilize the battery energy queue. To achieve this,
We define the drift-plus-penalty term $\Delta_{v,P}\!=\!{\Delta_{P}^{k}}-V_{P}\mathbb{E}\{pro^{k}\left| {B^{k}} \right.\}$.
%\begin{equation}
%\Delta_{v,P}={\Delta_{P}^{k}}-V_{P}\mathbb{E}\{pro^{k}\left| {B^{k}} \right.\},\
%\label{eq29}
%\end{equation}
%where $V_{P}$ is a constant which represents the tradeoff between energy queue stability and the increase of the profit of PME and its maximum value will be specified later.
%Specially, for any feasible strategy that can be implemented during slot $k$ under constraints \eqref{eq9} \eqref{eq11}, \eqref{eq10} and \eqref{eq15}, we have
Based on \eqref{eq26} and the definition of $L(B^{k})$, the upper bound of $\Delta_{v,P}$ is given by
\begin{equation}
{{\Delta }_{v,P}}\le \Omega _{P}^{\max }+\mathbb{E}\{{{B}^{k}}{{y}^{k}}\left| {{B}^{k}} \right.\}-{{V}_{P}}\mathbb{E}\{pr{{o}^{k}}\left| {{B}^{k}} \right.\},\
\label{eq30}
\end{equation}
where  $\Omega_{P}^{\max}=\tfrac{1}{2}\max \{{({u^{\rm cmax }})^{2}},{({u^{\rm dmax }})^{2}}\}$ and the weighting parameter $V_{P}$ are all constants.
%Besides, by ca
%In other words, the time-coupling constraint \eqref{eq11} is satisfied automatically under such operation as shown in Theorem~\ref{th2}.

By minimizing the upper bound of $\Delta_{v,P}$, the profit of PME is greedily maximized and queue $B^{k}$ is stabilized.
We can prove that the time-coupling constraints \eqref{eq9} and \eqref{eq11} are already satisfied under such operation in Theorem~\ref{th2}.
%Finally, the original problem \textbf{P2} can be rewritten \textbf{P4} as follows,
Finally, the original problem \textbf{P2} can be converted into the following problem \textbf{P4} over individual time-slot,
\begin{align}
%\min\limits_{p_{s}^{k}, p_{b}^{k}, y^{k}} \ &\; pro^{'}\label{eq31}\\
{\min}_{\bm{\chi}^{k}} &\;\, pro^{'}\label{eq31}\\
{\rm s.t.}\;\;\,&\eqref{eq10}, \eqref{eq14}, \eqref{eq333},\  \forall \textit{k} \notag
\end{align}
where the objective
%$pro^{'}\!=\!{B}^{k}{y}^{k}\!+\!{{V}_{P}}[m_{s}^{k}\!\cdot\!\max(\sum\nolimits_{i=1}^{n}{tp_{i}^{k}}\!-\!G_{T}^{k}\!+\!{{y}^{k}}, 0)\!+m_{b}^{k}\!\cdot\!\min(\sum\nolimits_{i=1}^{n}{tp_{i}^{k}}\!-\!G_{T}^{k}\!+\!{{y}^{k}}, 0)]\!-\!{{V}_{P}}[\sum\nolimits_{i=1}^{n}{p_{s}^{k}\!\cdot\!\max(tp_{i}^{k}, 0)}\!+\!\sum\nolimits_{i=1}^{n}{p_{b}^{k}\!\cdot\!\min (tp_{i}^{k}, 0)}]$.
$pro^{'}\!=\!{B}^{k}{y}^{k}\!-\!{{V}_{P}}[\sum\nolimits_{i=1}^{n}{p_{s}^{k}\!\cdot\!\max(tp_{i}^{k}, 0)}\!+\!\sum\nolimits_{i=1}^{n}{p_{b}^{k}\!\cdot\!\min (tp_{i}^{k}, 0)}]\!+\!{{V}_{P}}[m_{s}^{k}\cdot\max(\sum\nolimits_{i=1}^{n}{tp_{i}^{k}}\!-\!G_{T}^{k}\!+\!{{y}^{k}}, 0)\!+m_{b}^{k}\!\cdot\!\min(\sum\nolimits_{i=1}^{n}{tp_{i}^{k}}\!-\!G_{T}^{k}\!+\!{{y}^{k}}, 0)+\frac{1}{2}C_{b}(y^{k})^{2}]$, and the constraint \eqref{eq15} in \textbf{P2} is replaced by \eqref{eq333}.
Hence the feasible strategy set of PME is $\Omega_{\text{PME}}\!=\!\{{\bm{\chi}^{k}}\!=\![p_{s}^{k},p_{b}^{k},y^{k}]|p_{s}^{k},p_{b}^{k},y^{k}\in R, \text{subject to}~\eqref{eq10},\eqref{eq14},\eqref{eq333}\}$.

The solution analysis is deferred to Appendix~\ref{App4} by discussing two situations in detail.
In addition, note that it could not obtain the calculated expression directly due to the implicit strategy function of followers.
Hence, we develop a best response algorithm to derive solution strategies of problems \textbf{P3} and \textbf{P4} iteratively which is shown as Algorithm 1 in the later section.

After completing the above processes,
we do not need to consider the stochastic processes related with unknown factors such as distributed generations supply $RP_{i}^{k}$.
We can decide the strategy based on the observed current state at each slot to achieve the optimization in a long-term horizon without the need of forecasting any system parameters which makes the originally complicated energy management problems tractable.
%By utilising the above Lyapunov optimization approach, the original time-average problems can be transformed  into some per-slot subproblems.
Specifically, on each slot $t$, the controller of energy management system observes the current state of the distributed power generation and chooses the HVAC power demand from the decision space. This decision, together with the current status of ambient temperature, determines the vector of temperature queue/virtual queue.
%In this approach, the virtual queues (i.e., temperature queues, storage level queues) are needed to ensure that the related time average constraints are met.
Inefficient energy management decisions would incur a larger backlog in certain queues.
These backlogs will act as sufficient statistics on which the next energy management decision to base.
According to Theorem 4.8 in~\cite{neely2010stochastic}, such an approach yields an optimal performance within $O(1/V_{P})$ from the optimality which has used the complete information.
The advantage of this approach is that it uses both current states to stabilize the
system, and it does not require a-priori knowledge of random event probabilities.
%After completing the above processes, we can decide the strategy based on the observed current state at each slot to achieve the optimization in a long-term horizon without the need of forecasting any system parameters which makes the game model \eqref{eq32} tractable.
%%Then the optimal energy consumption of nanogrid and bidirectional prices and battery charge-discharge amount of the PME can be obtained by solving the relaxed problems at each time slot.
\subsection{Game between PME and Nanogrids}
Note that, the bidirectional pricing scheme set by PME will induce how nanogrids schedule their power consumption, which will conversely affect the planning of price mechanism through the total profit obtained by PME. Motivated by this observation, in this sub-section, the coupling decision-making process between nanogrids and PME is captured by a one-leader and multi-follower Stackelberg game, where PME is modeled as the leader, and nanogrids are modeled as followers according to their functionalities.
%In this game, leader designs rational price, while followers decide their energy management actions from
%In this game, followers decide their energy management actions from their feasible strategy sets in response to the bidirectional prices designed by the leader to optimize their respective objectives presented in~\eqref{eq23} and~\eqref{eq31}. Meanwhile, the leader is responsible for making a rational battery charging/discharging strategy and offsetting the unbalance energy with the main grid. Certainly, the proposed game is a two-stage optimization problem where followers optimize their utilities in stage \uppercase\expandafter{\romannumeral2} while in stage \uppercase\expandafter{\romannumeral1} leader determines its strategy by knowing the results of best demand responses of followers.
In this game, followers decide their energy management actions from their feasible strategy sets in response to the bidirectional prices designed by the leader to optimize their respective objectives presented in~\eqref{eq23} and~\eqref{eq31}. Meanwhile, the leader is responsible for making a rational battery charging/discharging strategy and offsetting the unbalance energy with the main grid. Certainly, the proposed game is a bilevel optimization problem where followers optimize their utilities in the lower-level while in the upper-level leader determines its strategy by knowing the results of best demand responses of followers.

%the strategy space $\{{{\mathbf{e}}_{\mathbf{1}}},...,{{\mathbf{e}}_{n}}\}$ in response to the prices and the battery charging strategy set by the leader (PME) from the space $\{{{\mathbf{p}}_{\mathbf{b}}};{{\mathbf{p}}_{\mathbf{s}}};\mathbf{y}\}$.

It is observed that the problem of seeking best strategies can be equivalent to sequentially optimizing the utility functions of nanogrids (followers) and the PME (leader) in a backward manner \cite{8715389}. The result at the end of each sequence of the game where neither PME nor nanogrids can obtain more benefits by a unilateral change of their strategy is called as SE.
%Hence a set of strategies $({\bm{\chi}^{k,*}}, e^{k,*})$ constitutes SE for the proposed Stackelberg game if it is a feasible solution of the following optimization problem $\mathcal{G}$,
Thus a set of strategies $({\bm{\chi}^{k,*}}, e^{k,*})$ constitutes an SE for the proposed Stackelberg game if it corresponds to a feasible solution of the following problem $\mathcal{G}$,
\begin{equation} \label{eq608}
  \begin{split}
  ({\bm{\chi}^{k,*}}, e^{k,*})&=\underset{\tiny{({\bm{\chi}^{k,*}}, e^{k,*})\in \Omega_{\text{PME}} \times \Omega_{ng,i}}}{\mathop{\min }}\,pro^{'}({\bm{\chi}^{k}}, e^{k,*}),    \\
  {\rm s.t.}\;\;e_{i}^{k,*}&=\underset{\tiny{e_{i}^{k}\in \Omega_{ng,i}}}{\mathop{\min }}\,UN_{i}^{'}(e_{i}^{k}),\;\forall i.
  \end{split}
\end{equation}

It is pointed out that an equilibrium in pure strategies might not always exist in a noncooperative game.
Therefore, we need to prove that there exists a unique SE for the proposed Stackelberg game. See Appendix~\ref{App4} for its proof.
\begin{theorem}\label{th3}
A unique SE exists for the proposed Stackelberg game if the following three conditions are met.
\begin{enumerate}[1)]
\item The strategy sets of PME and nanogrids are nonempty, compact and convex.
\item
    Once each nanogrid is notified of the strategy set of PME, it has a unique best-response strategy.
\item PME only has one optimal strategy given the identified optimal best-response strategies of all nanogrids.
\end{enumerate}
\end{theorem}

%Note that an equilibrium is not always guaranteed to exist in certain game model.
Theorem \ref{th3} guarantees that the proposed game can reach the equilibrium as soon as PME is able to find the unique optimal strategy while nanogrids select their optimal energy demand.
\subsection{Performance Analysis}
%In this section, we will demonstrate the feasibility of the proposed approach and the existence of the unique SE.
%To begin with, we introduce three mild assumptions,
%\begin{eqnarray}
%&T_{i,out}^{\max}\leq T_{i}^{\max},\label{eq33}\\
%&\eta_{i}e_{i}^{\max}+T_{i,out}^{\min}\geq T_{i}^{\min},\label{eq34}\\
%&T_{i}^{\max}-T_{i}^{\min}>\varphi_{i},
%\label{eq35}
%\end{eqnarray}
%where $\varphi_{i}=(1-\varepsilon_{i})(T_{i,out}^{\max}+\eta_{i}e_{i}^{\max}-T_{i,out}^{\min})$.
In this section, we will demonstrate the feasibility of the proposed approach.
To begin with, we introduce three mild assumptions: (a) $T_{i,out}^{\max}\!\leq\!T_{i}^{\max}$,
(b) $\eta_{i}e_{i}^{\max}\!+\!T_{i,out}^{\min}\!\geq\!T_{i}^{\min}$,
(c) $T_{i}^{\max}\!-\!T_{i}^{\min}\!>\!(1\!-\!\varepsilon_{i})(T_{i,out}^{\max}\!+\!\eta_{i}e_{i}^{\max}\!-\!T_{i,out}^{\min})$.
%where $\varphi_{i}\!=\!(1\!-\!\varepsilon_{i})(T_{i,out}^{\max}\!+\!\eta_{i}e_{i}^{\max}\!-\!T_{i,out}^{\min})$.
It is noted that these assumptions make sense in real scenarios.
For example, (a) is obviously valid in winter; (b) ensures that the indoor temperature can rise to comfort level even from the lowest outdoor temperature by injecting the full power of HVAC unit; and (c) is imposed to guarantee $V_{i}^{\max}$ is nonnegative.
Now, we can show that the proposed approach can guarantee the users' thermal comfort and stabilize the storage energy level summarized in the following two theorems. The proofs are in Appendices~\ref{App2} and \ref{App3}, respectively.
%Appendices B and C.
\begin{theorem}\label{th1}
 %If $\Gamma_{i}$ and $V_{i}$ satisfy
 For $\Gamma_{i}\in[{{\Gamma}_{i}^{\min }}, {{\Gamma}_{i}^{\max }}]$ and $V_{i}\in(0, V_{i}^{\max}]$, the users' temperature comfort level can be guaranteed, \textit{i.e.}, $T_{i}^{\min }\le T_{i}^{k}\le T_{i}^{\max }$, $\forall k$.
 \begin{eqnarray}
 &&\hspace{-2em}
 \Gamma_{i}^{\min}=\tfrac{V_{i}p_{s}^{\max}+{\max}_{k}\beta_{i}^{k}}{-\varepsilon_{i}(1-\varepsilon_{i}){\eta_{i}}}-\tfrac{T_{i}^{\max}-(1-\varepsilon_{i})(T_{i,out}^{\max}+e_{i}^{\max}{\eta_{i}})}{\varepsilon_{i}},
 \label{eq37}\\
 &&\hspace{-2em}
 \Gamma_{i}^{\max}=\tfrac{V_{i}p_{b}^{\min}+{\min}_{k}\alpha_{i}^{k}}{-\varepsilon_{i}(1-\varepsilon_{i}){\eta_{i}}}-\tfrac{T_{i}^{\min}-(1-\varepsilon_{i})T_{i,out}^{\min}}{\varepsilon_{i}},
 \label{eq38}\\
 &&\hspace{-2em}
 V_{i}^{\max}=\tfrac{(1-\varepsilon_{i}){\eta_{i}}(T_{i}^{\max}-T_{i}^{\min}-\varphi_{i})}{p_{s}^{\max}-p_{b}^{\min}+2\gamma_{i}(1-\varepsilon_{i}){\eta_{i}}[\varphi_{i}+\varepsilon_{i}(T_{i}^{\max}-T_{i}^{\min})+\Lambda_i]},
 \label{eq39}
 \end{eqnarray}
where \begin{small}{$\Lambda_i\!=\!\max_{k}{T_{i}^{opt,k}}\!-\!\min_{k}{T_{i}^{opt,k}}$}\end{small}, \begin{small}{$\varphi_{i}\!=\!(1\!-\!\varepsilon_{i})(T_{i,out}^{\max}\!+\!\eta_{i}e_{i}^{\max}\!-\!T_{i,out}^{\min})$}\end{small}.
%  \textcolor{blue}{where $\varphi_{i}\!=\!(1\!-\!\varepsilon_{i})(T_{i,out}^{\max}\!+\!\eta_{i}e_{i}^{\max}\!-\!T_{i,out}^{\min})$, $\Lambda_i\!=\!\max_{k}\{T_{i}^{opt,k}\}-\min_{k}\{T_{i}^{opt,k}\}$.}
\end{theorem}
\begin{theorem}\label{th2}
%For the PME, if its battery queue shift parameter $\theta$ and control parameter $V_{P}$ satisfy $\theta \in[\theta^{\min}, \theta^{\max}]$ and $V_{P}\in(0, V_{P}^{\max}]$ respectively, then the battery energy level can be guaranteed, \textit{i.e.}, ${{E}^{\min }}\le {{E}^{k}}\le {{E}^{\max }}$, $\forall k$, where
For the PME, if $\small{\theta \in[\theta^{\min}, \theta^{\max}]}$ and $V_{P}\in(0, V_{P}^{\max}]$, then the battery energy level can be guaranteed, \textit{i.e.}, $\small{{{E}^{\min }}\le {{E}^{k}}\le {{E}^{\max }}}$, $\forall k$, where
\begin{eqnarray}
&&\hspace{-2.7em}
\theta^{\min}=u^{\rm cmax}\!-\!E^{\max}\!-\!V_{P}\!\cdot\!{{\min}_{k}}m_{b}^{k}\!-\!V_{P}\!\cdot\!\mathcal{C}^{\min},
\label{eq40}\\
&&\hspace{-2.7em}
\theta^{\max}=\!-\!u^{\rm dmax}\!-\!E^{\min}\!-\!V_{P}\!\cdot\!{{\max}_{k}}m_{s}^{k}\!-\!V_{P}\!\cdot\!\mathcal{C}^{\max},
\label{eq41}\\
&&\hspace{-2.8em}
V_{P}^{\max}=\tfrac{E^{\max}-E^{\min}-(u^{\rm cmax}+u^{\rm dmax})}{{{\max}_{k}}{m_{s}^{k}}-{{\min}_{k}}{m_{b}^{k}}+\mathcal{C}^{\max}-\mathcal{C}^{\min}}.
\label{eq42}
\end{eqnarray}
%{\color{blue}The parameters $\mathcal{C}^{\max}$ and $\mathcal{C}^{\min}$ in above formula are given as $\mathcal{C}^{\min}\!=\!\min\{C_{b}u^{\rm cmax}, -C_{b}u^{\rm dmax}\}$, $\mathcal{C}^{\max}\!=\!\max\{C_{b}u^{\rm cmax}, -C_{b}u^{\rm dmax}\}$.}
Besides, parameters \begin{small}{$\mathcal{C}^{\max}$}\end{small} and \begin{small}{$\mathcal{C}^{\min}$}\end{small} are shown as \begin{small}{$\mathcal{C}^{\min}\!=\!\min\{C_{b}u^{\rm cmax},\!-\!C_{b}u^{\rm dmax}\}$}\end{small}, \begin{small}{$\mathcal{C}^{\max}\!=\!\max\{C_{b}u^{\rm cmax},\!-\!C_{b}u^{\rm dmax}\}$}\end{small}.
\end{theorem}

%In brief, Lemma \ref{le1} and Theorem \ref{th1}-\ref{th2} indicate the feasibility of the proposed approach by showing that the constraint \eqref{eq11} and \eqref{eq6} can be satisfied.
%Next, we prove that there exists a unique SE for the proposed Stackelberg game \eqref{eq32}. See \ref{App4} for its proof.
%\begin{theorem}\label{th3}
%A unique SE exists for the proposed Stackelberg game if the following three conditions are met.
%\begin{enumerate}[1)]
%\item The strategy sets of PME and nanogrids are nonempty, compact and convex.
%\item
%    Once each nanogrid is notified of the strategy set of PME, it has a unique best-response strategy.
%\item PME only has one optimal strategy given the identified optimal best-response strategies of all nanogrids.
%\end{enumerate}
%\end{theorem}
%
%%Note that an equilibrium is not always guaranteed to exist in certain game model.
%Theorem \ref{th3} guarantees that the proposed game can reach the equilibrium as soon as PME is able to find the unique optimal strategy while nanogrids select their optimal energy demand.
\section{Designed algorithm to reach SE}\label{sec4}
%\begin{figure}[htbp]
%\centering
%\includegraphics[width=3.28in]{revischart8.eps}
%%\includegraphics[width=3.28in]{fangdashiyan.eps}
%\caption{Information interactions between PME and nanogrids.}
%\label{fig2}
%%\vspace{-0.3cm}
%\end{figure}
%Although, a unique SE exists theoretically, it is difficult to obtain an explicit formulation of $e_{i}^{k,*}\!=\!f({\bm{\chi}}^{k})$.
Although, a unique SE exists theoretically, it is difficult to obtain an analytical solution directly for the bilevel complicated optimization problem.
%In this section, we will develop a heuristic algorithm mainly based on genetic algorithm to reach SE iteratively in a distributed way.
In this section, we will develop an iterative energy management algorithm with the bidirectional pricing scheme to reach SE in a distributed way.

\begin{algorithm}[!t]
     \caption{Algorithm to reach the SE point}
      \label{alg:Framwork}
      \KwIn{Parameters of all nanogrids, PME and prices of the main grid}
      \KwOut{Solution: in period $k$, the strategy $e_{i}^{k}$ for each nanogrid $i$ and the strategy set  $\{\bm{\chi}^{k}\}=\{p_{s}^{k},p_{b}^{k},y^{k}\}$ for PME}
        %Initialization:
        \textbf{Initialize}
        ${p}_{s}^{k,1}$, ${p}_{b}^{k,1}$, ${y}^{k,1}$ and set ${p}_{s}^{k,1}={p}_{s}^{k,0}+1$, ${p}_{b}^{k,1}={p}_{b}^{k,0}+1$, $y^{k,1}=y^{k,0}+1$, $m=1$\;
\While
{$(|p_{s}^{k,m}-p_{s}^{k,m-1}|\leq\varrho$, $|p_{b}^{k,m}-p_{b}^{k,m-1}|\leq\varrho$ and $|y^{k,m}-y^{k,m-1}|\leq\varrho)$}
{
PME release the strategy information $\{\bm{\chi}^{k,m}\}\!=\!\{p_{s}^{k,m},p_{b}^{k,m},y^{k,m}\}$\;
        \For {each nanogrid $i$}
        {
        Each nanogrid $i$ updates $e_{i}^{k,m}$ after receiving PME strategy $\{\bm{\chi}^{k,m}\}$ by solving \textbf{P3}\;
          Each nanogrid $i$ calculates $tp_{i}^{k,m}$ and sends this value to PME\;
          %\EndFor
          }
%          PME Calculates $\delta_{s}^{m}=\frac{1}{\delta_{s,0}+\delta_{s,1}m}$,  $\delta_{b}^{m}=\frac{1}{\delta_{b,0}+\delta_{b,1}m}$,  $\delta_{y}^{m}=\frac{1}{\delta_{y,0}+\delta_{y,1}m}$\;
          PME Calculates the adjust parameters $\delta_{s}^{m}$, $\delta_{b}^{m}$, $\delta_{y}^{m}$\;
          Based on the received $tp_{i}^{k,m}$, PME updates its strategies: $p_{s}^{k,m\!+\!1}\!=\!{{P_{+}[p_{s}^{k,m}\!-\!\delta _{s}^{m}{g_{p_s^k}^{m}}]}}$, $p_{b}^{k,m\!+\!1}\!=\!{{P_{+}[p_{b}^{k,m}\!-\!\delta _{b}^{m}{g_{p_b^k}^{m}}]}}$,  $y^{k,m\!+\!1}\!=\!{{P_{+}[y^{k,m}\!-\!\delta _{y}^{m}{g_{y^k}^{m}}]}}$\;
          $m=m+1$\;
          %\EndWhile
          }
        %Update $B^{k+1}$ by PME according to \eqref{eq26}\;
        Update $T_{i}^{k+1}, H_{i}^{k+1}$ by nanogrid $i$ according to \eqref{eq1}, \eqref{eq17}\;
        Update $B^{k+1}$ by PME according to \eqref{eq26}\
\end{algorithm}
%\vspace{-0.25cm}
The detailed procedure is shown in Algorithm \ref{alg:Framwork} which is separated into two main parts respectively executed by the PME (steps~1-3 and 7-9, 11) and each nanogrid (steps 4-6 and 10) at each slot.
First, PME arbitrarily generates its strategy set including two-way prices and battery charge-discharge amount before the iteration.
The iterative loop in steps~2-9 illustrates the interaction between PME and nanogrids.
Within the $m$th iteration, each nanogrid $i$ receives the strategy set \begin{small}{$\{\bm{\chi}^{k,m}\}$}\end{small} from PME, and determines the HVAC power consumption by minimizing \textbf{P3} with nonlinear programming tools in step~5.
Then, each nanogrid~$i$ calculates its injection power \begin{small}{$tp_{i}^{k,m}$}\end{small} according to \begin{small}{$tp_{i}^{k,m}\!=\!D_{i}^{k}\!+\!e_{i}^{k,m}\!-\!RP_{i}^{k}$}\end{small}, and uploads this value to PME (step~6).
After that, with the collected information \begin{small}{$tp_{i}^{k,m} (i\in{1,...,n})$}\end{small}, PME updates the bidirectional prices and battery charging value based on the subgradient projection method~\footnote{The objective functions are all convex.} \cite[Sec.6.3]{bertsekas2009convex}\cite{liu2017convergence}.
In step~8, \begin{small}{$P_{+}$}\end{small} is the projection operator which has the variables map to the feasible regions defined by constraints \eqref{eq10} and \eqref{eq14}.
%$g_{p_s^k}^{(m)}=-V_{P}\sum\limits_{i=1}^{n}{\max(tp_{i}^{k},0)}+\sum\limits_{i\in{tp_i^k}}{\frac{V_Pp_{s}^{k}}{\hbar_i}}$,
$\scriptstyle{g_{p_s^k}^{m}\!=\!-V_{P}\!\sum\limits_{i=1}^{n}\!{\max(tp_{i}^{k\!,\!m},0)}\!+\!\sum\nolimits_{i\!\in\!\{tp_{i}^{k\!,\!m}\geq0\}}({V_Pp_{s}^{k,m}\hbar_i\!-\!\{V_Pm_{s}^{k}\hbar_i, V_Pm_{b}^{k}\hbar_i\}})}$, $\scriptstyle{g_{p_b^k}^{m}\!=\!-V_{P}\!\sum\limits_{i=1}^{n}\!{\min(tp_{i}^{k\!,\!m},0)}\!+\!\sum\nolimits_{i\!\in\!\{tp_{i}^{k\!,\!m}<0\}}({V_Pp_{b}^{k,m}\hbar_i\!-\!\{V_Pm_{s}^{k}\hbar_i, V_Pm_{b}^{k}\hbar_i\}})}$ (where $\scriptstyle{\hbar_i=\frac{1}{2\gamma_{i}{(1-\varepsilon_{i})^{2}}{{\eta_{i}}^{2}}}}$),
and $\scriptstyle{g_{y^k}^{m}=B^{k,m}+C_{b}y^{k,m}+\{V_Pm_{s}^{k}, V_Pm_{b}^{k}\}}$ denote the subgradients of the optimization function \begin{small}{$pro^{'}$}\end{small} with respect to \begin{small}{$p_{s}^{k}$}\end{small}, \begin{small}{$p_{b}^{k}$}\end{small} and \begin{small}{$y^{k}$}\end{small} during iteration \begin{small}{$m$}\end{small}, respectively.
%These subgradients are expressed as $g_{p_s^k}^{(m)}=-{{V}_{P}}\sum\nolimits_{i=1}^{n}{\max (tp_{i}^{k,(m)},0)}$, $g_{p_b^k}^{(m)}=-{{V}_{P}}\sum\nolimits_{i=1}^{n}{\min (tp_{i}^{k,(m)},0)}$ and ${g_{y^k}^{(m)}}\in \{B^{k}+V_{P}m_{s}^{k}, B^{k}+V_{P}m_{b}^{k}\}$.
%\begin{eqnarray}
%&&\hspace{1.5em}
%g_{p_s^k}^{(m)}=-{{V}_{P}}\sum\nolimits_{i=1}^{n}{\max (tp_{i}^{k,(m)},0)},
%\label{eq703}\\
%&&\hspace{1.5em}
%g_{p_b^k}^{(m)}=-{{V}_{P}}\sum\nolimits_{i=1}^{n}{\min (tp_{i}^{k,(m)},0)},
%\label{eq704}\\
%&&\hspace{-3.5em}
%{g_{y^k}^{(m)}}=\left\{ \begin{aligned}
%& B^{k}+V_{P}m_{s}^{k}\quad&\text{if}\;y^{k,(m)}\geq G_{T}^{k}-\sum\nolimits_{i=1}^{n}{tp_{i}^{k,(m)}};\\
%& B^{k}+V_{P}m_{b}^{k}\quad&\text{if}\,y^{k,(m)}<G_{T}^{k}-\sum\nolimits_{i=1}^{n}{tp_{i}^{k,(m)}}.\\
%\end{aligned} \right.
%\end{eqnarray}
We point out that, in Algorithm \ref{alg:Framwork}, the adjustment parameters for two-way prices and battery charging are adopted as $\scriptstyle{\delta_{s}^{m}=\frac{1}{\delta_{s,0}+\delta_{s,1}m}}$, $\scriptstyle{\delta_{b}^{m}=\frac{1}{\delta_{b,0}+\delta_{b,1}m}}$, $\scriptstyle{\delta_{y}^{m}=\frac{1}{\delta_{y,0}+\delta_{y,1}m}}$ where $\scriptstyle{\delta_{s,0}}$, $\scriptstyle{\delta_{s,1}}$, $\scriptstyle{\delta_{b,0}}$, $\scriptstyle{\delta_{b,1}}$, $\scriptstyle{\delta_{y,0}}$ and $\scriptstyle{\delta_{y,1}}$ are constants. Under such application, the convergence of algorithm can be guaranteed and found in \cite{bertsekas2009convex,wang2016incentivizing}.
The algorithm will turn to the next iteration until the distance between two consecutive iterations is smaller than a specified value \begin{small}{$\varrho$}\end{small}.
Finally, nanogrids and PME will update queue status for the optimization in next time slot.
A simple computation complexity analysis of the proposed algorithm is presented. In fact, the computation complexity of the PME side optimization problem is \begin{small}{$O(n)$}\end{small} and the computation complexity of the one of followers is \begin{small}{$O(1)$}\end{small} respectively, where \begin{small}{$n$}\end{small} is the number of nanogrids.

Actually, the proposed algorithm is executed iteratively in the EMS of nanogrids and PME sides.
The equilibrium of Stackelberg game would be reached in a distributed way naturally in the broader sense.
It can be seen that PME does not need to know the detailed information about power generations, demands, temperature and weighting parameter preferences of all nanogrids and only requires the result of injection power \begin{small}{$tp_{i}^{k,m}$}\end{small} for each nanogrid.
In this way, with less information interchange and only local computation resources, our algorithm can find optimal strategies independently, which helps preserve the users' privacy.
For more detail, the information interaction within the loop steps of Algorithm~\ref{alg:Framwork} is briefly described as follows.
Before time slot~\begin{small}{$k$}\end{small}, the EMS of PME will receive market prices (\begin{small}{$m_{s}^{k}$}\end{small} and \begin{small}{$m_{b}^{k}$}\end{small}) from the main grid.
In each iteration, the EMS of PME updates the pricing strategy set \begin{small}{$\bm{\chi}^{k,m}$}\end{small} and sends them to nanogrids
for their power consumption updates.
After receiving action information of two-way transaction price from the PME, the EMS of nanogrids will react and select its best response strategies.
%It can be seen that PME does not need to know the detailed information about power generations, demands, temperature and weighting parameter preferences of all nanogrids and only requires the result of injection power \begin{small}{$tp_{i}^{k,m}$}\end{small} for each nanogrid.
%In this way, with less information interchange and only local computation resources, our algorithm can find optimal strategies independent of any center, which helps preserve the users' privacy.
%Besides, the proposed algorithm
On the other hand, when the algorithm is compared with the centralized
method based on swarm optimization, our experience shows that the centralized one usually
could converge to the optimum value at a faster speed.
%\begin{figure}[!t]
%\centering
%\includegraphics[width=3.28in]{EPESrererevischart1.eps}
%\caption{Information interactions between PME and nanogrids.}
%\label{fig2}
%\vspace{-0.3cm}
%\end{figure}
\section{Numerical experiments}\label{sec5}
%In this section, we provide the experiment results by applying the proposed algorithm.
In this section, we provide the experiment results by applying the proposed algorithm corresponding to the bilevel energy management problem.
The simulation is performed
on a desktop with an Intel Core i5-7200 CPU 2.50 GHz and 8 GB of RAM
using MATLAB.
%In this section, we provide the experiment results by applying the proposed algorithm for the bilevel energy management problem.
\vspace{-0.25cm}
\subsection{Simulation Setup}
\begin{figure*}[tbp]
\subfloat[Renewable energy generation of nanogrids]{
\label{fig3:mini:subfig:a} %% label for first subfigure
\begin{minipage}[t]{0.33\textwidth}
\centering
\includegraphics[width=2.23in,height=1.24in]{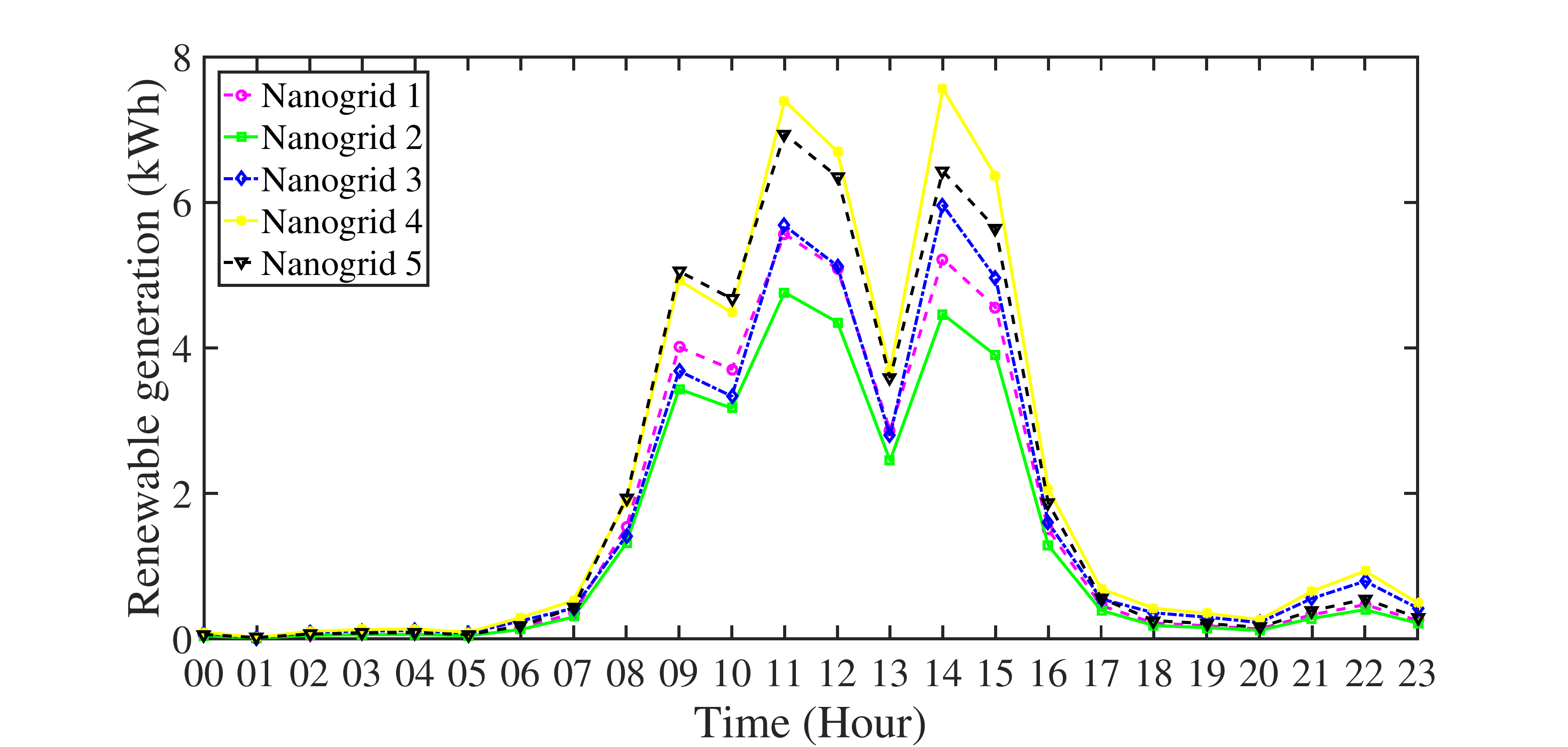}
\end{minipage}}%
\subfloat[Basic inelastic load of nanogrids]{
\label{fig3:mini:subfig:b} %% label for first subfigure
\begin{minipage}[t]{0.33\textwidth}
\centering
\includegraphics[width=2.23in,height=1.24in]{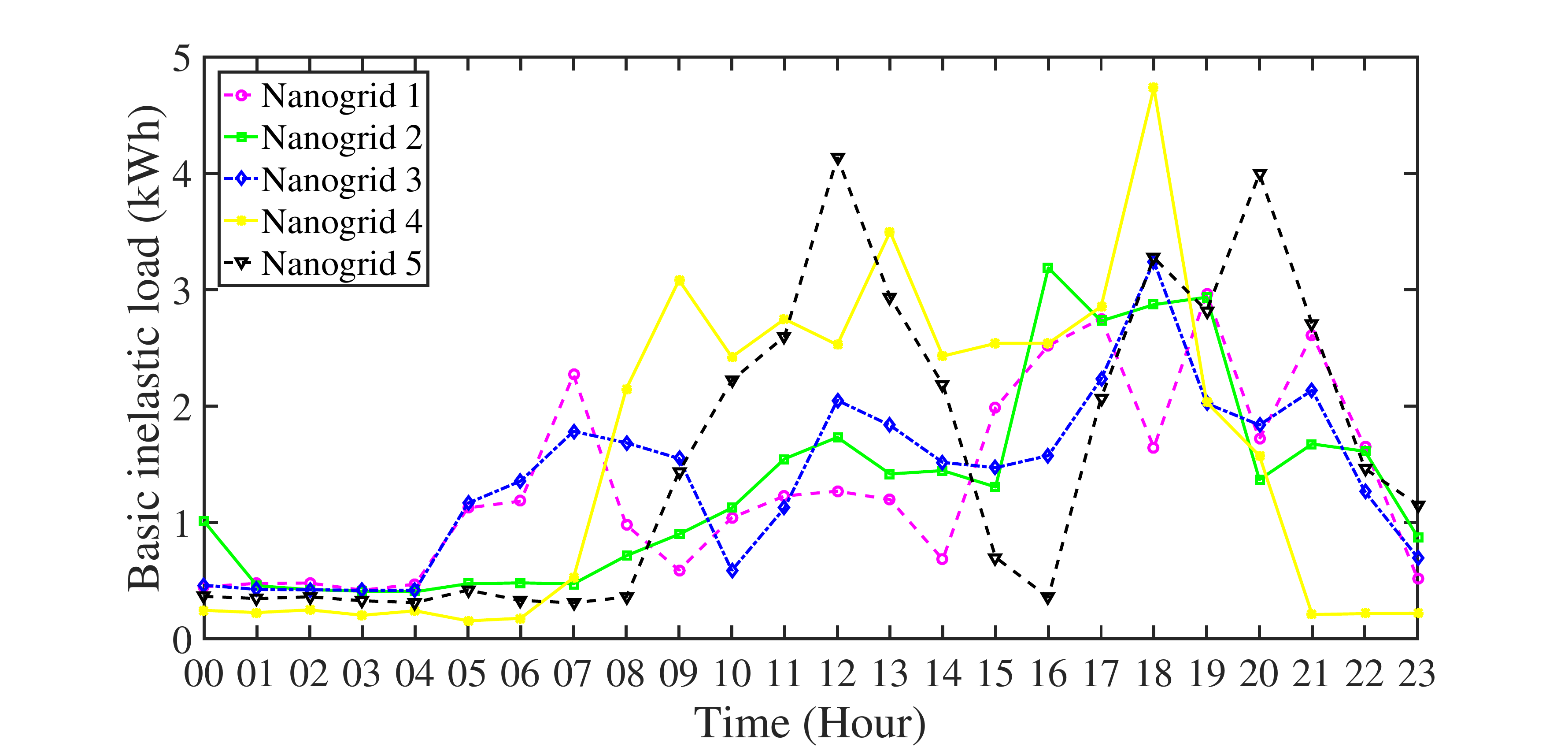}
\end{minipage}}%
\subfloat[Outdoor environment temperature]{
\label{fig3:mini:subfig:c} %% label for first subfigure
\begin{minipage}[t]{0.33\textwidth}
\centering
\includegraphics[width=2.23in,height=1.24in]{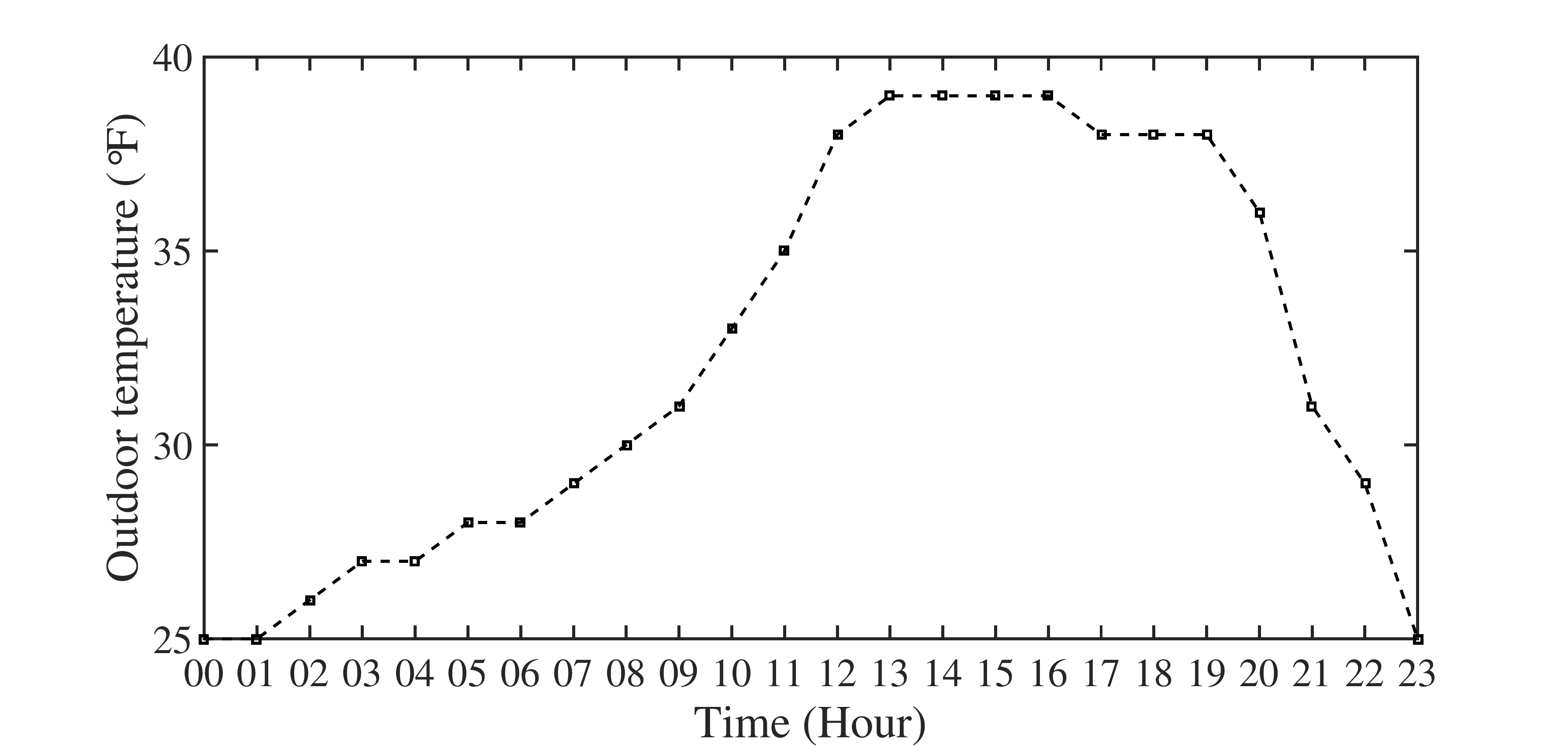}
\end{minipage}}%
\caption{Experiment environment setup.}
\label{fig3:mini:subfig} %% label for entire figure
%\vspace{-0.5cm}
\end{figure*}

In simulation experiment, five nanogrids, a PME and a main grid are considered. Each nanogrid is equipped with basic loads, an HVAC unit and DGs (including rooftop solar photovoltaic panels and small wind turbines).
For the renewable output of DGs in nanogrids, the data given in Fig.~\ref{fig3:mini:subfig}~\subref{fig3:mini:subfig:a} are generated with a typical wind turbine power curve in~\cite{wood2011small} and a photovoltaic generation model in~\cite{li2017helos} using the wind speed and solar radiation data from the websites~\cite{data6} and~\cite{data5}.
The basic loads of nanogrids shown in Fig.~\ref{fig3:mini:subfig}~\subref{fig3:mini:subfig:b} are obtained from~\cite{data3}.
The outdoor temperature data are collected from the online weather website~\cite{data2} as shown in Fig.~\ref{fig3:mini:subfig}~\subref{fig3:mini:subfig:c}.
The inertial coefficient $\varepsilon_{i}$ is set to $[0.93, 0.98]$ which is randomized for different HVAC systems in nanogrids.
As for the parameters in Theorems~\ref{th1} and~\ref{th2}, for the purpose of the largest reduction in the nanogrid's comprehensive energy cost and temperature queue backlog, we adopt $V_{i}=V_{i}^{\max}$, ${{\Gamma}_{i}}={{\Gamma}_{i}^{\min}}$, $V_{P}=V_{P}^{\max}$ and ${\theta}={{\theta}^{\min}}$.
Moreover, we assume that $G_{T}^{k}$ in each slot takes value from $[-15, 25]$~kW uniformly at random.
As for the selling price of main grid, we have used the data from~\cite{data4}.
Besides, the buying price is set to three \textcent/kWh for simplicity. We set the battery cost parameter $C_{b}=0.01$\textcent/(kWh)$^2$. We adopt one hour as the algorithm control slot. Other main parameters
are shown in Table~1.
\begin{table}[tbp]
\centering
\setlength{\abovecaptionskip}{0.cm}
\setlength{\belowcaptionskip}{-0.9cm}
\footnotesize

\textbf{Table 1}~~Simulation parameters.\\
%\label{Tab1}
\setlength{\tabcolsep}{1.25mm}{%

\begin{tabular}{|c|c||c|c||c|c|}

\hline
Parameter&value&Parameter&value&Parameter&value\\
\hline
$e_{i}^{\max}$&5kW&$\eta_{i}$&15$^{\circ}$F/kWh&$\gamma_{i}$&0.01\textcent/($^{\circ}$F)$^{2}$\\
\hline
$T_{i}^{\min}$&66$^{\circ}$F&$T_{i}^{\max}$&77$^{\circ}$F&$E^{\min}$&2kWh\\
\hline
$u^{\rm dmax}$&1kW&$u^{\rm cmax}$&1kW&$E^{\max}$&16kWh\\
\hline
\end{tabular}
}
\label{Tab1}
\vspace{-0.392cm}
\end{table}
\vspace{-0.2cm}
\subsection{Results and Analysis}
\subsubsection{Results of pricing and energy management}
\begin{figure*}
\subfloat[]{
\label{fig20:mini:subfig:a} %% label for first subfigure
\begin{minipage}[t]{0.5\textwidth}
\centering
\includegraphics[width=3in]{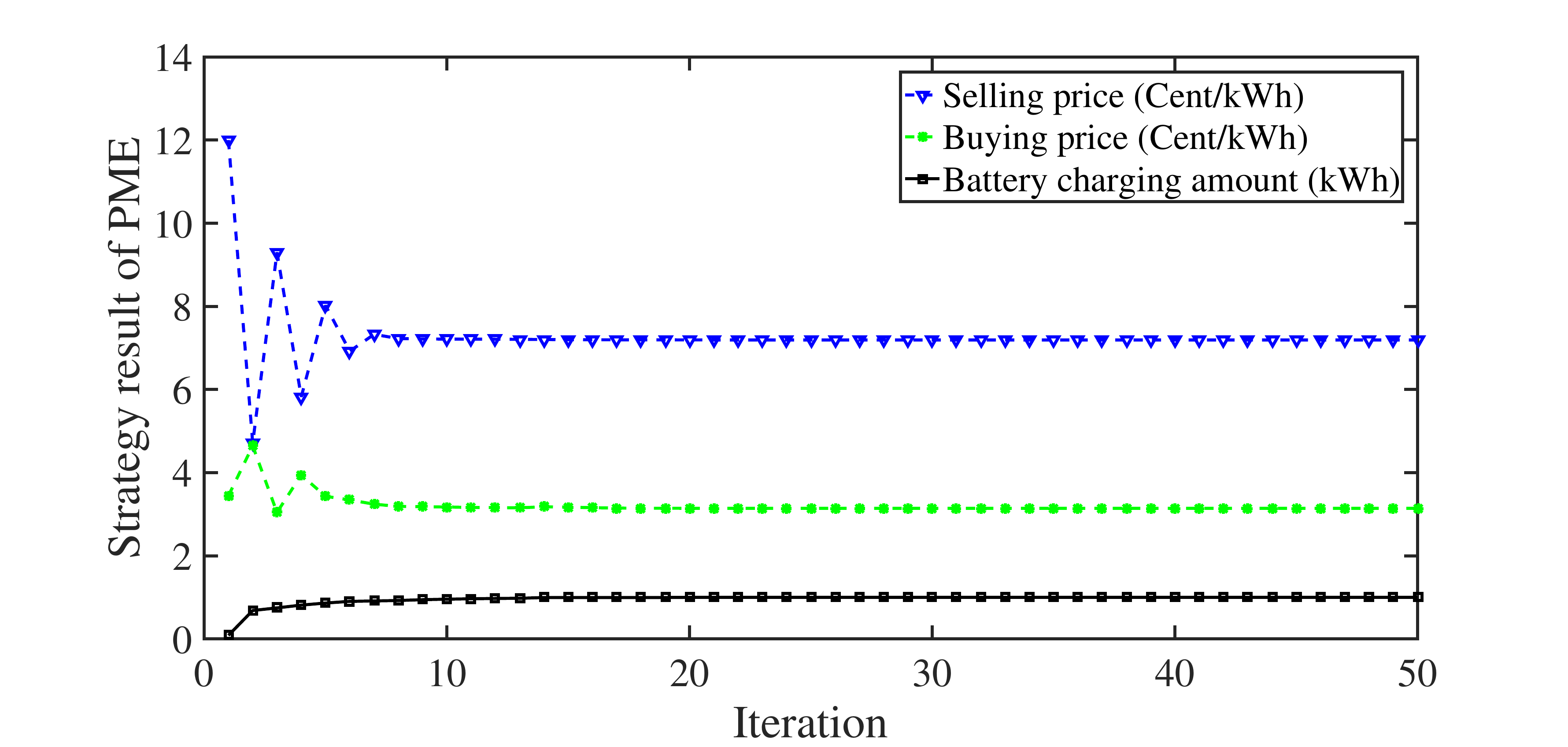}
\end{minipage}}%
\subfloat[]{
\label{fig20:mini:subfig:b} %% label for first subfigure
\begin{minipage}[t]{0.5\textwidth}
\centering
\includegraphics[width=3in]{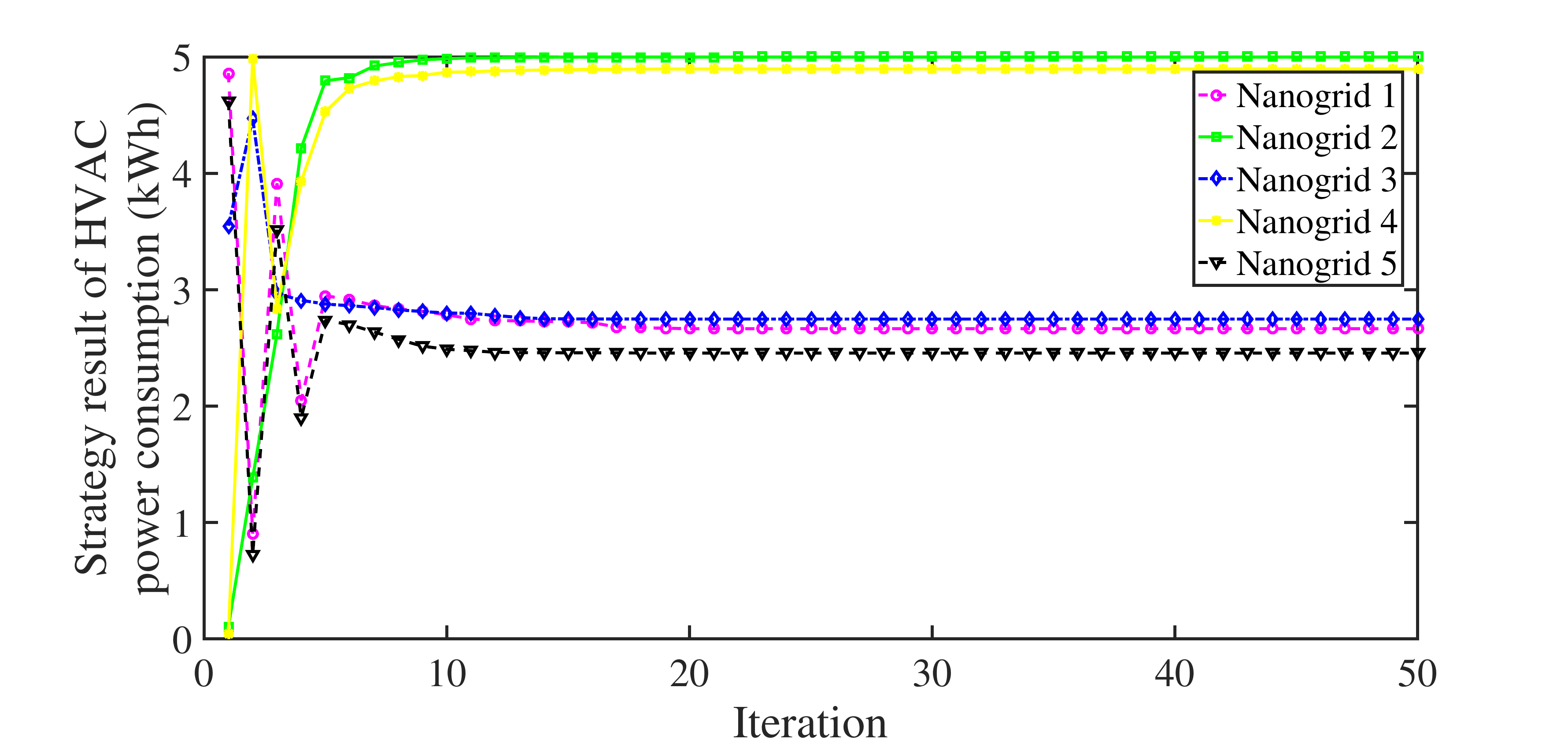}
\end{minipage}}%

\caption[A set of two subfigures.]{Iteration process:
\subref{fig20:mini:subfig:a} Strategy result of PME,
\subref{fig20:mini:subfig:b} Strategy result of nanogrids.}
\label{fig20:mini:subfig} %% label for entire figure
\vspace{-0.392cm}
\end{figure*}
First, based on the algorithm described in Section~\ref{sec4}, the optimization iterative processes are given in Fig.~\ref{fig20:mini:subfig}. It is observed that, from different initial values, the bidirectional prices, battery charging amount of PME and HVAC power consumptions of nanogrids are converged to the equilibrium after about 35 iterations.
\begin{figure*}[!t]
\subfloat[Optimized bidirectional prices of PME]{
\label{fig4:mini:subfig:a} %% label for first subfigure
\begin{minipage}[t]{0.5\textwidth}
\centering
\includegraphics[width=3in]{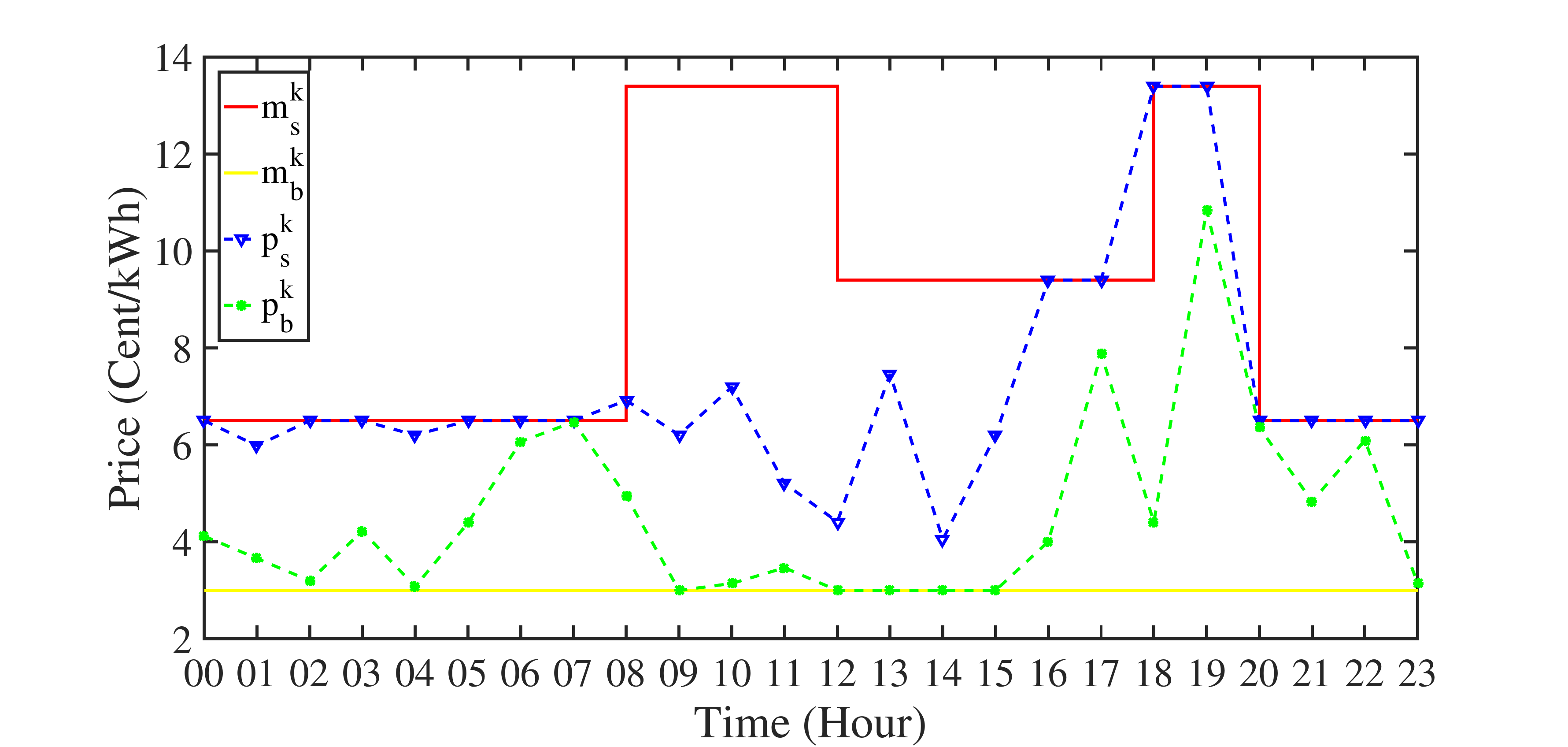}
\end{minipage}}%
\subfloat[Power consumption of HVAC units in nanogrids]{
\label{fig4:mini:subfig:b} %% label for first subfigure
\begin{minipage}[t]{0.5\textwidth}
\centering
\includegraphics[width=3in]{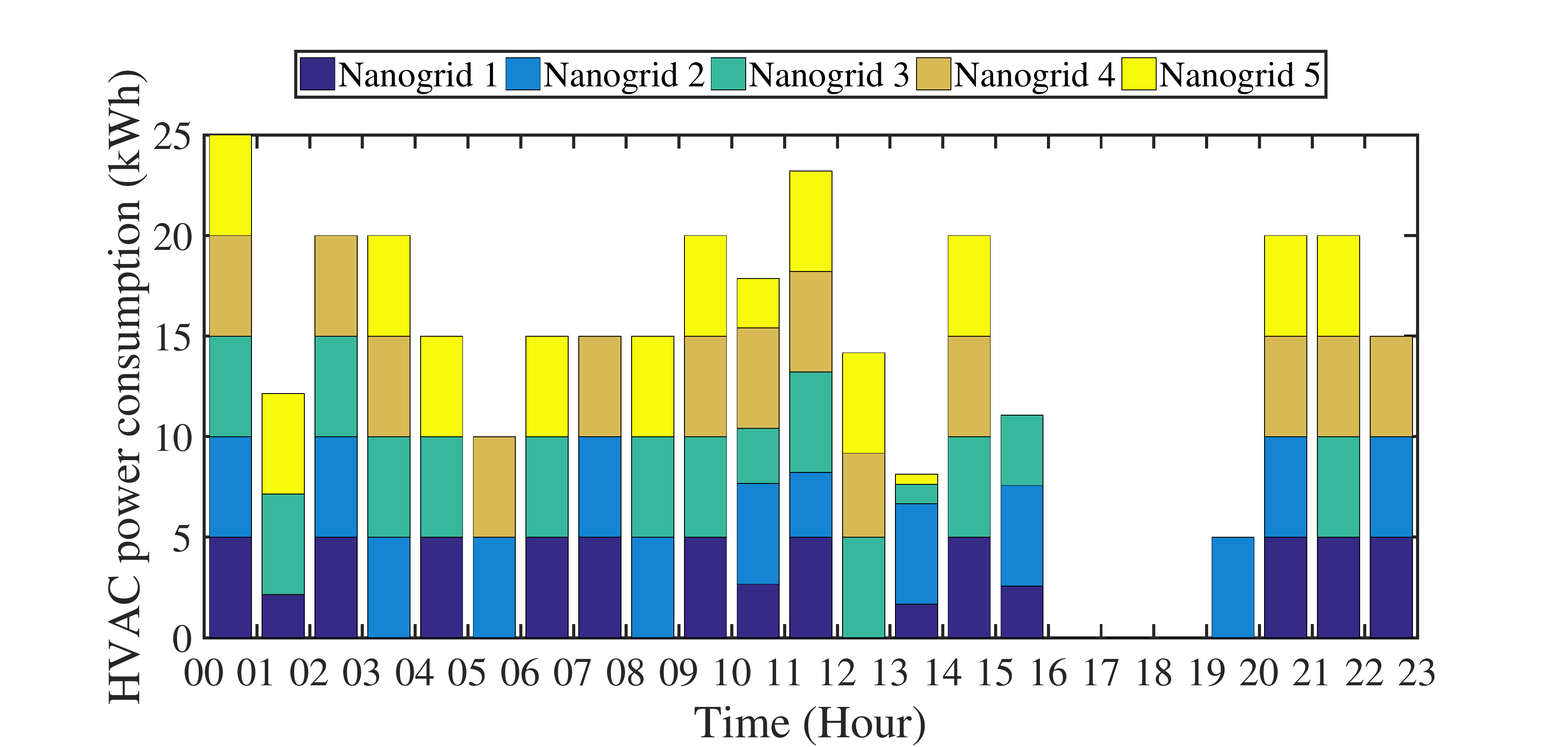}
\end{minipage}}%

\subfloat[Indoor temperature of nanogrids]{
\label{fig4:mini:subfig:c} %% label for first subfigure
\begin{minipage}[t]{0.5\textwidth}
\centering
\includegraphics[width=3in]{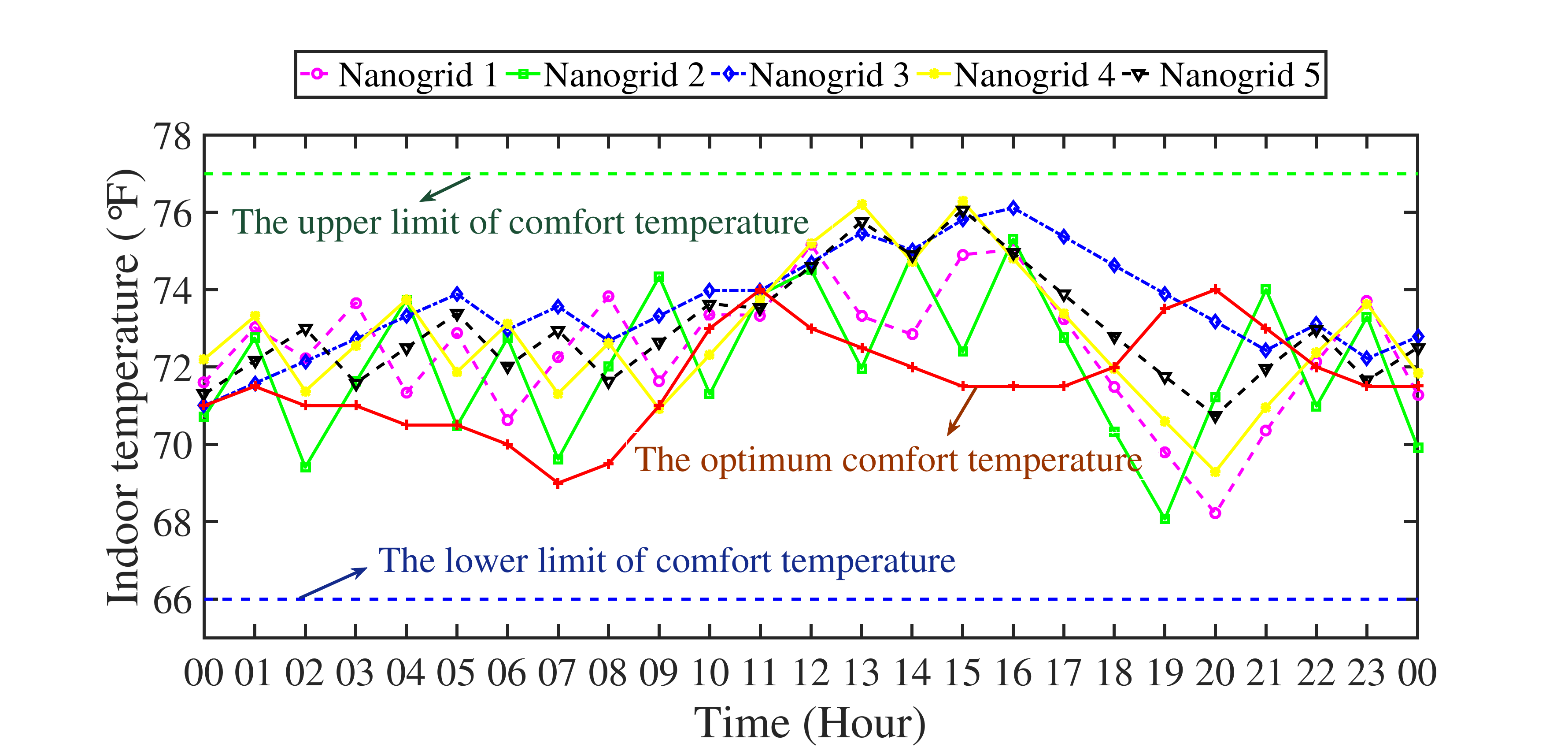}
\end{minipage}}%
\subfloat[Battery energy level of PME]{
\label{fig4:mini:subfig:d} %% label for first subfigure
\begin{minipage}[t]{0.5\textwidth}
\centering
\includegraphics[width=3in]{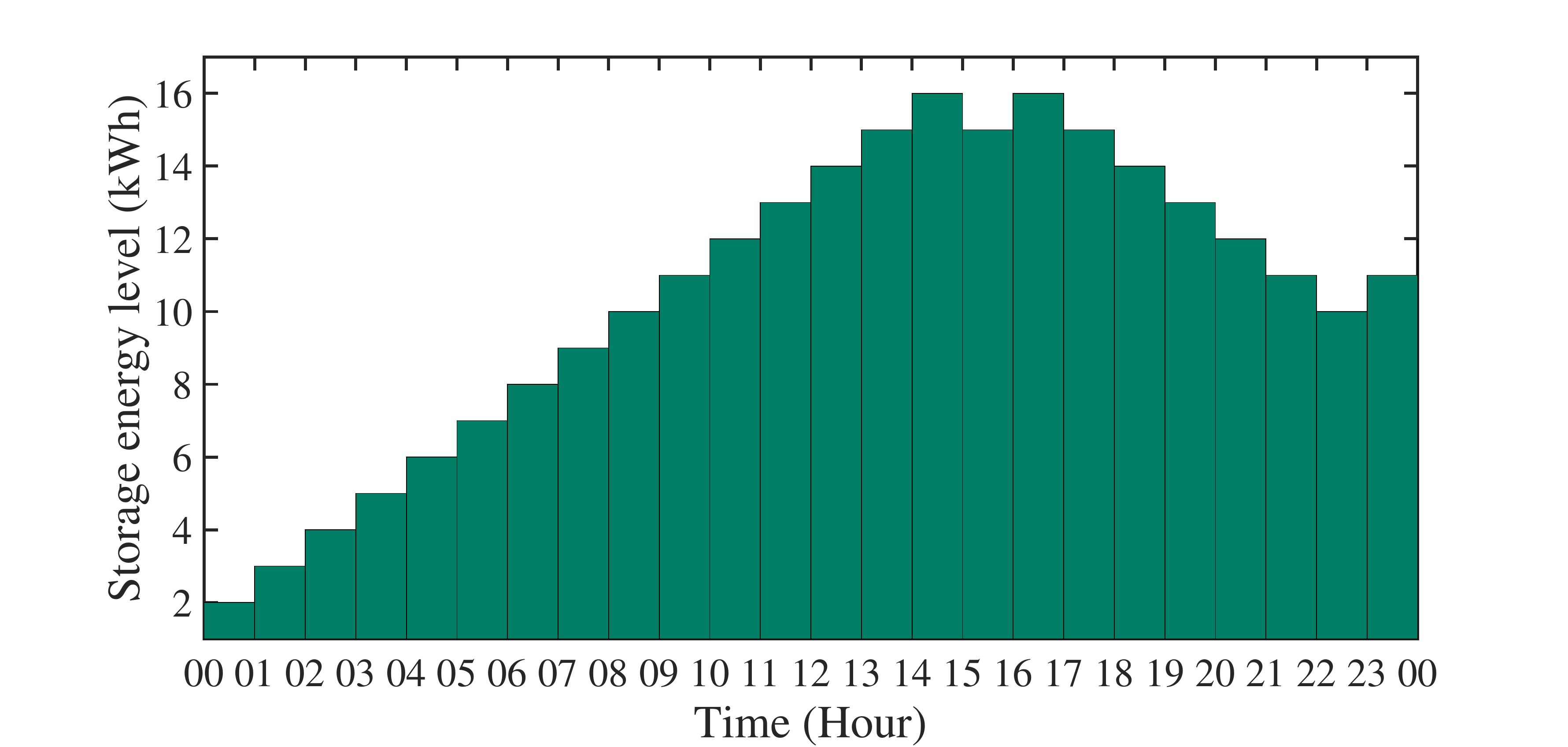}
\end{minipage}}%
\caption{Optimized results of pricing and energy management.}
\label{fig4:mini:subfig}
 %% label for entire figure
\vspace{-0.282cm}
\end{figure*}

The optimized selling and buying prices of PME are presented with the blue and green dashed line in Fig.~\ref{fig4:mini:subfig}~\subref{fig4:mini:subfig:a} respectively.
It is observed that the selling prices of PME are not higher than the selling prices of the main grid across the total time horizon. Besides, the purchasing prices of PME are not lower than the ones of the main grid. Thus, instead of trading with the main grid directly, nanogrids can benefit from this trading pattern. Simultaneously, the PME can also obtain more revenue because its purchasing prices are lower than the selling prices of the main grid.
Besides, the optimal power consumptions of HVAC units in nanogrids are given in Fig.~\ref{fig4:mini:subfig}~\subref{fig4:mini:subfig:b}.
Specifically, when selling (buying) prices of PME become large, the HVAC power consumption is decreased to reduce the nanogrids' energy purchasing cost (to increase the gain from power selling).

In addition, we check the optimized results of indoor temperature and the battery energy level to validate Theorem~\ref{th1} and Theorem~\ref{th2}.
In Fig.~\ref{fig4:mini:subfig}~\subref{fig4:mini:subfig:c}, a time-varying optimum comfort temperature is adopted and shown as the red solid line.
It can be found that the indoor temperatures of all nanogrids fall between the upper and lower bounds of comfort temperature which proves that the desired temperature constraints can be met by the proposed algorithm under the time-varying optimum comfort temperature.
Likewise, in Fig.~\ref{fig4:mini:subfig}~\subref{fig4:mini:subfig:d}, it is observed that the battery energy level varies within $[2, 16]$~kWh which verifies Theorem~\ref{th2}.
\subsubsection{Economic benefit evaluation}
We further evaluate the economic performance of proposed algorithm with other cases:
(1) Case~1 is similar to~\cite{adhikari2016simulation} which employs a fixed-point temperature control method to maintain the optimum indoor comfort temperature for residents in nanogrids.
(2) Case 2 based on~\cite{hao2017transactive} also tends to pursue the optimum temperature. The main difference between these two cases is that the second adopts optimized real-time pricing while the first is based on the forecast of the balancing market prices.
(3) Case 3 based on the game model proposed in~\cite{ye2017game} aims to minimize the energy cost at each time without taking account of the future optimization.
(4) Case 4 is the proposed algorithm in Section~\ref{sec4}.
%The fourth is the proposed algorithm in this work.
(5) In case 5, a modified algorithm is proposed with a social welfare scenario to optimize the aggregate cost\footnote{The aggregate cost $=$ nanogrids' discomfort cost $+$ nanogrids' energy trading cost $-$ trading profit of the PME.} of PME and nanogrids. In this scenario, the HVAC power consumption and battery charging amount are regulated concurrently under the premise that all the participants are cooperative (i.e., no pricing and charges for PME and nanogrids). The social welfare in time slot $k$ is formulated as $C_{s}^{k}\!=\!\tfrac{1}{2}C_{b}(y^{k})^{2}\!+\!m_{s}^{k}\!\cdot\!\max(\sum_{i=1}^{n}{tp_{i}^{k}\!-\!G_{T}^{k}+{{y}^{k}}},0)\!+\!m_{b}^{k}\!\cdot\!\min (\sum_{i=1}^{n}{tp_{i}^{k}\!-\!G_{T}^{k}+{{y}^{k}}},0)\!+\!\sum_{i=1}^{n}\{\gamma_{i} {{(T_{i}^{k+1}\!-\!T_{i}^{opt,k+1})}^{2}}\}$. Thus the corresponding long-term social welfare optimization problem is given as follows,
\begin{align}
\min \quad &\overline{C_{s}}=\underset{T\to \infty}{\mathop{\lim }}
\tfrac{1}{T}\sum\nolimits_{k=0}^{T-1}{\mathbb{E}\left\{ {{C}_{s}^{k}}\right\}}\label{eq607}\notag\\
{\rm s.t.}\quad&(1)-(8) \quad \forall k. \notag
\end{align}
%\vspace{-0.2cm}

\newcommand{\tabincell}[2]{
\begin{tabular}{@{}#1@{}}#2\end{tabular}
}

\begin{table}[tbp]
\centering
\footnotesize
\textbf{Table 2}~~Numerical comparison results (given unit: cent).\\
\setlength{\tabcolsep}{1.8mm}{%

\begin{tabular}{@{}cccccc@{}}

\toprule

  & Case 1 & Case 2 & Case 3 & Case 4 & Case 5\\[3pt] \midrule

\tabincell{c}{Trading profit\\of the PME} & 743.194 & 779.48 & 1652.556 & 1738.771 & $\smallsetminus$\\[5pt]

\tabincell{c}{Energy cost\\of nanogrids} & 2928.653 & 2871.029 & 2264.928 & 2229.346 & $\smallsetminus$ \\[5pt]

\tabincell{c}{Discomfort cost\\of nanogrids} & 0.158 & 0.158 & 38.315 & 5.454 & 5.768 \\[5pt]

Aggregate cost & 2185.617 & 2091.707 & 650.687 & 496.029 & 359.736 \\ \bottomrule

\end{tabular}%

}
\label{Tab2}
\vspace{-0.182cm}
\end{table}
The comparison results are given in Table~2.
By comparing case~1 with case 2, we find that algorithm with real-time pricing can increase revenue of the PME and reduce the energy trading cost of users in nanogrids.
It is observed that case 3 can further reduce the aggregate cost by taking part in the game. However, its thermal discomfort cost is remarkably increased by 38.157 cents.
By optimizing the utility in a long-term horizon with two-way pricing, the discomfort cost of case~4 has decreased by 85.77\% from case 3. And the aggregate cost of case~4 is further reduced by 154.658 cents.
Besides, compared with case~2, the profit of PME in the proposed algorithm is increased by 959.291 cents and users' energy trading cost is reduced by 641.683 cents.
Furthermore, the aggregate costs of case 4 and case 5 have gone down by 76.29\% and 82.8\% from the case~2.
To sum up, the last two cases can provide effective approaches to scheduling the consumptions of HVAC units when users in nanogrids pay attention to both thermal discomfort and aggregate cost.
\subsubsection{The impact of comfort temperature range}
\begin{figure*}[!t]%%%% Discomfort cost under the variation of $T_{i}^{\min}$
\subfloat[Discomfort cost under different $T_{i}^{\min}$]{
\label{fig21:mini:subfig:a} %% label for first subfigure
\begin{minipage}[t]{0.33\textwidth}
\centering
\includegraphics[width=2in]{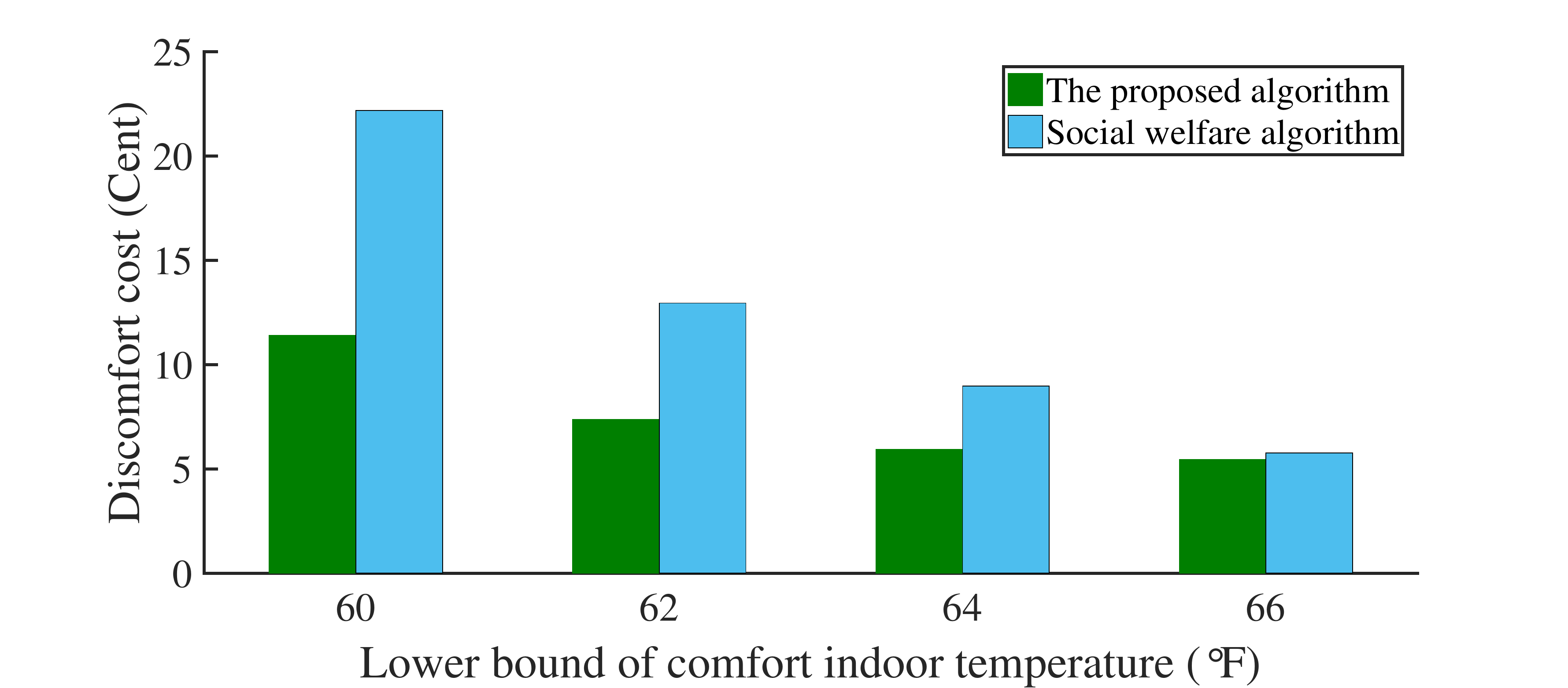}
\end{minipage}}%
\subfloat[Aggregate cost under different $T_{i}^{\min}$]{
\label{fig21:mini:subfig:b} %% label for first subfigure
\begin{minipage}[t]{0.33\textwidth}
\centering
\includegraphics[width=2in]{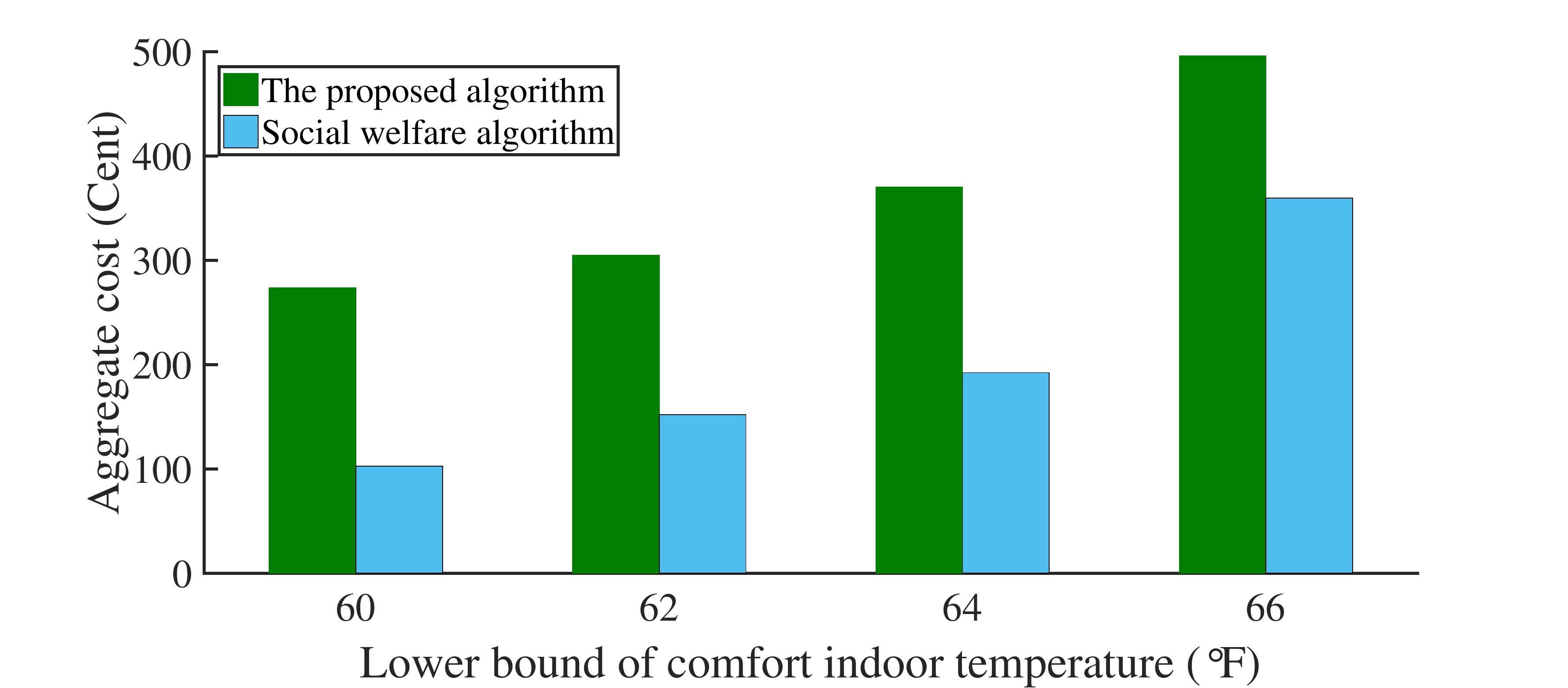}
\end{minipage}}%
\subfloat[Power consumption of HVAC units under different $T_{i}^{\min}$]{
\label{fig21:mini:subfig:c} %% label for first subfigure
\begin{minipage}[t]{0.33\textwidth}
\centering
\includegraphics[width=2in]{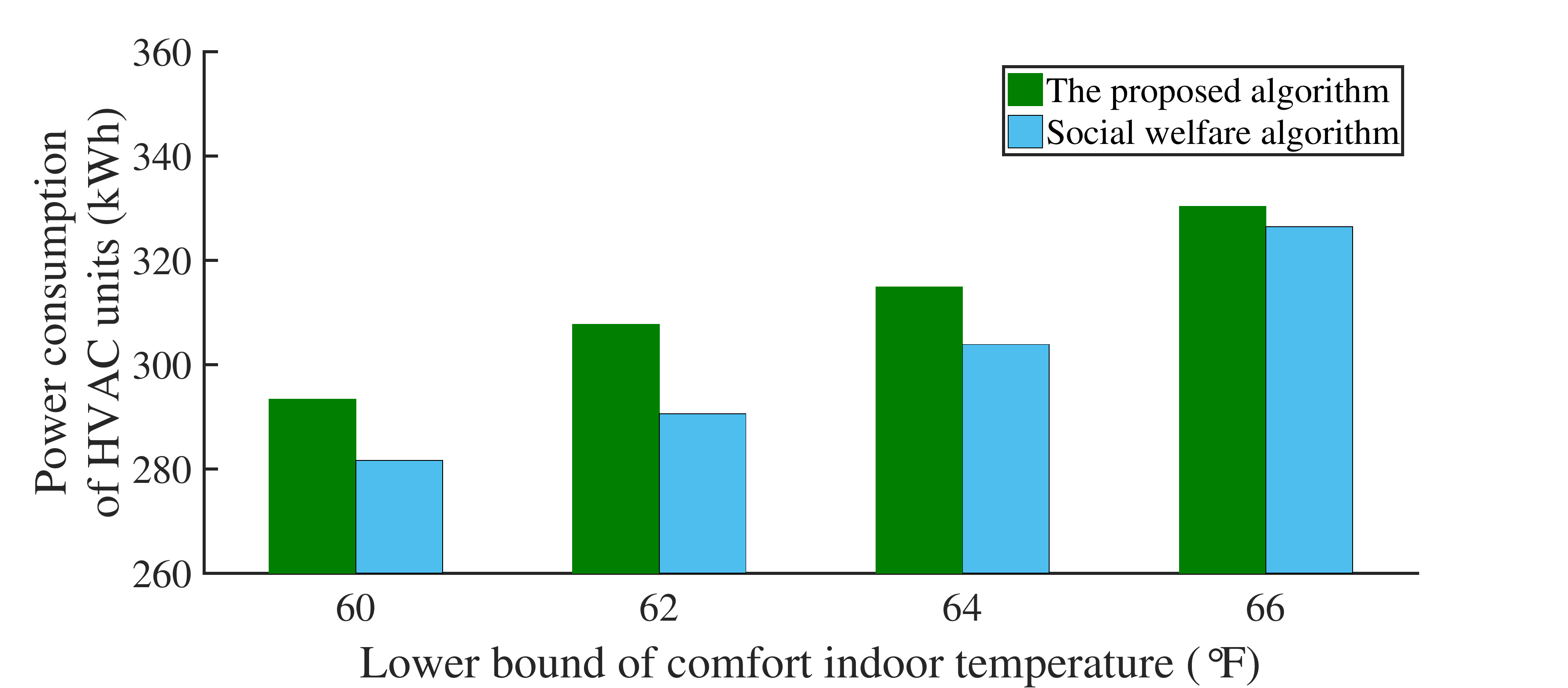}
\end{minipage}}%
\caption{The impact of $T_{i}^{\min}$.}
\label{fig21:mini:subfig}
%\vspace{-0.382cm}
\end{figure*}
\begin{figure*}[!t]%%%% two clomun
\subfloat[Discomfort cost under different $T_{i}^{\max}$]{
\label{fig22:mini:subfig:a} %% label for first subfigure
\begin{minipage}[t]{0.33\textwidth}
\centering
\includegraphics[width=2in]{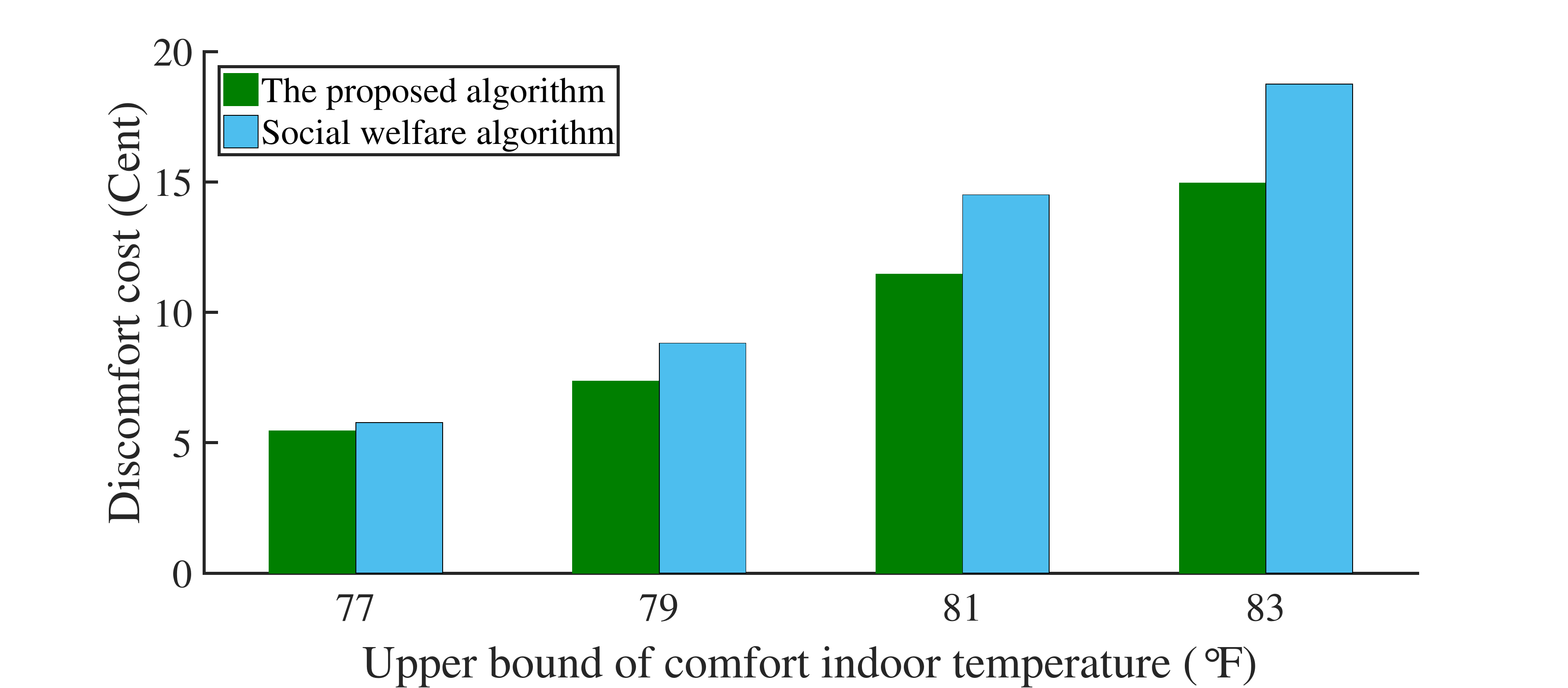}
\end{minipage}}%
\subfloat[Aggregate cost under different $T_{i}^{\max}$]{
\label{fig22:mini:subfig:b} %% label for first subfigure
\begin{minipage}[t]{0.33\textwidth}
\centering
\includegraphics[width=2in]{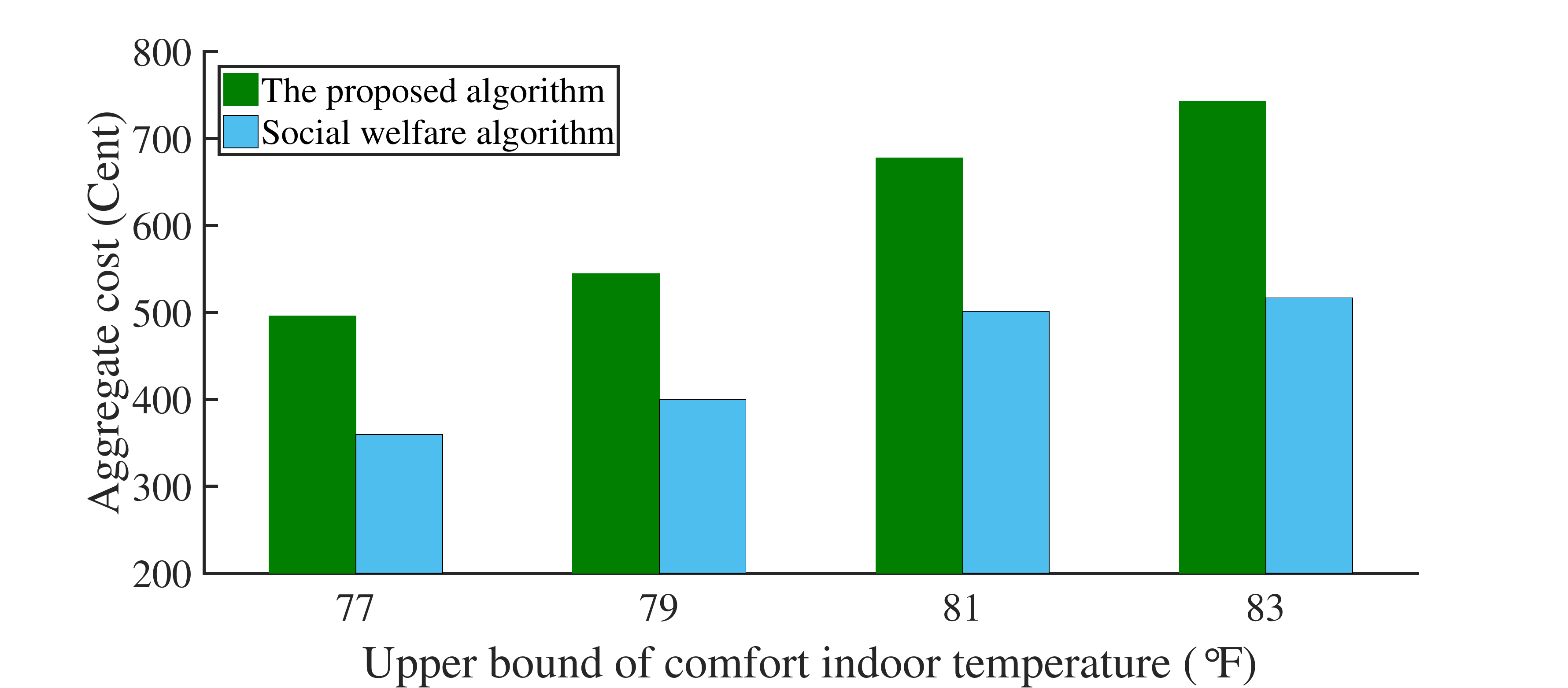}
\end{minipage}}%
\subfloat[Power consumption of HVAC units under different $T_{i}^{\max}$]{
\label{fig22:mini:subfig:c} %% label for first subfigure
\begin{minipage}[t]{0.33\textwidth}
\centering
\includegraphics[width=2in]{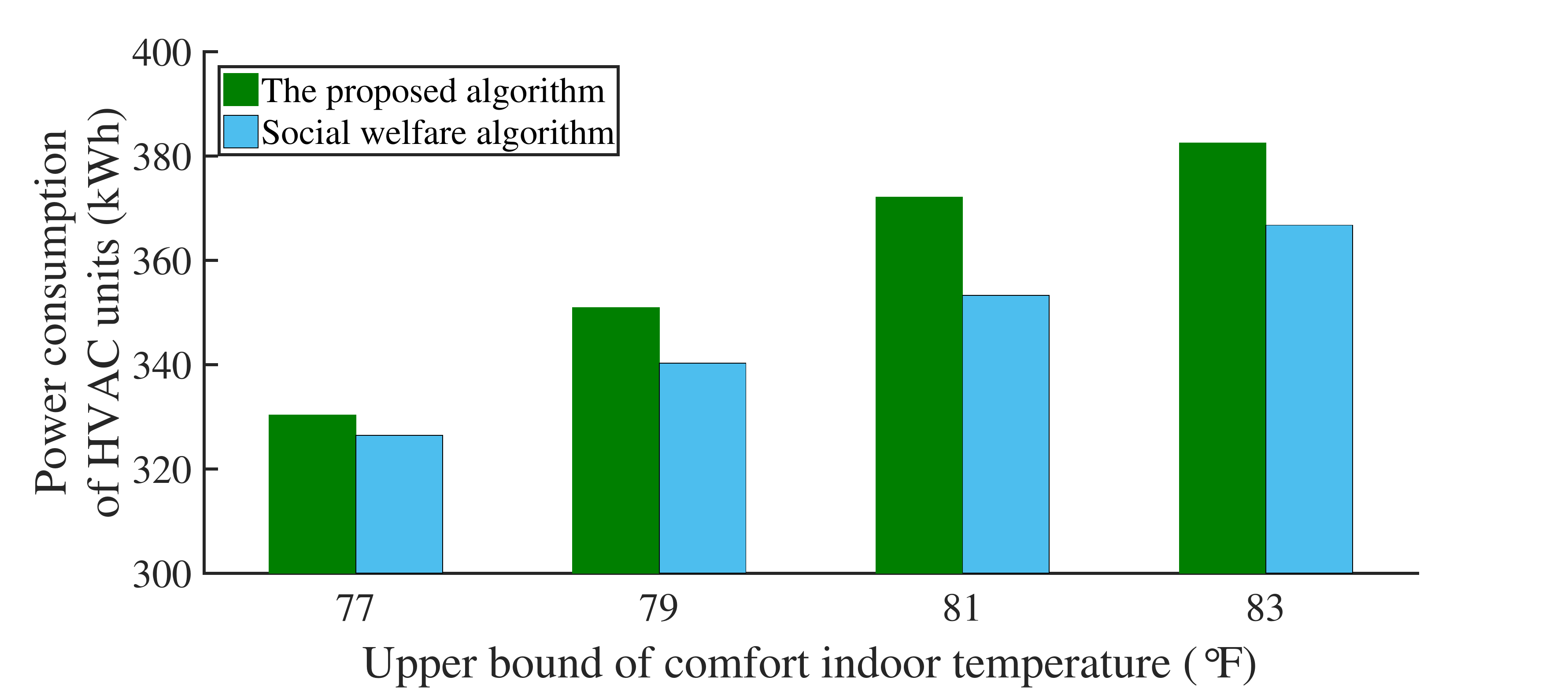}
\end{minipage}}%
\caption{The impact of $T_{i}^{\max}$.}
\label{fig22:mini:subfig}
%\vspace{-0.382cm}
\end{figure*}
The impact of larger comfort temperature range is investigated by reducing/rasing the lower/upper limit of comfort temperature, individually.
From Fig.~\ref{fig21:mini:subfig}~\subref{fig21:mini:subfig:a} and Fig.~\ref{fig22:mini:subfig}~\subref{fig22:mini:subfig:a}, the discomfort cost is elevated along with the decrease of $T_{i}^{\min}$ and the increase of $T_{i}^{\max}$. It demonstrates that a larger comfort temperature range will lead to a higher thermal discomfort cost.
Fig.~\ref{fig21:mini:subfig}~\subref{fig21:mini:subfig:b} and Fig.~\ref{fig22:mini:subfig}~\subref{fig22:mini:subfig:b} show that the aggregate cost of the proposed approach is larger than the value of modified social welfare scenario owing to the selfishness of the players in Stackelberg game. By comparing Fig.~\ref{fig21:mini:subfig}~\subref{fig21:mini:subfig:b} with Fig.~\ref{fig22:mini:subfig}~\subref{fig22:mini:subfig:b}, we find that the aggregate cost is reduced along with the decrease of $T_{i}^{\min}$ and rises along with increasing $T_{i}^{\max}$.
The intuition behind such result is that when increasing $T_{i}^{\max}$, on one hand, the discomfort cost increases. On the other hand, the indoor temperature tends to maintain a higher level compared with a smaller $T_{i}^{\max}$ since $T_{i}^{\max}$ is the upper bound of average indoor temperature.
Consequently, there is a larger power consumption of the HVAC unit in heating mode, which results in a higher energy cost.
The optimized total power consumptions of HVAC units in nanogrids
have been provided in Fig.~\ref{fig21:mini:subfig}~\subref{fig21:mini:subfig:c} and Fig.~\ref{fig22:mini:subfig}~\subref{fig22:mini:subfig:c}. The results verify that the HVAC power consumptions have lowered along with the decrease in $T_{i}^{\min}$ and increased along with the increase in $T_{i}^{\max}$.
\subsubsection{The economic profit of battery}
\begin{figure}[t]
\centering
\includegraphics[width=3.5in]{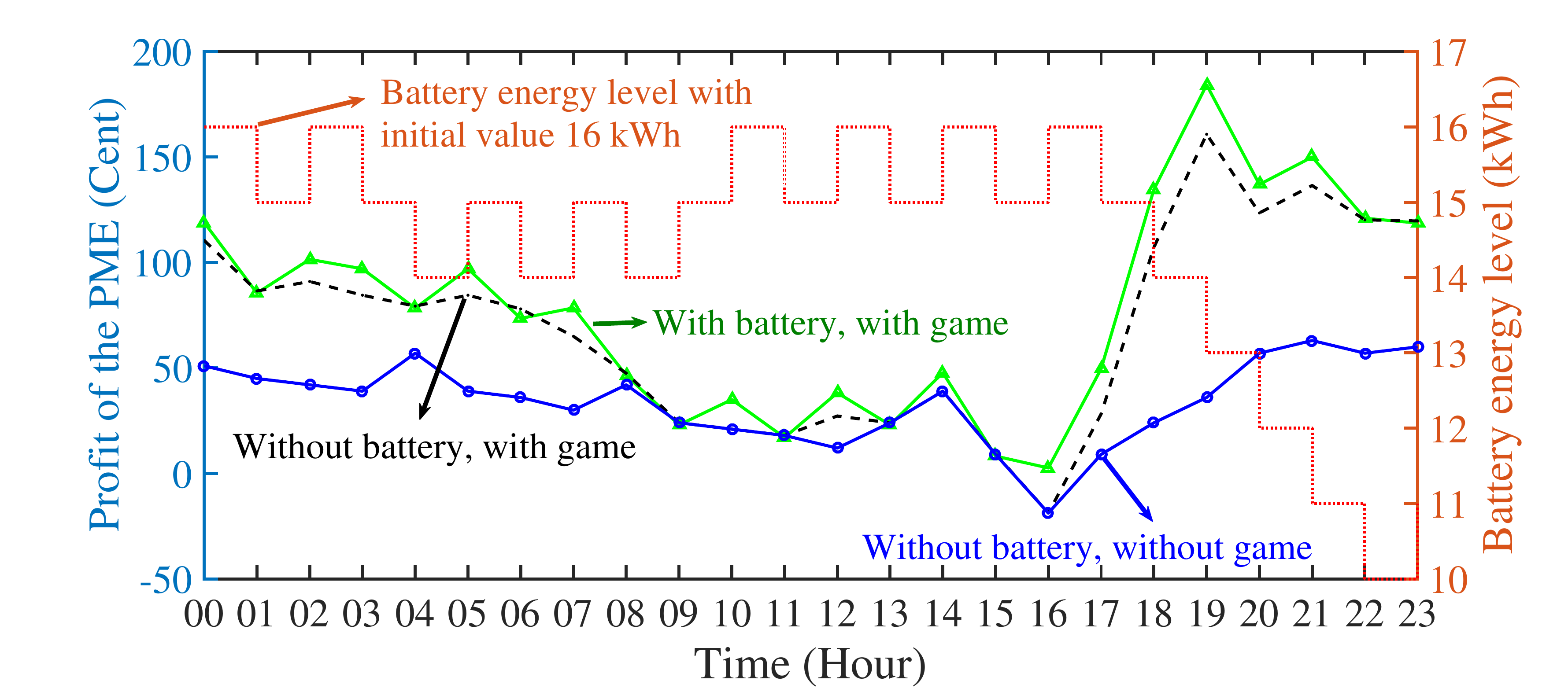}
\caption{Comparison results of trading profit procured by PME.}
\label{fig23}
\vspace{-0.36cm}
\end{figure}
The impact of battery storage system in the proposed Stackelberg game is evaluated by comparing the trading profit of PME with the other two scenarios: (\romannumeral1) A PME without any energy storage device and does not participate in the Stackelberg game; (\romannumeral2) A PME without any energy storage device but participates in the Stackelberg game. As shown in Fig.~\ref{fig23}, the profit of PME under the proposed algorithm (corresponding with the green solid line) is usually higher than the other scenarios. On one hand, by participating in the game, there is a significant increase in the trading revenue for PME. This is because the PME in the game can optimize its profit by selling a portion of energy to nanogrids at a higher price as compared with the buying price of the main grid (i.e., $p_{s}^{k}>m_{b}^{k}$). In addition, the PME can also procure a part of energy from nanogrids cheaply considering the higher selling prices of the main grid (i.e., $p_{b}^{k}<m_{s}^{k}$). On the other hand, when the battery is discharged during the peak period (e.g. 17 hour-20 hour in Fig.~\ref{fig23}) the profit of PME under the proposed algorithm becomes higher as compared with the second scenario. It is because that less amount of electricity will be purchased from the main grid with the pre-stored energy.
%It is because that with the pre-stored energy, less amount of electricity will be purchased from the main grid in this circumstance.
\subsubsection{The impact of discomfort weighting coefficient}
\begin{figure}[tbp]
\begin{minipage}[t]{0.5\textwidth}
\centering
\includegraphics[width=3.6in]{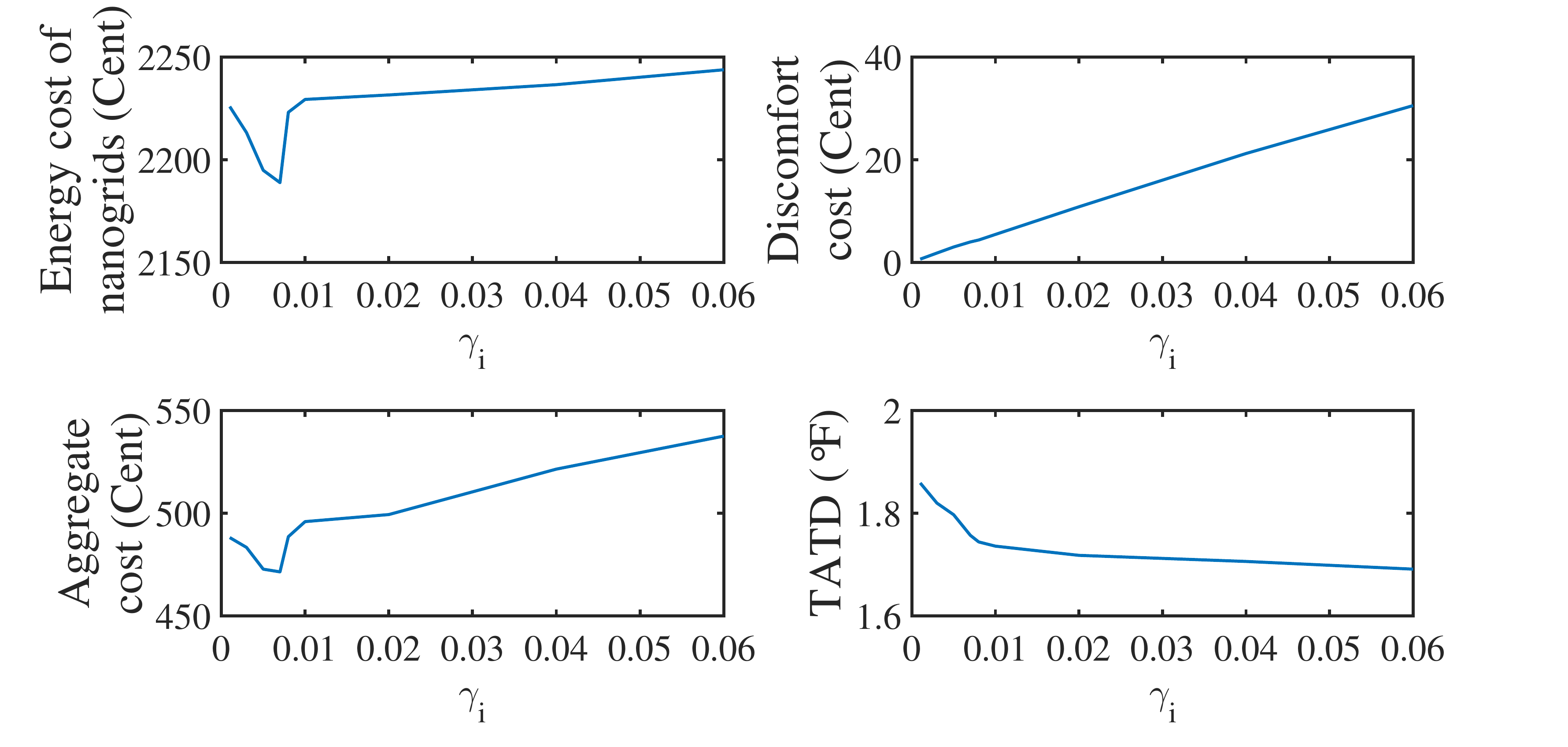}
\caption{The impact of varying $\gamma_{i}$.}
\label{fig24}
\end{minipage}%
\hfill
\begin{minipage}[t]{0.5\textwidth}
\centering
\includegraphics[width=3.6in]{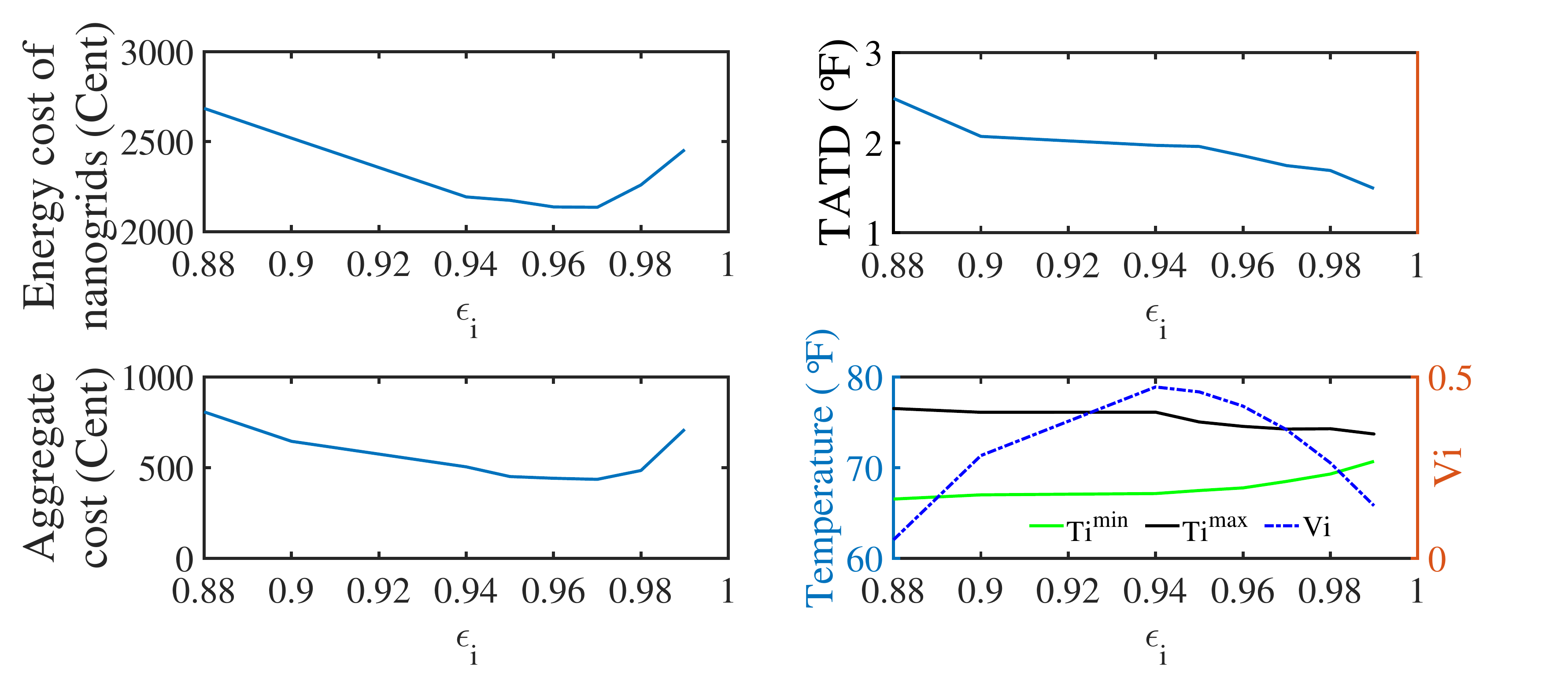}
\caption{The impact of varying $\varepsilon_{i}$.}
\label{fig25}
\end{minipage}
%\vspace{-0.39cm}
\vspace{-0.19cm}
\end{figure}

As shown in Fig.~\ref{fig24}, the influence of the varying cost weighting coefficient $\gamma_{i}$
on the performance of the proposed algorithm is illustrated.
It is observed that the proposed algorithm can obtain the minimum aggregate cost and nanogrids' energy cost when $\gamma_{i}$ is located at $[0.007, 0.008]$.
Besides it is found that the thermal discomfort cost increases near linearly with $\gamma_{i}$.
The total average temperature deviation (TATD) from the optimum comfort temperature is decreased with the increasing $\gamma_{i}$ (
$\text{TATD}=\tfrac{1}{nT}\sum\nolimits_{i=1}^{n}{\sum\nolimits_{k=0}^{T-1}{\left| T_{i}^{k+1}-T_{i}^{opt,k+1} \right|}}$).
And the descent rate slows down when $\gamma_{i}$ increases to a certain value (i.e., about $0.01$).
%Besides, it is found that the proposed algorithm can obtain the minimum aggregate cost and nanogrids' energy cost when $\gamma_{i}$ located at $[0.007, 0.008]$.
%\begin{figure}[t]
%\centering
%\includegraphics[width=3.6in]{gammma2.eps}
%\caption{The impact of varying $\gamma_{i}$.}
%\label{fig24}
%\end{figure}
\subsubsection{The impact of HVAC inertial coefficient}
The performance of the proposed algorithm under varying inertial coefficient of the HVAC unit is investigated as shown in Fig.~\ref{fig25}. We find that a smaller nanogrids' energy cost and a smaller aggregate cost can be procured given a bigger $\varepsilon_{i}$ within a certain range. Besides, when $\varepsilon_{i}$ exceeds $0.97$, the nanogrids' energy cost and aggregate cost will increase instead. The reason can be found from the fourth subfigure. When $\varepsilon_{i}$ is large enough, the weighting parameter $V_{i}$ becomes smaller and the actual indoor temperature range becomes more narrow. Recall that $V_{i}$ denotes a tradeoff between the decrease of comprehensive energy cost and the indoor temperature queue stability. Therefore, when $\varepsilon_{i}$ is bigger than a certain value, the energy cost and aggregate cost will increase. In addition, under a narrow indoor temperature range, the TATD is decreased along with the increase of $\varepsilon_{i}$ which is shown in the second subfigure of Fig.~\ref{fig25}.
\subsubsection{The impact of number of nanogrids}
\begin{figure}[t]
\centering
\includegraphics[width=3.25in]{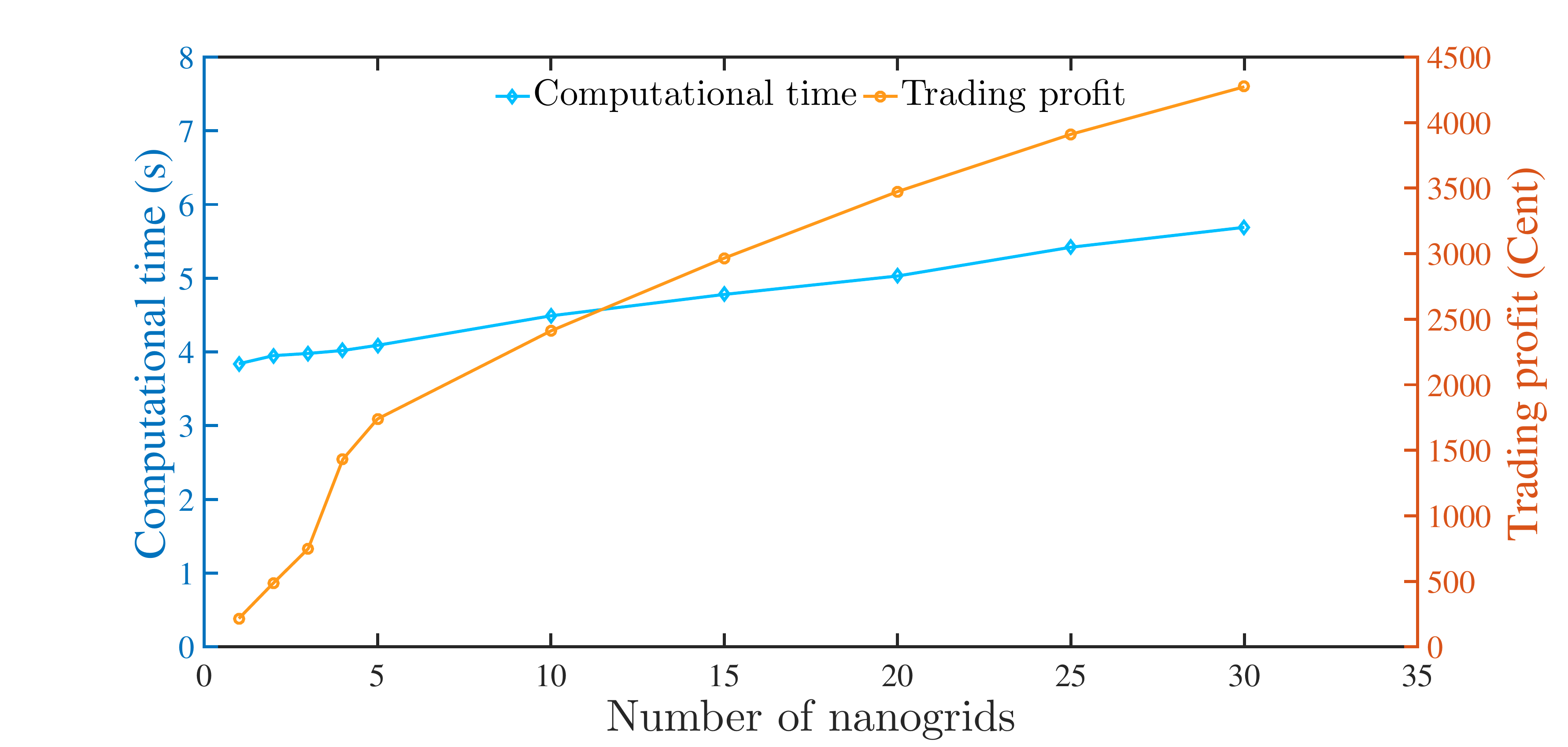}
\caption{Optimal results with different number of nanogrids.}
\label{fig1231}
\vspace{-0.36cm}
\end{figure}
In this subsection, the impact of the different number of nanogrids on computational time is demonstrated. The amount of nanogrids $n$ is increased from 1 to 30. Fig.~\ref{fig1231} presents the average computational time for each optimization problem. It is shown that the total computational time grows near polynomially with the increase in the number of follower players (nanogrids).
%Moreover, compared with the general computation requirements (usually setting $900$s ahead), the computational time is appropriately short and can boot its scalability in a larger amount of followers.
Moreover, compared with the adopted 1-hour time-scale (the strategy optimization is commonly required to be completed $15$min, i.e., $900$s ahead),
the computational time is appropriately short and can boot its scalability in a larger amount of followers.
Meanwhile, we can notice from Fig.~\ref{fig1231} that the trading profit of PME also rises progressively with more nanogrids for extended market share.
In this case, the proposed algorithm is sufficient in time complexity and privacy preservation for the optimization of energy transactions.
\section{Conclusion}\label{sec6}
In this paper, to stimulate the consumption of renewable energy as well as the long-term profits, a three-layer trading framework including the main grid, PME and nanogrids is devised where the energy transactions between different levels work both ways.
A bidirectional pricing scheme and novel DR problems are proposed in order to make a joint-optimization for PME and nanogrids with HVAC units in a long-term horizon.
%And trading interactions between PME and nanogrids are modeled by the Stackelberg game.
Considering the time-coupling properties of temperature and battery queue constraints, we resolve the time-averaged stochastic utility optimization problems by using the Lyapunov optimization technique.
The trading interactions between PME and nanogrids are modeled by the Stackelberg game.
The existence and uniqueness of SE are analyzed and the sufficient condition is also obtained.
%We have analyzed the existence and uniqueness of the SE with providing its sufficient condition.
Furthermore, we develop an optimization algorithm which is guaranteed to reach the unique SE. The simulation results with experimental dataset have shown that the proposed pricing scheme and energy management strategies can improve the economic utility for both parties involved and without affecting the satisfaction of residents compared with naive strategies.
% or

\appendices
\section{Formulation process of \textbf{P3}}\label{App0}
With the aforementioned optimization technique described in Section~\ref{sec3}, we can formulate the following problem \textbf{P5},
\begin{equation}
\begin{aligned}
\min\limits_{e_{i}^{k}}\quad&\Omega_{i}^{\max}+{{\varepsilon}_{i}}(1-{{\varepsilon}_{i}})H_{i}^{k}({{\Gamma}_{i}}+T_{i,out}^{k}+\eta_{i}e_{i}^{k})\\
&+{{V}_{i}}[p_{s}^{k}\cdot \max (tp_{i}^{k},0)+p_{b}^{k}\cdot \min (tp_{i}^{k},0)\\
&+{{\gamma }_{i}}{{(T_{i}^{k+1}-T_{i}^{opt,k+1})}^{2}}],\\
{\rm s.t.}\quad&\eqref{eq1}, \eqref{eq6}, \eqref{eq7}, \eqref{eq2}, \eqref{eq8},\  \forall k. \notag
\end{aligned}
\end{equation}

Incorporating \eqref{eq1} and \eqref{eq2} into the above objective function,
%we can note that %it is equivalent to optimizing
it is simplified as
the objective $UN_{i}^{'}$ in \textbf{P3} since $\Omega_{i}^{\max}$ is a constant, and $T_{i}^{k}$, $H_{i}^{k}$ $T_{i,out}^{k}$, $T_{i}^{opt,k+1}$ have been specified at the beginning of each slot.
Besides, \eqref{eq7} and \eqref{eq8} are integrated into a new constraint as shown in \eqref{eq24}.
In addition, the time-coupling constraint \eqref{eq6} is met automatically under the framework of virtual queue as demonstrated in Theorem~\ref{th1}. Hence, \eqref{eq6} can be omitted here.
Finally, we conclude that \textbf{P5} is equivalent to \textbf{P3}.
%%\section{Proof of Theorem \ref{th11}}
%%\vspace{-0.1cm}
%\subsection{Proof of Theorem \ref{th11}} \label{App1}
\section{Proof of Theorem \ref{th11}}\label{App1}
The proving process is mainly divided into two parts.

In the first part, we prove
the former two results in \eqref{eq36}
by using reduction to absurdity.
Denote $\{e_{i}^{k,*}, tp_{i}^{k,*}\}$ by the optimal solution to \textbf{P3} with optimal objective value $\Omega_{1}^{*}$. We first suppose $e_{i}^{k,*}\!>\!0$ if $V_{i}p_{b}^{\min}\!>\!-\varepsilon_{i}(1\!-\!\varepsilon_{i})H_{i}^{k}\eta_{i}\!-\!\alpha_{i}^{k}$ is satisfied. Then we construct another solution $\{e_{i}^{k,\dagger}, tp_{i}^{k,\dagger}\}$ with objective value $\Omega_{2}^{\dagger}$, where $e_{i}^{k,\dagger}\!=\!0$ and $tp_{i}^{k,\dagger}\!=\!tp_{i}^{k,*}\!-\!e_{i}^{k,*}$ according to \eqref{eq2}. Obviously, there exists $\Omega_{1}^{*}\!-\!\Omega_{2}^{\dagger}\!<\!0$.
Next, we will discuss the optimal solution from three situations.

\emph{1):} When $tp_{i}^{k,*}\!<\!0$,
%based on $tp_{i}^{k,\dagger}\!=\!tp_{i}^{k,*}\!-\!e_{i}^{k,*}$ and $e_{i}^{k,*}\!>\!0$
%we have $tp_{i}^{k,\dagger}\!<\!0$.
based on~\eqref{eq23}, we derive $\Omega_{1}^{*}-\Omega_{2}^{\dagger}=V_{i}\gamma_{i}(1-\varepsilon_{i})^{2}{\eta_{i}}^{2}(e_{i}^{k,*})^{2}+[V_{i}p_{b}^{k}+\varepsilon_{i}(1-\varepsilon_{i})H_{i}^{k}\eta_{i}+2V_{i}\gamma_{i}(1-\varepsilon_{i})^{2}\eta_{i}(T_{i,out}^{k}+\tfrac{\varepsilon_{i}T_{i}^{k}-T_{i}^{opt,k+1}}{1-\varepsilon_{i}})]e_{i}^{k,*}\!>\![V_{i}p_{b}^{\min}+\varepsilon_{i}(1-\varepsilon_{i})H_{i}^{k}\eta_{i}+\alpha_{i}^{k}]e_{i}^{k,*}\!>\!0$ which conflicts with $\Omega_{1}^{*}-\Omega_{2}^{\dagger}\!<\!0$.

\emph{2):} When $tp_{i}^{k,*}\!>\!0$, $tp_{i}^{k,\dagger}\!<\!0$, we derive $\Omega_{1}^{*}\!-\!\Omega_{2}^{\dagger}\!=\!V_{i}\gamma_{i}(1-\varepsilon_{i})^{2}{\eta_{i}}^{2}(e_{i}^{k,*})^{2}\!+\![V_{i}p_{b}^{k}+\varepsilon_{i}(1\!-\!\varepsilon_{i})H_{i}^{k}\eta_{i}\!+\!2V_{i}\gamma_{i}(1\!-\!\varepsilon_{i})^{2}\eta_{i}(T_{i,out}^{k}\!+\!\tfrac{\varepsilon_{i}T_{i}^{k}\!-\!T_{i}^{opt,k+1}}{1-\varepsilon_{i}})]e_{i}^{k,*}\!+\!V_{i}(p_{s}^{k}-p_{b}^{k})tp_{i}^{k,*}\!>\![V_{i}p_{b}^{\min}+\varepsilon_{i}(1-\varepsilon_{i})H_{i}^{k}\eta_{i}+\alpha_{i}^{k}]e_{i}^{k,*}\!>\!0$
%$\Omega_{1}^{*}\!-\!\Omega_{2}^{\dagger}\!>\![V_{i}p_{b}^{\min}+\varepsilon_{i}(1-\varepsilon_{i})H_{i}^{k}\eta_{i}\!+\!\alpha_{i}^{k}]e_{i}^{k,*}\!>\!0$,
which contradicts with $\Omega_{1}^{*}\!-\!\Omega_{2}^{\dagger}\!<\!0$.

\emph{3):} When $tp_{i}^{k,*}\!>\!0$, $tp_{i}^{k,\dagger}\!>\!0$,
%similarly we have
%$\Omega_{1}^{*}\!-\!\Omega_{2}^{\dagger}\!>\!~0$.
we derive $\Omega_{1}^{*}\!-\!\Omega_{2}^{\dagger}\!=\!V_{i}\gamma_{i}(1\!-\!\varepsilon_{i})^{2}{\eta_{i}}^{2}(e_{i}^{k,*})^{2}+[V_{i}p_{s}^{k}+\varepsilon_{i}(1-\varepsilon_{i})H_{i}^{k}{\eta_{i}}\!+\!2V_{i}\gamma_{i}(1-\varepsilon_{i})^{2}{\eta_{i}}(T_{i,out}^{k}\!+\!\tfrac{\varepsilon_{i}T_{i}^{k}-T_{i}^{opt,k+1}}{1-\varepsilon_{i}})]e_{i}^{k,*}\!>\![V_{i}p_{b}^{\min}\!+\!\varepsilon_{i}(1\!-\!\varepsilon_{i})H_{i}^{k}{\eta_{i}}\!+\!\alpha_{i}^{k}]e_{i}^{k,*}\!>\!0$ which contradicts with $\Omega_{1}^{*}\!-\!\Omega_{2}^{\dagger}\!<\!0$.
%It also contradicts with $\Omega_{1}^{*}-\Omega_{2}^{\dagger}<0$.

\textbf{Conclusion:}  When  %$V_{i}p_{b}^{\min}>-\varepsilon_{i}(1-\varepsilon_{i})H_{i}^{k}{\eta_{i}}-\alpha_{i}^{k}$,
$V_{i}p_{b}^{\min}\!>\!-\varepsilon_{i}(1-\varepsilon_{i})H_{i}^{k}{\eta_{i}}-\alpha_{i}^{k}$,
%the optimal strategy of nanogrid $i$ is
$e_{i}^{k,*}\!=\!0$.

In a similarly way, we can also prove that if %$V_{i}p_{s}^{\max}<-\varepsilon_{i}(1-\varepsilon_{i})H_{i}^{k}\tfrac{\eta_{i}}{A_{i}}-\beta_{i}^{k}$, $e_{i}^{k,*}=e_{i}^{max}$.
$V_{i}p_{s}^{\max}\!<\!-\varepsilon_{i}(1\!-\!\varepsilon_{i})H_{i}^{k}{\eta_{i}}\!-\!\beta_{i}^{k}$ is satisfied, $e_{i}^{k,*}\!=\!e_{i}^{max}$.

%\begin{figure}[!t]
%\centering
%\includegraphics[width=3.16in]{fig6a.eps}
%\caption{Analysis of curve trend related with nanogrid function.}
%\label{fig6:mini:subfig}
%\end{figure}
%In the second part, several possible scenarios are discussed by combining with the pricing condition \eqref{eq14}.
In the second part, several possible scenarios are discussed with~\eqref{eq14}.

%To begin with, we can further simplify \eqref{eq23} when $tp_{i}^{k}<0$,
%and its first derivative with regard to $e_{i}^{k}$ is derived as follows
%    \begin{equation}
%    \begin{aligned}
%    \tfrac{\delta UN_{i}^{'}}{\delta
%    e_{i}^{k}}=&\varepsilon_{i}(1-\varepsilon_{i}){H_{i}^{k}}{\eta_{i}}+V_{i}{p_{b}^{k}}+2V_{i}{\gamma_{i}}(1-\varepsilon_{i}){{\eta_{i}}}\\
%    &[(1-\varepsilon_{i})(T_{i,out}^{k}+{\eta_{i}})+\varepsilon_{i}{T_{i}^{k}}-{T_{i}^{opt}}].
%    \end{aligned}
%    \label{eq43}
%    \end{equation}
To begin with, we further simplify \eqref{eq23} when $tp_{i}^{k}\!<\!0$.
Its first derivative with regard to $e_{i}^{k}$ is derived as $\tfrac{\delta UN_{i}^{'}}{\delta
    e_{i}^{k}}=\varepsilon_{i}(1-\varepsilon_{i}){H_{i}^{k}}{\eta_{i}}+V_{i}{p_{b}^{k}}+2V_{i}{\gamma_{i}}(1-\varepsilon_{i}){{\eta_{i}}}
    [(1-\varepsilon_{i})(T_{i,out}^{k}+{\eta_{i}})+\varepsilon_{i}{T_{i}^{k}}-{T_{i}^{opt,k+1}}]$.
 By setting it equal to zero, we obtain the best-response function of nanogrid $i$ as ${e_{i\_1}^{k,*}}=\vartheta_{i}^{k}-\tfrac{p_{b}^{k}}{2\gamma_{i}{(1-\varepsilon_{i})^{2}}{{\eta_{i}}^{2}}}$, where ${\vartheta_{i}^{k}}=\tfrac{T_{i}^{opt,k+1}-\varepsilon_{i}{T_{i}^{k}}-(1-\varepsilon_{i}){T_{i,out}^{k}}}{(1-\varepsilon_{i}){\eta_{i}}}-\tfrac{\varepsilon_{i}H_{i}^{k}}{2V_{i}\gamma_{i}(1-\varepsilon_{i}){\eta_{i}}}$.

    %When $tp_{i}^{k}>0$, similarly, we obtain the first derivative with regard to $e_{i}^{k}$,
%    then by setting it equal to zero we derive
%    \begin{equation}
%    {e_{i\_2}^{k,*}}=\vartheta_{i}-\tfrac{p_{s}^{k}}{2\gamma_{i}{(1-\varepsilon_{i})^{2}}{{\eta_{i}}^{2}}}.
%    \label{eq45}
%    \end{equation}
     When $tp_{i}^{k}\!>\!0$, similarly, the corresponding best-response function of nanogrid $i$ is derived as ${e_{i\_2}^{k,*}}\!=\!\vartheta_{i}^{k}\!-\!\tfrac{p_{s}^{k}}{2\gamma_{i}{(1-\varepsilon_{i})^{2}}{{\eta_{i}}^{2}}}$.
    Then, according to \eqref{eq14}, we have ${e_{i\_1}^{k,*}}>{e_{i\_2}^{k,*}}$.
    %Based on this, the discussion can be divided into three categories with the corresponding trend curves as shown in Fig.~\ref{fig6:mini:subfig}.

\begin{figure}[!t]
\centering
\includegraphics[width=3.39in]{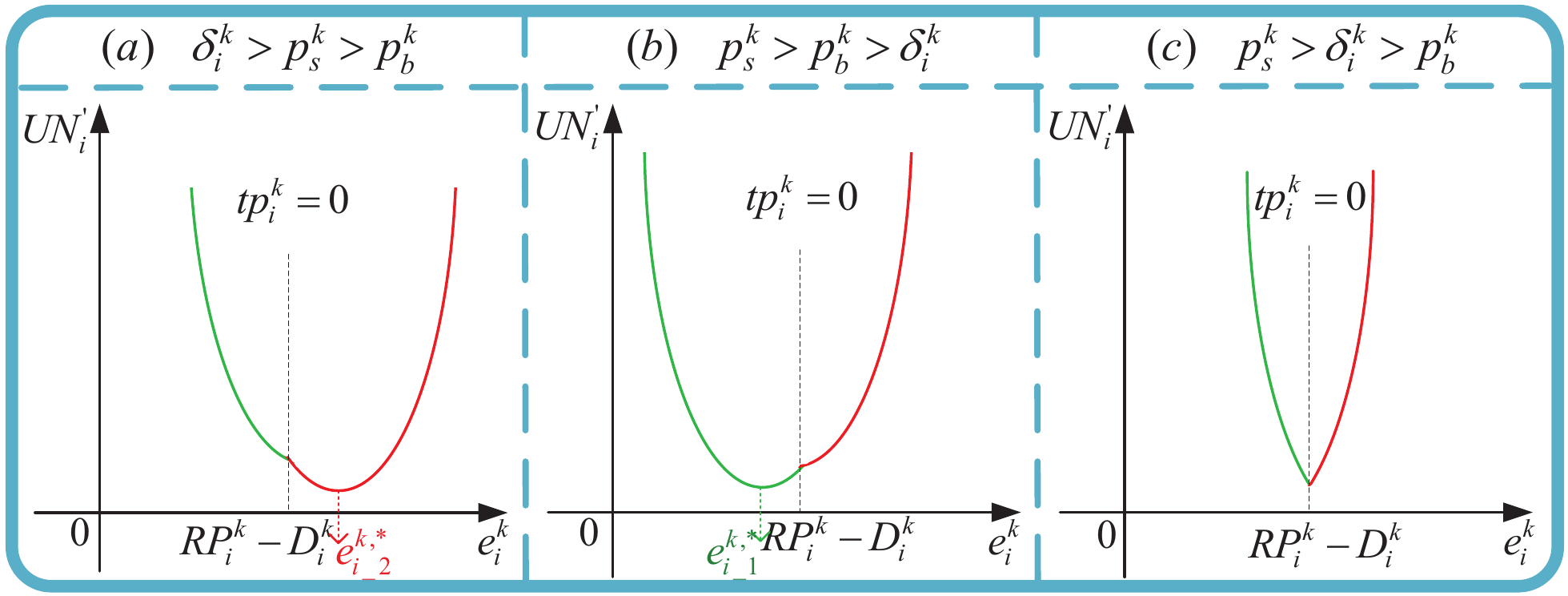}
\caption{Analysis of curve trend related with nanogrid function.}
\label{fig6:mini:subfig}
\vspace{-0.2cm}
\end{figure}

    Based on this, Fig.~\ref{fig6:mini:subfig} has provided insight into the
trend curves that the nanogrid function might be under several situations.
%The corresponding discussion is divided into the following three categories.
The corresponding discussions are given as follows.

In the first curve, corresponding to the scenario $\delta_{i}^{k}>p_{s}^{k}>p_{b}^{k}$ where $\delta_{i}^{k}=2\gamma_{i}(1-\varepsilon_{i}){\eta_{i}}[T_{i}^{opt,k+1}-\varepsilon_{i}{T_{i}^{k}}-(1-\varepsilon_{i}){T_{i,out}^{k}}]-\tfrac{\varepsilon_{i}(1-\varepsilon_{i}){H_{i}^{k}}{\eta_{i}}}{V_{i}}-2\gamma_{i}{(1-\varepsilon_{i})^{2}}{{\eta_{i}}^{2}}(RP_{i}^{k}-D_{i}^{k})$,
we have ${e_{i\_1}^{k,*}}>{e_{i\_2}^{k,*}}>RP_{i}^{k}-D_{i}^{k}$. The minimal point is located at ${e_{i\_2}^{k,*}}$.

For the second curve related to the scenario $p_{s}^{k}\!>\!p_{b}^{k}\!>\!\delta_{i}^{k}$, we derive $RP_{i}^{k}-D_{i}^{k}\!>\!{e_{i\_1}^{k,*}}\!>\!{e_{i\_2}^{k,*}}$. The minimal point is located at ${e_{i\_1}^{k,*}}$.

In the last scenario $p_{s}^{k}\!>\!\delta_{i}^{k}\!>\!p_{b}^{k}$, we have ${e_{i\_1}^{k,*}}\!>\!RP_{i}^{k}\!-\!D_{i}^{k}\!>\!{e_{i\_2}^{k,*}}$. The minimal point is located between ${e_{i\_2}^{k,*}}$ and ${e_{i\_1}^{k,*}}$.

    Specifically, the optimal value selection of $f({\bm{\chi}}^{k})$ is determined by the relationship of
    %$\delta_{i}^{k}$, $p_{s}^{k}$, $p_{b}^{k}$,
    ${e_{i\_1}^{k,*}}$, ${e_{i\_2}^{k,*}}$,
    $RP_{i}^{k}$ and $D_{i}^{k}$.
\section{Proof of Theorem \ref{th3}} \label{App4}
The proof process is divided into three stages which correspond to the steps 1-3 in Theorem \ref{th3}, respectively.

\emph{1):}
Obviously, the strategy sets of PME and nanogrids given in \eqref{eq10}, \eqref{eq14} and \eqref{eq24} satisfy
the requirement.

\emph{2):}
It is sufficient to point out that the optimization problem of nanogrid defined in this work always has only one best-response given the strategy of PME according to the discussion of Theorem~\ref{th11}.

\emph{3):}
At last, we need to prove that the strategy of PME is also unique once informed all the nanogrids' best-response strategies. For the optimization problem of PME, it is analyzed in two situations (i.e., ${\sum\nolimits_{i=1}^{n}{tp_{i}^{k}}}\!-\!G_{T}^{k}+y^{k}\!>\!0$ and ${\sum\nolimits_{i=1}^{n}{tp_{i}^{k}}}\!-\!G_{T}^{k}+y^{k}\!<\!0$) to further simplify its objective function \eqref{eq31}.

    For the situation ${\sum\nolimits_{i=1}^{n}{tp_{i}^{k}}}\!-\!G_{T}^{k}\!+\!y^{k}\!>\!0$, we assume that the nanogrid quantity corresponding to scenario (a)-(c) in Fig.~\ref{fig6:mini:subfig} is $x$, $y$ and $z$, respectively. Then $pro^{'}$ can be reduced as
    \begin{equation}
    {\min}_{p_{s}^{k}, p_{b}^{k}, y^{k}}\ \; pro_{1}^{'}+pro_{2}^{'},
    \label{eq50}
    \end{equation}
    where $pro_{1}^{'}\!=\!V_{P}m_{s}^{k}(\sum\nolimits_{i=1}^{x}{tp_{i}^{k}}\!+\!\sum\nolimits_{j=1}^{y}{tp_{j}^{k}})\!-\!V_{P}(\sum\nolimits_{i=1}^{x}{p_{s}^{k}tp_{i}^{k}}+\sum\nolimits_{j=1}^{y}{p_{b}^{k}tp_{j}^{k}})$, $pro_{2}^{'}\!=\!(B^{k}\!+\!V_{P}m_{s}^{k})y^{k}\!+\!\tfrac{1}{2}V_{P}C_{b}(y^{k})^{2}$.

Note that $pro_{2}^{'}$ is easy to solve. Specially, if $B^{k}\!+\!V_{P}m_{s}^{k}\!\geq\!V_{P}C_{b}u^{\rm dmax}$, its solution satisfies $y^{k,*}\!=\!-u^{\rm dmax}$; and if $B^{k}\!+\!V_{P}m_{s}^{k}\!\leq\!\!-\!V_{P}C_{b}u^{\rm cmax}$, $y^{k,*}\!=\!u^{\rm cmax}$; otherwise, $y^{k,*}\!=\!(B^{k}\!+\!V_{P}m_{s}^{k})/(\!-\!V_{P}C_{b})$.

  Incorporating \eqref{eq333} with the discussion results of scenarios (a)-(c) in Fig.\ref{fig6:mini:subfig}, we can reformulate $tp_{i}^{k}$ and $tp_{j}^{k}$ in $pro_{1}^{'}$. Then we derive the first-order partial derivative of $pro_{1}^{'}$ with respect to $p_{s}^{k}$ and $p_{b}^{k}$
%as follows,
%\begin{equation}
%  \resizebox{0.894\hsize}{!}{$\tfrac{\delta pro_{1}^{'}}{\delta
%  p_{s}^{k}}=-V_{P}{\sum\nolimits_{i=1}^{x}[\vartheta_{i}-\tfrac{p_{s}^{k}}{\gamma_{i}{(1-\varepsilon_{i})^{2}}{{\eta_{i}}^{2}}}+D_{i}^{k}-RP_{i}^{k}+\varrho_{i}^{k}]},$}
%  \label{eq53}
%  \end{equation}
%  \begin{equation}
%  \resizebox{0.8941\hsize}{!}{$\tfrac{\delta pro_{1}^{'}}{\delta
%  p_{b}^{k}}=-V_{P}{\sum\nolimits_{j=1}^{y}[\vartheta_{j}-\tfrac{p_{b}^{k}}{\gamma_{j}{(1-\varepsilon_{j})^{2}}{{\eta_{j}}^{2}}}+D_{j}^{k}-RP_{j}^{k}+\varrho_{j}^{k}]},$}
%  \label{eq54}
%  \end{equation}
as $\tfrac{\delta pro_{1}^{'}}{\delta
  p_{s}^{k}}\!=\!-V_{P}{\sum\nolimits_{i=1}^{x}[\vartheta_{i}^{k}\!-\!\tfrac{p_{s}^{k}}{\gamma_{i}{(1\!-\!\varepsilon_{i})^{2}}{{\eta_{i}}^{2}}}\!+\!D_{i}^{k}\!-\!RP_{i}^{k}\!+\!\varrho_{i}^{k}]}$ and
$\tfrac{\delta pro_{1}^{'}}{\delta p_{b}^{k}}\!=\!-V_{P}{\sum\nolimits_{j=1}^{y}[\vartheta_{j}^{k}\!-\!\tfrac{p_{b}^{k}}{\gamma_{j}{(1\!-\!\varepsilon_{j})^{2}}{{\eta_{j}}^{2}}}\!+\!D_{j}^{k}\!-\!RP_{j}^{k}\!+\!\varrho_{j}^{k}]}$,
where $\varrho_{i}^{k}\!=\!\tfrac{m_{s}^{k}}{2\gamma_{i}{(1-\varepsilon_{i})^{2}}{{\eta_{i}}^{2}}}$.

  And then, the Hessian matrix of $pro_{1}^{'}$ is given as
  \begin{equation}
  \resizebox{0.88\hsize}{!}{$\text{Hess} (pro_{1}^{'})=
\left[ \begin{matrix}
   {{V}_{P}}\sum\nolimits_{i=1}^{x}{\frac{1}{{{\gamma }_{i}}{{(1-{{\varepsilon }_{i}})}^{2}}{{\eta_{i}}^{2}}}} & 0  \\
   0 & {{V}_{P}}\sum\nolimits_{j=1}^{y}{\frac{1}{{{\gamma }_{j}}{{(1-{{\varepsilon }_{j}})}^{2}}{{\eta_{j}}^{2}}}}  \\
\end{matrix} \right].\ $}
  \label{eq56}
  \end{equation}

  Evidently
  %$\text{Hess} (pro_{1}^{'})$
  %it is a positive definite matrix
  $\text{Hess}(pro_{1}^{'})$ is positive definite
  and there is only one minimum point. So, PME has a unique optimal strategy.

  In a similarly way, we can prove that the optimal strategy of PME is also unique under the condition of ${\sum\nolimits_{i=1}^{n}{tp_{i}^{k}}}\!-\!G_{T}^{k}+y^{k}\!<\!0$.
This completes the proof of Theorem \ref{th3}.
%\subsection{Proof of Theorem \ref{th1}} \label{App2}
\section{Proof of Theorem \ref{th1}} \label{App2}
We prove Theorem \ref{th1} by mathematical induction with assumption
%that the initial temperature
of $T_{i}^{0}\in[T_{i}^{\min}, T_{i}^{\max}]$. To begin with, for slot~$k$, we suppose that $T_{i}^{\min }\!\le\!T_{i}^{k}\!\le\!T_{i}^{\max }$ holds under the given parameters, then we show it still holds in the next slot $k\!+\!1$. Specifically, the proof is divided into the following three cases.

\emph{1):}
When $T_{i}^{\min}\!\leq\!T_{i}^{k}\!\leq\! {(V_{i}p_{s}^{\max}\!+\!\beta_{i}^{k})/[-\varepsilon_{i}(1\!-\!\varepsilon_{i}){\eta_{i}}]}\!-\!\Gamma_{i}$, based on Theorem~\ref{th11}, we obtain $e_{i}^{k,*}\!=\!e_{i}^{\max}$. Then for slot $k\!+\!1$,
with assumption (b) i.e., $\eta_{i}e_{i}^{\max}\!+\!T_{i,out}^{\min}\!\geq\!T_{i}^{\min}$,
we have $T_{i}^{k+1}\!\geq\! \varepsilon_{i}T_{i}^{\min}\!+\!(1\!-\!\varepsilon_{i})(T_{i,out}^{\min}\!+\!{\eta_{i}}e_{i}^{\max})\!\geq\! T_{i}^{\min}$, and similarly, $T_{i}^{k+1}\!\leq\! \varepsilon_{i}[{\tfrac{V_{i}p_{s}^{\max}\!+\!\beta_{i}^{k}}{-\varepsilon_{i}(1-\varepsilon_{i}){\eta_{i}}}}\!-\!\Gamma_{i}]\!+\!(1-\varepsilon_{i})(T_{i,out}^{\max}+{\eta_{i}}e_{i}^{\max})\!\leq\! T_{i}^{\max}$, where $\Gamma_{i}^{\min}$ is shown in~\eqref{eq37}.

\emph{2):}
When ${(V_{i}p_{b}^{\min}\!+\!\alpha_{i}^{k})/[-\varepsilon_{i}(1\!-\!\varepsilon_{i}){\eta_{i}}]}\!-\!\Gamma_{i}\!\leq\! T_{i}^{k}\!\leq\!T_{i}^{\max}$, $e_{i}^{k,*}\!=\!0$. Then for slot $k\!+\!1$, with assumption (a) i.e., $T_{i,out}^{\max}\!\leq\!T_{i}^{\max}$, we have $T_{i}^{k+1}\!\leq\! \varepsilon_{i}T_{i}^{\max}\!+\!(1\!-\!\varepsilon_{i}T_{i,out}^{\max})\!\leq\!T_{i}^{\max}$. $T_{i}^{k+1}\!\geq\! \varepsilon_{i}[{\tfrac{V_{i}p_{b}^{\min}\!+\!\alpha_{i}^{k}}{-\varepsilon_{i}(1-\varepsilon_{i}){\eta_{i}}}}\!-\!\Gamma_{i}]\!+\!(1-\varepsilon_{i})T_{i,out}^{\min}\!\geq\! T_{i}^{\min}$ is derived with $\Gamma_{i}^{\max}$ shown as~\eqref{eq38}.

\emph{3):}
%When ${\tfrac{V_{i}p_{s}^{\max}+\beta_{i}^{k}}{-\varepsilon_{i}(1-\varepsilon_{i}){\eta_{i}}}}-\Gamma_{i}\leq T_{i}^{k}\leq {\tfrac{V_{i}p_{b}^{\min}+\alpha_{i}^{k}}{-\varepsilon_{i}(1-\varepsilon_{i}){\eta_{i}}}}-\Gamma_{i}$,
Otherwise, we get $T_{i}^{\min}\!\leq\! \varepsilon_{i}[{\tfrac{V_{i}p_{s}^{\max}+{\max_{k}}\beta_{i}^{k}}{-\varepsilon_{i}(1-\varepsilon_{i}){\eta_{i}}}}\!-\!\Gamma_{i}]\!+\!(1\!-\!\varepsilon_{i}){T_{i,out}^{\min}}\!\leq\! T_{i}^{k+1}\!\leq\! \varepsilon_{i}[{\tfrac{V_{i}p_{b}^{\min}+{\min_{k}}\alpha_{i}^{k}}{-\varepsilon_{i}(1-\varepsilon_{i}){\eta_{i}}}}\!-\!\Gamma_{i}]\!+\!(1\!-\!\varepsilon_{i})(T_{i,out}^{\max}\!+\!{\eta_{i}}e_{i}^{max})\!\leq\! T_{i}^{\max}$ where we can obtain the upper bound of $V_{i}$,and assumption (c) i.e., $T_{i}^{\max}\!-\!T_{i}^{\min}\!>\!(1\!-\!\varepsilon_{i})(T_{i,out}^{\max}\!+\!\eta_{i}e_{i}^{\max}\!-\!T_{i,out}^{\min})$,
is incorporated to guarantee $V_{i}^{\max}\!>\!0$.
%\subsection{Proof of Theorem \ref{th2}} \label{App3}
\section{Proof of Theorem \ref{th2}} \label{App3}
%\vspace{-0.1cm}
%\section{}
%Theorem \ref{th2} can be proved by mathematical induction. First, we assume that the initial battery energy level $E^{0} \in [E^{\min}, E^{\max}]$, i.e., for slot $k$, the inequality $E^{\min}\leq E^{k}\leq E^{\max}$ holds, then we just need to verify that it still holds in the next slot. In detail, the proof is composed of three situations.
Based on mathematical induction, firstly, we assume that $E^{0}\in[E^{\min}, E^{\max}]$, i.e., for slot $k$, $E^{\min}\leq E^{k}\leq E^{\max}$ holds. Then we just need to verify that it still holds in next slot. In detail, the proof is composed of three situations.

\emph{1):} When $E^{\min}\!\leq\!E^{k}\!<\!-(V_{P}m_{s}^{k}\!+\!\theta\!+\!V_{P}\mathcal{C}^{\max})$, the partial derivative of object function in \textbf{P4} with regard to $y^{k}$ is always negative ($B^{k}\!+\!V_{P}m_{b}^{k}\!+\!V_{P}\mathcal{C}^{\min}\!<\!B^{k}\!+\!V_{P}m_{s}^{k}\!+\!V_{P}\mathcal{C}^{\max}\!<\!0$), then $y^{k,*}\!=\!u^{\rm cmax}$. %Therefore, according to \eqref{eq9},
In slot~$k\!+\!1$ we have
$E^{\min}\!\leq\!E^{k}\!<\!E^{k+1}\!<\!-(V_{P}m_{s}^{k}\!+\!\theta\!+\!V_{P}\mathcal{C}^{\max})\!+\!u^{\rm cmax}\!\leq\!E^{\max}$, where $\theta^{\min}$ is derived as shown in \eqref{eq40}.

\emph{2):} When $\!-\!(V_{P}m_{b}^{k}\!+\!\theta\!+\!V_{P}\mathcal{C}^{\min})\!<\!E^{k}\!\leq\!E^{\max}$, the partial derivative
%of the objective function in \textbf{P4}
with regard to $y^{k}$ is always positive (i.e., $B^{k}\!+\!V_{P}m_{s}^{k}\!+\!V_{P}\mathcal{C}^{\max}\!>\!B^{k}\!+\!V_{P}m_{b}^{k}\!+\!V_{P}\mathcal{C}^{\min}\!>\!0$), then
%the optimal charge-discharge value is
$y^{k,*}\!=\!-u^{\rm dmax}$.
%Therefore, according to \eqref{eq9},
In slot $k\!+\!1$ we have
%the inequality
$E^{\min}\!\leq\!-(V_{P}m_{b}^{k}\!+\!\theta\!+\!V_{P}\mathcal{C}^{\min})\!-\!u^{\rm dmax}\!<\!E^{k+1}\!<\!E^{k}\!\leq\!E^{\max}$, where $\theta^{\max}$ is derived as shown in \eqref{eq41}.

\emph{3):} When $-(V_{P}m_{s}^{k}\!+\!\theta\!+\!V_{P}\mathcal{C}^{\max})\!\leq\!E^{k}\!\leq\!-(V_{P}m_{b}^{k}\!+\!\theta\!+\!V_{P}\mathcal{C}^{\min})$, combining with \eqref{eq9} and \eqref{eq10}, for slot $k\!+\!1$, $E^{\min}\!\leq\! -(V_{P}{{\max}_{k}}m_{s}^{k}\!+\!\theta\!+\!V_{P}\mathcal{C}^{\max})-u^{\rm dmax}\!\leq\!E^{k+1}\!\leq\! -(V_{P}{{\min}_{k}}m_{b}^{k}+\theta\!+\!V_{P}\mathcal{C}^{\min})+u^{\rm cmax}\!\leq\!E^{\max}$ holds where we have used the upper limit of $V_{P}$ as shown in \eqref{eq42}.
%\section*{Acknowledgment}
%This work was supported by National Key Research and Development Program of China (2016YFB0901900), National Natural Science Foundation of China (61573245, 61174127, 61521063, U1405251, and 61633017). This work was also partially supported by Shanghai Rising-Star Program under Grant 15QA1402300 and Shanghai Municipal Commission of Economy and Informatization under SH-CXY-2016-003.

\ifCLASSOPTIONcaptionsoff
  \newpage
\fi



%\bibliographystyle{IEEEtran}
%\begin{thebibliography}{1}


%\end{thebibliography}

%\bibliographystyle{IEEEtran}
%\bibliography{mybibfile1}

%\begin{CJK*}{GBK}{song}
%\hspace*{\fill} \\%空一整行
%\noindent
%{\CJKfamily{kai} 作者简介:}\\
% {\scriptsize
%\indent
%\textbf{曹嘉馨}~~~~(1996-)，女，博士生, 目前研究方向为不确定环境下综合能源系统能量管理和交易优化, E-mail: jiaxincao@sjtu.edu.cn;\\
%\indent
%\textbf{杨~~~博}~~~~(1980-)，男，教授，博士生导师, 目前研究方向为能源互联网的优化运行与控制，工业物联网及车联网的优化分析, E-mail: bo.yang@sjtu.edu.cn.}
%\end{CJK*}
\end{document}